\documentclass[11pt]{article}
\pdfoutput=1

\usepackage{jheppub}
\usepackage{url,comment}
\usepackage{times}
\usepackage{latexsym}
\usepackage{graphicx, graphics, hyperref, amsmath, amssymb, slashed, color}
\usepackage{bbm,amsthm}
 
\newcommand{\nc}{\newcommand}

\nc{\beq}{\begin{equation}}
\nc{\eeq}{\end{equation}}
\nc{\barray}{\begin{eqnarray}}
\nc{\earray}{\end{eqnarray}}
\newcommand{\bea}{\begin{eqnarray}\begin{aligned}}
\newcommand{\eea}{\end{aligned}\end{eqnarray}}
\nc{\barrayn}{\begin{eqnarray*}}
\nc{\earrayn}{\end{eqnarray*}}
\nc{\bcenter}{\begin{center}}
\nc{\ecenter}{\end{center}}
\nc{\mc}{\mathcal}
\nc{\er}[1]{(\ref{eq:#1})}
\nc{\onehalf}{\frac{1}{2}}
\nc{\partialbar}{\bar{\partial}}
\nc{\psit}{\widetilde{\psi}}
\nc{\Tr}{\mbox{Tr}}
\nc{\hc}{\mbox{H.c.}}
\nc{\ev}{\;\mathrm{eV}}
\nc{\mev}{\;\mathrm{MeV}}
\nc{\gev}{\;\mathrm{GeV}}
\nc{\tev}{\;\mathrm{TeV}}

\newcommand{\abs}[1]{\left| #1\right|}
\newcommand{\vev}[1]{\left< #1\right>}

\newcommand{\braket}[3]{\left< #1\right| #2 \left| #3\right>}

\def\chii0{\chi_i^0}
\def\chij0{\chi_j^0}

\newcommand{\gsim}{\lower.7ex\hbox{$\;\stackrel{\textstyle>}{\sim}\;$}}
\newcommand{\lsim}{\lower.7ex\hbox{$\;\stackrel{\textstyle<}{\sim}\;$}}
\nc{\ttbar}{t\bar t}

\nc{\Lag}{\mathcal{L}}

\newcommand{\subsubsubsection}[1]{\vspace{2mm}{\noindent\bf #1 --- }}
\newcommand{\lp}{\left(}
\newcommand{\rp}{\right)}
\newcommand{\order}[1]{\mathcal{O}\lp#1\rp}

\nc{\feff}{f_{\mathrm{eff}}}
\nc{\too}{\leftrightarrow}
\nc{\infinity}{\infty}

\newcommand{\HSFOH}{HSFO-HP}
\newcommand{\HSFOV}{HSFO-VP}

\title{
  Looking for the WIMP Next Door
}

\author[a,b]{Jared A.~Evans,}
\author[b]{Stefania Gori,}
\author[a]{and Jessie Shelton}
\affiliation[a]{Department of Physics, University of Illinois at Urbana-Champaign, Urbana, IL 61801, USA}
\affiliation[b]{Department of Physics, University of Cincinnati, Cincinnati, Ohio 45221, USA}

\emailAdd{jaredaevans@gmail.com}
\emailAdd{stefania.gori@uc.edu}
\emailAdd{sheltonj@illinois.edu}

\abstract{
We comprehensively study experimental constraints and prospects for a class of minimal hidden sector dark matter (DM) models, highlighting 
how the cosmological history of these models informs the experimental signals.  We study simple `secluded' models, where the DM freezes out into  unstable dark mediator states, and consider the minimal cosmic history of this dark sector, where coupling of the dark mediator to the SM was sufficient to keep the two sectors in thermal equilibrium at early times.  In the well-motivated case where the dark mediators couple to the Standard Model (SM) via renormalizable interactions, the requirement of thermal equilibrium provides a minimal, UV-insensitive, and predictive cosmology for hidden sector dark matter. 
 We call  DM that freezes out of a dark radiation bath in thermal equilibrium with the SM a {\em  WIMP next door}, and demonstrate that the parameter space for such WIMPs next door is sharply defined, bounded, and in large part potentially accessible. This parameter space, and the corresponding signals, depend on the leading interaction between the SM and the dark mediator; we establish it for both Higgs and vector portal interactions.   In particular, there is a cosmological lower bound on the portal coupling strength necessary to thermalize the two sectors in the early universe.  We determine this thermalization floor as a function of equilibration temperature for the first time. We demonstrate that direct detection experiments are currently probing this cosmological lower bound in  some regions of parameter space, while indirect detection signals and terrestrial searches for the mediator cut further into the viable parameter space. We present regions of interest for both direct detection and dark mediator searches, including  motivated parameter space for the direct detection of sub-GeV DM. 
}

\arxivnumber{nnnn.nnnn}

\begin{document}

\maketitle

\section{Introduction}\label{sec:intro}

The existence of some form of dark matter (DM) constituting 26$\%$ of
the present-day energy budget of our universe is well-established
through its gravitational imprint on baryonic matter
\cite{Ade:2015xua}. No evidence to date indicates that DM must
interact in any way beyond gravitationally.  The cosmological history
of DM, however, will typically require DM to have some
non-gravitational interaction(s) responsible for establishing its
observed relic abundance, and these interactions can leave potentially
observable footprints. For instance, the cosmic coincidence of the
``weakly interacting massive particle (WIMP) miracle'' implies that
new stable weak-scale particles with weak interactions that freeze out
of the thermal Standard Model (SM) plasma in the early universe can
provide a good DM candidate.  The cosmic abundance of WIMPs is
directly determined by their coupling to the SM, and thus this class
of models makes sharp predictions for   
signals accessible to a variety of experiments. While the
parameter space for thermal WIMPs is now acutely limited by the
interplay of null results at direct and indirect detection experiments
and at the Large Hadron Collider (LHC) \cite{Escudero:2016gzx}, thermal
DM that freezes out directly to SM particles via new beyond-the-SM
(BSM) mediator(s) similarly has a cosmological abundance directly set
by the strength of its interactions with the SM, and has thus driven
the terrestrial DM discovery program in recent years. These models, too, are becoming increasingly challenged by the lack of signals to date 
\cite{Abercrombie:2015wmb, Alexander:2016aln}.

This class of thermal relics, however, represents only a fraction of
possible identities for dark matter. Hidden sector freezeout (HSFO
\cite{Pospelov:2007mp, Feng:2008ya, Feng:2008mu,ArkaniHamed:2008qn}),
where the DM relic abundance is chiefly determined by interactions
internal to a thermal dark sector with little to no involvement of the
SM, provides a much broader class of models.  In this paper, we survey
the current constraints and future discovery prospects in the simplest
exemplars of hidden sector freezeout.  In these simple and minimal
models, DM is a thermal relic that annihilates not to SM states, but
to pairs of dark {\em mediators} that subsequently decay via small
couplings into the SM.  We take these small couplings to be the
leading interaction between the HS and the SM, and consider the
well-motivated and generic case where this interaction is
renormalizable.

Any theory where DM arises from an internally thermalized dark sector
must also address the question: how was this dark sector populated in
the early universe?  The most minimal cosmological history for a dark
sector is for it to interact strongly enough with the SM that the two
sectors were in thermal equilibrium at early times.  In this case, the
existence of a thermal SM plasma in the early universe guarantees the
population of the dark sector.  We call DM that freezes out from a
thermal dark radiation bath in thermal equilibrium with the SM a {\em
  WIMP next door}.  Mandating this cosmological history for the dark
sector imposes a {\it lower bound} on the interactions between the
dark sector and the SM today, the thermalization floor.  The parameter 
space for WIMPs next door is bounded: the DM mass must lie between 
$\sim 1 \mev$ (to preserve the successful predictions of Big Bang 
Nucleosynthesis (BBN) and $\sim \mathrm{few}
\tev$ (from perturbative unitary), while the coupling between the SM
and the HS must be sufficiently strong to thermalize the dark sector
with the SM prior to DM freezeout.

The aim of this paper is to establish this bounded parameter space for
two minimal models of HS freezeout and systematically map out how this
parameter space can be tested in indirect detection, accelerator, and
direct detection through a variety of experiments spanning the cosmic,
energy, and intensity frontiers.  The characteristic signatures of
hidden sector freezeout are largely dictated by the Lorentz quantum
numbers of the DM and the mediator, together with the choice of portal
operator.  We focus here on dark sectors which have a leading
renormalizable coupling with the SM, through either the vector portal
interaction,
\beq
 \frac{\epsilon}{2\cos\theta} \, Z_ {D\mu\nu}\, B^{\mu\nu},
 \label{eq:VP}
\eeq
or the Higgs portal interaction,
\beq
\frac{\epsilon}{2} S^2 |H|^2 .
 \label{eq:HP}
\eeq
We will use two simple reference models in this work,
\begin{itemize}
\item {\em \HSFOV:} fermionic DM $\chi$, annihilating to vector
  mediators, $Z_D$, that couple to the SM through the vector portal; and
\item {\em \HSFOH:} fermionic DM $\chi$, annihilating to scalar
  mediators $s$ that couple to the SM through the Higgs portal.
\end{itemize}
These models of HSFO can be probed via complementary methods across
different experimental frontiers.  Direct searches for the dark
mediator are the most sensitive test at accelerator-based experiments,
far outpacing more traditional collider searches for DM that rely on a
missing energy signature.  Direct detection experiments can access the
cosmological lower bound on the portal coupling in significant
portions of the parameter space. Indirect detection remains a powerful
probe, provided the DM has an appreciable $s$-wave annihilation
cross-section, as in our minimal vector portal model. Our minimal
Higgs portal model, on the other hand, freezes out through $p$-wave
interactions, placing traditional cosmic ray signals largely out of
reach.  The constraints on our simple reference models provide a
reasonably general guide to the physics of more complicated hidden
sectors, as we discuss below.

We begin with a discussion of WIMPs next door in
Sec.~\ref{sec:general}, where we establish the physical parameter
space of our models. 
 In Sec.~\ref{sec:expt} we discuss different
experimental avenues to test this parameter space.  In
Secs.~\ref{sec:vv} and~\ref{sec:ss} we show the consequences for
vector and Higgs portal models respectively, and in
Sec.~\ref{sec:conclusions} we summarize our results.
Three Appendices describe details of our
calculations of thermal scattering rates, Sommerfeld
enhancements, and bounds from dwarf galaxies.

\section{Parameter Space for Minimal Hidden Sector Freezeout and the WIMP Next Door}
\label{sec:general}

In hidden sector freezeout, DM is part of a larger dark sector that is
thermally populated in the early universe.  As the universe expands
and cools, the relic abundance of DM is determined by the freezeout of
its annihilations to a dark {\em mediator} state,
$\chi\chi\to\phi\phi$, with little to no involvement of SM particles
\cite{Pospelov:2007mp, Feng:2008ya, Feng:2008mu, ArkaniHamed:2008qn}.
In the simplest realizations of hidden sector freezeout, these dark
mediators, $\phi$, are cosmologically unstable, decaying into the SM
through a small coupling.  These decays must occur sufficiently
rapidly to avoid disrupting the successful predictions of BBN, thus generally requiring $\tau_\phi\lesssim 1$
s, and providing a cosmological lower bound on the strength of the
coupling of the mediators to the SM.  When the interaction that allows
$\phi$ to decay to the SM is the leading interaction between the two
sectors, it will additionally control the thermalization of the dark
sector and the SM in the early universe.

Requiring the SM and the dark sector to be in thermal equilibrium
prior to DM freezeout is the simplest and most minimal cosmology for
the origin of the dark sector.  DM freezing out from a thermal dark
radiation bath in equilibrium with the SM radiation bath is what we
define as a {\em WIMP next door}. We will focus here on the
well-motivated cases where the vector or scalar portal operators
(\ref{eq:VP}--\ref{eq:HP}) mediate the leading interactions between
sectors, and establish the observable consequences.  
If the portal coupling $\epsilon$ is sufficiently large to ensure that
the SM and the dark sector were in thermal equilibrium for some
temperature above the DM freezeout temperature, $T_{eq}>T_f$, the
existence of the SM thermal bath is then sufficient to guarantee the
population of the dark sector.  If, on the other hand, the portal
interactions cannot thermalize the two sectors prior to DM freezeout,
then some other mechanism, such as asymmetric reheating
\cite{Hodges:1993yb, Berezhiani:1995am, Adshead:2016xxj}, must be
invoked to populate the dark thermal bath in the early universe.

When the leading interaction between sectors is renormalizable, this
minimal cosmology is additionally {\em UV--insensitive}: the
scattering rates controlling thermalization obey $\Gamma\propto T$,
and become more important in comparison with $H\propto T^2/M_{Pl}$ as
the temperature drops.  Thus $\epsilon_{min}(T_{eq})$, the minimum
value of $\epsilon$ consistent with thermalization at a temperature
$T_{eq}$, does not depend on the unknown reheating temperature of the
universe (provided $T_{RH}>T_{eq}$) or other unknown UV particle
content.  This cosmic origin for DM also significantly sharpens
predictivity by limiting the degree to which the temperature of the
dark sector can differ from the temperature of the SM.

In order to determine the thermalization floor $\epsilon_{min}(T_{eq})$, we have to distinguish
between two cases: first, when the hidden sector contains (at least)
one relativistic species at $T_{eq}$, and second, when {\em all}
species in the hidden sector are already nonrelativistic at
thermalization, $T_{eq}< m/2.46$ for all masses.  The value of
$T=m/2.46$ is the point where a bosonic species contribution to $g_*$ drops by
a factor of 2, and is our definition for when a species transitions 
from relativistic to non-relativistic.  In the first case, the energy 
in the hidden sector radiation bath is the same per degree of freedom
as in the SM, and thermalization requires that inter-sector reactions
are efficient enough to transfer a sizable amount of energy per SM
degree of freedom.  In the second case, all hidden sector species have
exponentially suppressed number densities at $T_{eq}$, and the energy
that must be transferred from the SM to thermally populate the hidden
sector is thus exponentially reduced.  The resulting bounds on minimal
coupling strengths are correspondingly much weaker.

    \begin{figure}[t]
\begin{center}
\raisebox{0.1cm}{\includegraphics[scale=0.58]{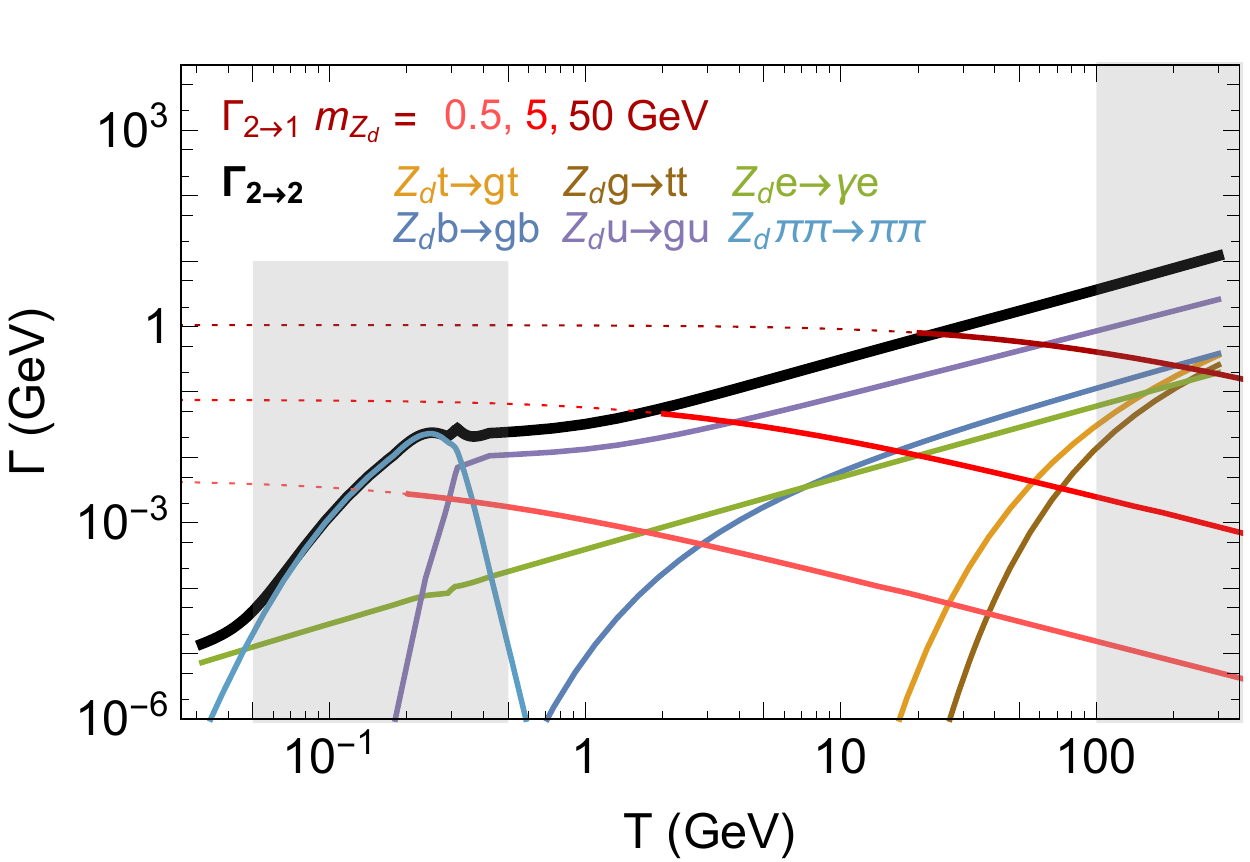}}~~~~
\includegraphics[scale=0.60]{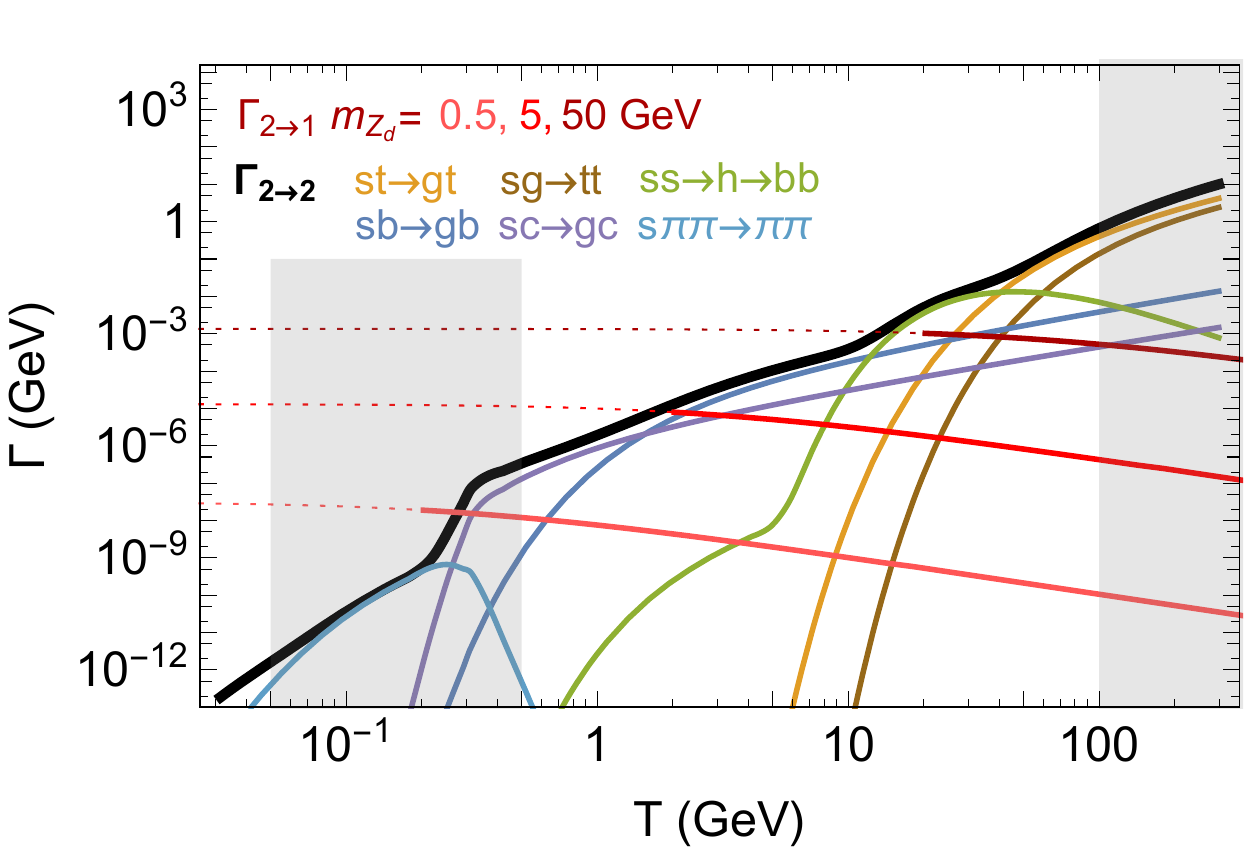}
\end{center}
\caption{Rates for $2\too 2$ (in black) vs.~$2\too 1$ transitions (in
  different shades of red depending on the mediator mass) as a function of the temperature. Dotted
  lines correspond to the $2\too 1$ rates after the mediator becomes
  non-relativistic, $T < m / 2.46$, with $m$ the mediator mass.  The
  gray region at low temperatures corresponds to the 
  uncertain regions near $\Lambda_{QCD}\equiv 300$ MeV, above which
  $\alpha_s$ becomes large, and the leading order calculation is unreliable, 
  and below which $3\too 2$ pion processes (included in the $2\too2$ 
  rate, shown in light blue) dominate the equilibration. The gray region 
  at high temperature is near the electroweak phase transition $T_c=160$ 
  GeV; see Appendix \ref{sec:KD} for more details.  {\bf Left:} 
  HSFO-VP.  In this model the $2\too 2$ scattering rate is nearly linear 
  with temperature above the chiral phase transition.  After pion processes
  become ineffective and QED processes dominate, the scaling is nearly linear 
  again.  {\bf Right:} HSFO-SP.  In this model the $2\too 2$ rate is more 
  sensitive to mass thresholds.  It drops sharply after the chiral phase 
  transition (kaon processes have been neglected).}
\label{fig:rateexample}
\end{figure}

We focus here on the first case where the HS has a radiation bath at
$T_{eq}$.  In this cosmology the lower bound of the thermalization floor
is typically far more stringent than the lower bound from requiring
mediator decays to occur prior to BBN.
We require that the two sectors thermalize at least at $T_{eq}=T_f$.
For simplicity, we consider minimal models that consist only
of a dark matter species $\chi$ and a dark mediator $\phi$.  In order
to have a dark radiation bath at DM freezeout, we thus require the
mediator to have $m_\phi < 2.46 \,T_f$.  

\begin{figure}[!t]
\begin{center}
\raisebox{0.1cm}{\includegraphics[scale=0.58]{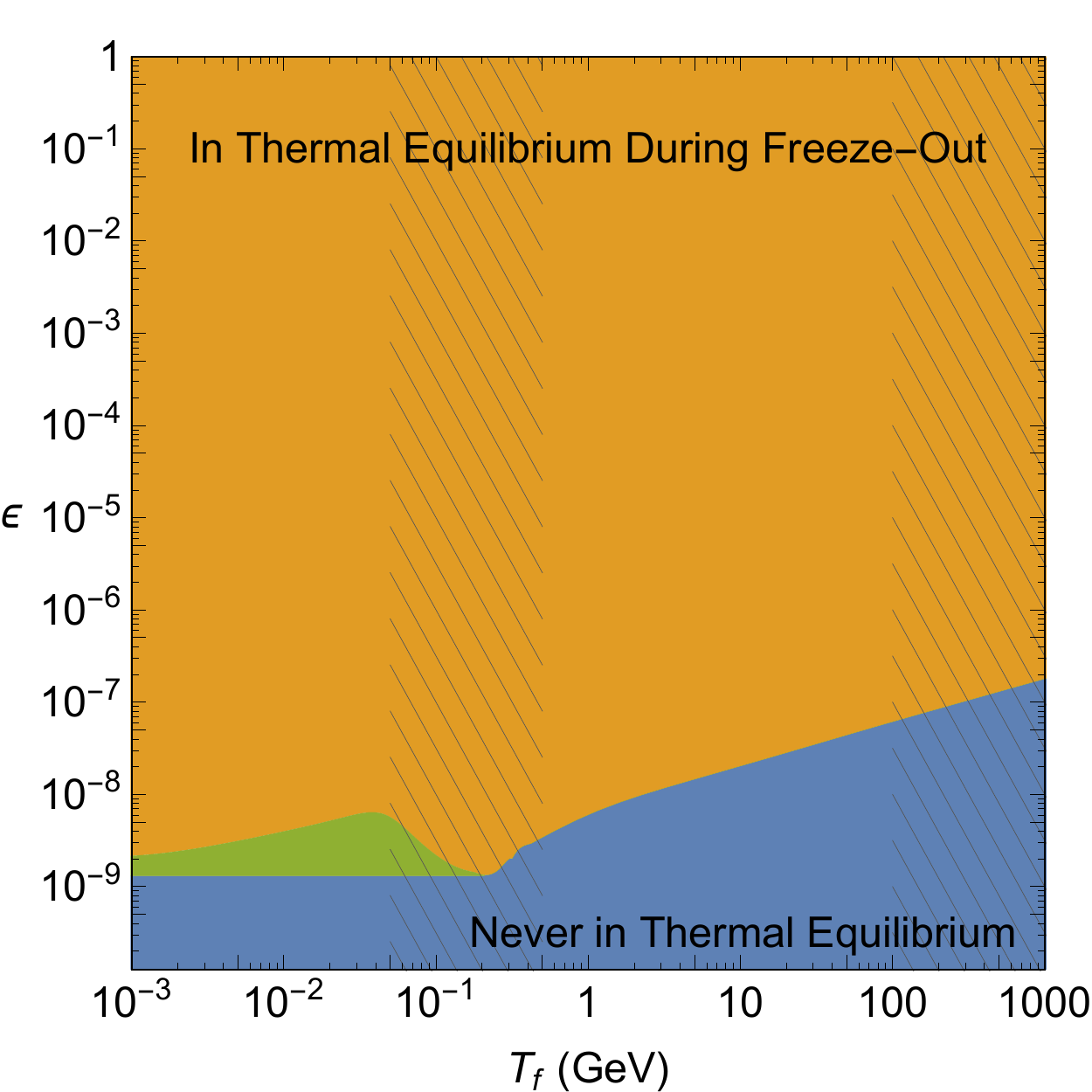}}~~~~
\includegraphics[scale=0.60]{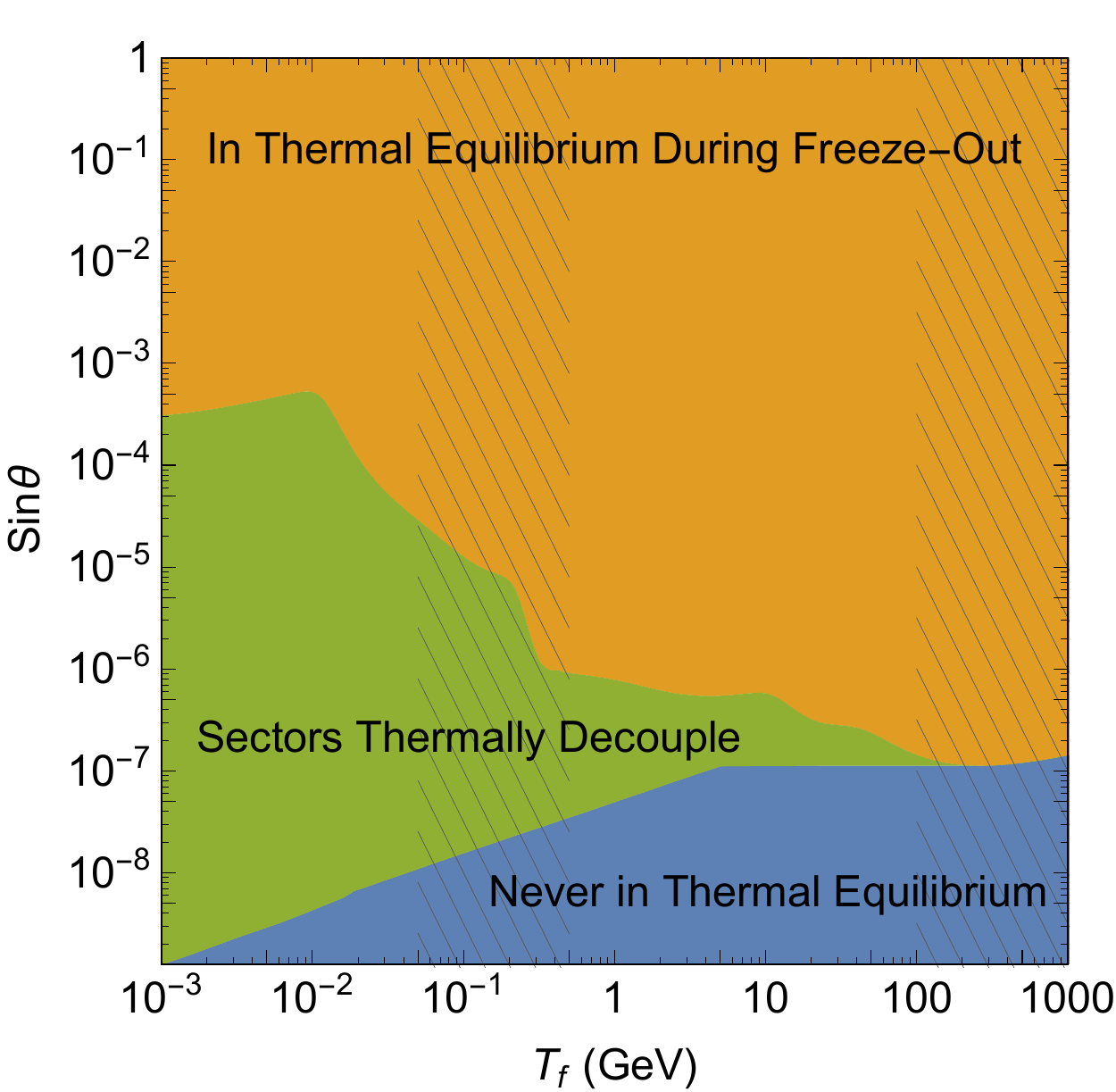}
\end{center}
\caption{{\bf Left:} Thermal coupling regions for the vector model as a function of the freezeout temperature, $T_f$.
  The orange region is where the dark sector is in thermal equilibrium
  with the SM at freezeout, while the blue region has the two sectors never
  in thermal equilibrium.   In the green region, the SM and hidden
  sector were in equilibrium at some higher temperature (here near 
  the QCD phase transition), but fell out of equilibrium by $T_f$, 
  so that the temperatures of the two sectors may drift apart 
  (\ref{eq:Tdrift}).  The hatched regions are near either the chiral or 
  electroweak phase transitions, where our calculation is less 
  reliable.  {\bf Right:} Thermal coupling regions for
  the scalar model as a function of the freezeout temperature, $T_f$.  Colors and hatching are as in the vector model. 
  The kink in the blue region near 4 GeV is when the effectiveness 
  of the equilibration from the $ss\to f \bar f$ process near 30 GeV 
  exceeds that from top processes near 200 GeV.  For more details, 
  see Appendix~\ref{sec:KD}.}
\label{fig:KD}
\end{figure}

When $T\gg m_\phi$, $2\too 2$ scatterings $(\mathrm{SM})\phi\too
(\mathrm{SM})(\mathrm{SM})$ are the dominant process responsible for
equilibrating the two sectors.  When $T\sim m_\phi$, $1\too 2$
scatterings $\phi \too (\mathrm{SM})(\mathrm{SM})$ become
dominant. This temperature scaling is evident in
Fig.~\ref{fig:rateexample}, where we show $1\too 2$ and $2\too 2$
scattering rates in each of our models as a function of
the temperature. In the absence of mass thresholds, $\Gamma_{2\too
  2}\propto T$ at high temperatures, while $\Gamma_{1\too 2} \propto
m_\phi^2/T$. The SM has many mass thresholds, which makes the
temperature dependence of the net scattering rates less transparent.
Full details of the calculation of these scattering rates are
presented in Appendix~\ref{sec:KD}; as discussed there, the thermalization floor that we obtain is an initial estimate,
computed up to a factor of $\sim 2$.  The resulting new cosmological lower bound on portal couplings is shown in
Fig.~\ref{fig:KD} as a function of $T_f$,  in the regime where $m_\phi \lesssim 0.1\, m_\chi$.  The thermalization floor is insensitive to the mediator
mass as long as $2\too 2$ rates dominate the scattering, a condition that holds generically (but not always) when the mediator is relativistic at the time of
freezeout.

Both our minimal models can
be described by four independent parameters, namely the DM mass, the
mediator mass, the portal coupling $\epsilon$, and the coupling
$\alpha_D$ between DM and the mediator.  Simplified model approaches
can be effective at highlighting the key physical features of classes
of DM theories \cite{Cheung:2013dua, deSimone:2014pda, Berlin:2015wwa,
  Abdallah:2015ter,Abercrombie:2015wmb}, and, in that spirit, our simple
HSFO models can be taken as useful guides to the physics of a general
WIMP next door, as we discuss further below.  We emphasize, however,
that our minimal HSFO models are, themselves, UV-complete and
self-consistent.

WIMPs next door have a sharply defined and bounded parameter space.
The dark matter-dark mediator coupling, $\alpha_D$, is fixed by the dark
matter relic abundance, while the coupling $\epsilon$ of the dark
sector to the SM is bounded from below by the 
thermalization floor.  Previous estimates of these thermalization floors (e.g. \cite{Feng:2010zp, Krnjaic:2015mbs}) have considered a subset of processes and/or studied equilibration at a fixed temperature.

As for standard WIMPs, the upper limit on the mass of DM is TeV-scale,
arising when the interaction governing freezeout becomes
non-perturbative.  The precise value of this upper bound will depend
in detail on the particle content of the dark sector.  For instance,
for DM freezing out via annihilations to massive dark photons, the
upper bound depends on the structure of $U(1)$ symmetry-breaking in
the dark sector \cite{Cline:2014dwa}.  Perturbative unitarity
constraints in specific models can further tighten the upper bounds on
the DM mass (e.g.,~\cite{Hedri:2014mua}).  We will indicate in
our parameter spaces where obtaining the correct relic abundance in
our simple models requires the dark matter-dark mediator coupling to
become non-perturbative, $\alpha_D \geq 1$.  This occurs for
$m_\chi\sim 10-150$ TeV in both simplified models, where the lower end
of the mass range is for small DM-dark mediator mass splitting, and
the upper end is for large splitting.  The Sommerfeld enhancement
(discussed in Appendix~\ref{sec:SE}) included in our freezeout
calculation heavily sculpts this range.  When the Sommerfeld effect
becomes very large, our numerical freezeout calculation becomes less
reliable, and we will further indicate these regions in presenting our
parameter space.  However, as the phenomenology does not undergo
qualitative changes in this $m_{DM}\gg$ TeV region of parameter space,
we will not discuss it in detail.

Meanwhile, the number of relativistic degrees of freedom that can be
present at temperatures $T\lesssim 2 m_e \sim$ MeV are restricted by
BBN, which mandates that $2.3 < N_{\mathrm{eff}}
< 3.4 $ \cite{Cyburt:2015mya}.  When the dark sector is in thermal
contact with the SM at temperatures $T_f\lesssim$ MeV, we must then
have both DM and the mediator be nonrelativistic by $T\sim$ MeV.  We
here impose the simple requirement $m_{DM}, m_{med} >$ MeV.  A more
careful treatment of the regions shown in green in Fig.~\ref{fig:KD}
where the dark sector has departed from equilibrium with the SM prior
to DM freezeout would relax these bounds slightly.  A detailed
treatment of this region is interesting, but beyond the scope of this
paper.

\section{Direct, Indirect, and Accelerator Constraints}
\label{sec:expt}

WIMPs next door give rise to signals in many different kinds of
experiments.  In this section, we briefly discuss the relevant
experimental results and their application to our simple models,
highlighting how signatures can differ from traditional WIMP models.

\begin{figure}[!t]
\begin{center}
\includegraphics[scale=1.2]{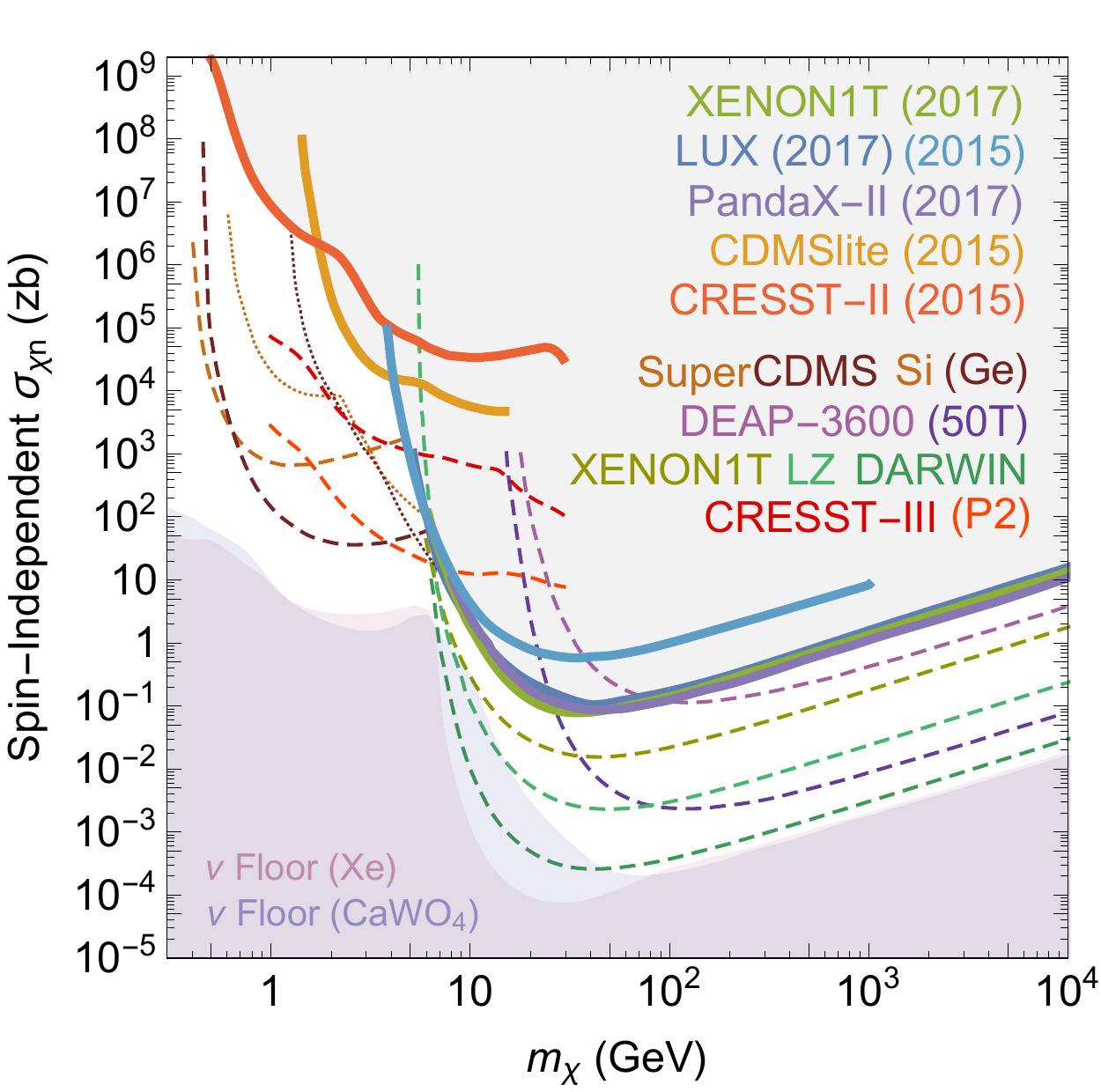}
\end{center}
\caption{ The shaded gray region represents the current exclusion from
  the combined results of XENON1T \cite{Aprile:2017iyp}, LUX
  \cite{Akerib:2016vxi,Akerib:2015rjg}, PandaX-II \cite{Cui:2017nnn}
  CDMSlite \cite{Agnese:2015nto}, and CRESST-II
  \cite{Angloher:2015ewa}.  Also shown in dashed lines are the
  projected limits from argon-based DEAP-3600 as well as its proposed
  50 ton-year upgrade \cite{Amaudruz:2014nsa}; the xenon-based
  experiments XENON1T \cite{Aprile:2015uzo}, LUX-ZEPLIN (LZ)
  \cite{Akerib:2015cja}, and DARWIN \cite{Schumann:2015cpa}; the
  CaWO$_4$-based CRESST-III and its Phase 2 upgrades
  \cite{Strauss:2016sxp} and the projected limits from SuperCDMS
  \cite{Calkins:2016pnm} for both silicon and germanium for both the
  interleaved Z-sensitive Ionization and Phonon (iZIP) detectors
  (thin, dotted) and those run in high voltage (HV) mode (dashed).
  The coherent neutrino scattering floor is shown for both CaWO$_4$
  and xenon \cite{Billard:2013qya,Ruppin:2014bra}.  The floor for
  silicon, germanium, and argon is very similar to xenon, while more
  complex materials, like the CaWO$_4$ in CRESST and the proposed
  EURECA \cite{Angloher:2014bua} experiment, can have a substantially
  different $\nu$ floor. }
\label{fig:DD}
\end{figure}

\paragraph{Direct detection.} Both our vector portal and Higgs portal
models have a leading spin-independent scattering cross-section with
nuclei.  Unlike for traditional WIMPs, the size of this cross-section
is not directly related to the dark matter annihilation cross-section:
it is proportional to the square of the portal coupling and can be
parametrically small.  We will demonstrate that both current and
proposed direct detection experiments have the sensitivity to test
cosmologically interesting values of the portal coupling.
Currently, the best constraints on spin-independent DM-nucleus
scattering come from XENON1T \cite{Aprile:2017iyp}, LUX
\cite{Akerib:2016vxi,Akerib:2015rjg} and PandaX-II \cite{Cui:2017nnn}
at higher masses, while CDMSlite \cite{Agnese:2015nto} and CRESST-II
\cite{Angloher:2015ewa} set the strongest limits at lower
masses.\footnote{While this work was being completed, the CRESST
  collaboration published limits on dark matter in the 140 MeV -- 500
  MeV mass range \cite{Angloher:2017sxg, Petricca:2017zdp}.  These constraints are not
  treated in this work.}  We show the current limits, along with
projections for several future experiments
\cite{Amaudruz:2014nsa,Aprile:2015uzo,Akerib:2015cja,Schumann:2015cpa,Strauss:2016sxp,Calkins:2016pnm},
in Figure~\ref{fig:DD}. In the figure, we also present the neutrino floor for both xenon and calcium tungstate ($\rm{CaWO}_4$) \cite{Billard:2013qya,Ruppin:2014bra}.

\paragraph{Indirect detection.}  In contrast to direct detection,
results from indirect detection searches are insensitive to the
(small) portal coupling, and test the dark matter annihilation
cross-section directly.  There are multiple sensitive probes of dark
matter annihilation in the universe.  The most important for our
models are the Fermi-LAT limits on dark matter annihilation in dwarf
galaxies \cite{Ackermann:2015zua,Fermi-LAT:2016uux} 
and Planck constraints on DM annihilations near recombination
\cite{Ade:2015xua,Slatyer:2015jla}.  Charged cosmic rays are another
important source of information about galactic DM annihilation, but
are subject to much larger systematic uncertainties arising from their
propagation within the galaxy.  While AMS-02 measurements of the
cosmic antiproton flux \cite{Aguilar:2015ooa} can potentially give
more powerful constraints on hadronic annihilation channels than
searches with gamma rays \cite{Giesen:2015ufa}, the difficulty in
accurately determining propagation parameters remains a serious
hurdle.  We follow \cite{Elor:2015bho} in considering AMS-02 positron
results \cite{Aguilar:2013qda}, which can place bounds on leptonic
channels where searches in photons have little reach, but neglecting
antiproton searches, as they constrain channels for which the far less
uncertain gamma-ray searches of
\cite{Ackermann:2015zua,Fermi-LAT:2016uux} have good sensitivity.
Meanwhile, CMB limits are mainly sensitive to the net energy deposited
in the $e^-$-$\gamma$ plasma by DM annihilations near recombination
\cite{Padmanabhan:2005es}, and are thus robust and nearly
model-independent.  The HAWC experiment can place constraints on very
high dark matter masses \cite{Albert:2017vtb} in the highly
Sommerfeld-enhanced regime; these constraints are currently exceeded
by the CMB constraints everywhere, but may become more important as
HAWC collects more data, or our understanding of the Triangulum II
dwarf galaxy, which dominates HAWC's sensitivity, improves
\cite{Laevens:2015una,Kirby:2015bxa}.  In principle, H.E.S.S. should have sensitivity to our DM models when $m_\chi \sim$ TeV, but they do not provide enough information to allow their results to be reliably reinterpreted.\footnote{The results of Ref.~\cite{Profumo:2017obk}, which appeared while this work was being completed,  indicate HESS is likely to be slighty more sensitive than dwarfs in this regime.}

\paragraph{Accelerator.} 
On the collider front, there are several potential discovery avenues
for hidden sector dark matter.  The direct production of DM (or of
an invisible mediator) in events
with large missing energy is no longer the leading signal, as we will
demonstrate below.  Rather, the leading accelerator signal is the
direct production of the dark mediator, followed by its decay back to
visible SM states.  Mediators can be produced through rare Kaon and
B-meson decays, directly through their interaction with electrons and
quarks at LEP and LHC, at lower energy colliders such as Babar, 
and at
beam dump and other intensity frontier experiments such as NA62.  They can also be produced in exotic Higgs decays \cite{Curtin:2013fra,Martin:2014sxa}.  Precision tests of $Z$
and Higgs couplings can also constrain the mixing between dark and
visible states. 

\paragraph{Astrophysical and cosmological constraints on dark
  mediators.}  Beyond the standard suite of DM search strategies,
models with long-lived dark mediator states face several additional
constraints from astrophysical and cosmological observations.  As the
requirement that the dark sector be thermalized with the SM places
lower bounds on the coupling of the dark mediator, these
constraints will largely be important for the \HSFOH\ model in the
sub-GeV regime where small Yukawa couplings help increase the mediator
lifetime.  Most constraining here are cooling in Supernova 1987A
\cite{Krnjaic:2015mbs,Chang:2016ntp}, and early universe limits on the
dark scalar lifetime coming from potential disruptions of isotope
abundances produced during BBN or dilutions
of neutrino and/or baryon abundances \cite{Flacke:2016szy}.

\section{Vector portal}
\label{sec:vv}

We first consider a simple vector portal model, containing a fermionic
DM, $\chi$, and a dark photon, $Z_D$. This type of model has been studied
extensively in the literature, especially to address cosmic ray
anomalies (HEAT, PAMELA, and ATIC first
\cite{Cholis:2008vb,ArkaniHamed:2008qn}, and more recently the
Galactic Center excess
\cite{Hooper:2012cw,Abdullah:2014lla,Martin:2014sxa,Cline:2014dwa}).

In the following, we define our model and establish notation. We
introduce a massive dark photon $\hat Z_D$, the gauge boson for a new dark
$U(1)_{Z_D}$ symmetry, that interacts with the SM through kinetic mixing
with SM hypercharge \cite{Galison:1983pa,Holdom:1985ag}.  The dark
photon mass could arise from the Stueckelberg mechanism
\cite{Stueckelberg:1938zz,Feldman:2007wj} or from a dark Higgs
mechanism.  For the sake of minimality, we will assume a Stueckelberg
origin, so that the only dark sector particles in our model are the
dark vector and the dark matter.  Including a dark Higgs boson could
open up additional annihilation channels, such as $\chi\chi\to Z_D
h_D$, which could become the leading process in the regime 
$m_{DM}\gg m_{Z_D}, m_{h_D}$
\cite{Cline:2014dwa}; we discuss this possibility further in Sec.~\ref{sec:beyondminimalZD}. 
The dark vector Lagrangian is thus given by 
\beq\label{eq:KM}
\mathcal{L_{Z_D}}=  -\frac{1}{4} \,\hat B_{\mu\nu}\, \hat B^{\mu\nu} - \frac{1}{4} \,\hat Z_{D\mu\nu}\, \hat Z_D^{\mu\nu}  + \frac{\epsilon}{2\cos\theta_W} \,\hat Z_ {D\mu\nu}\,\hat B^{\mu\nu} + \frac{1}{2}\, m_{Z_{D0}}^2\, \hat Z_D^\mu \, \hat Z_{D\mu}\, ,
\eeq
where $\theta_W$ is the Weinberg angle and $\epsilon$ is the
dimensionless kinetic mixing parameter.  Additionally, we introduce a
Dirac fermion $\chi$ with unit charge under $U(1)_{Z_D}$ and with mass
$m_\chi$ to serve as DM. Making the standard field redefinition to
diagonalize the hypercharge and $Z_D$ boson kinetic terms rescales the
dark coupling $g_D= \hat g_D/\sqrt{1-\epsilon ^ 2/\cos^ 2\theta_W}$, and
results in the following mass matrix for the neutral gauge bosons after
electroweak symmetry breaking,
\beq
\mathcal{M}^2_V = m_{Z,0}^2 \left(\begin {array}{ccc} 0  & 0 &  0 \\  0 & 1 & -\eta \sin\theta_W \\
                          0 &-\eta\sin\theta_W & \eta ^ 2\sin ^ 2\theta_W +\delta ^ 2 \end{array}\right)
\eeq
in the basis $(A, Z_0, Z_{D,0}) $.  Here $\eta\equiv
\frac{\epsilon}{\cos\theta_W\sqrt{1-\epsilon^2/\cos^2\theta_W}}$ and
$\delta\equiv m_{Z_{D0}}/m_{Z_0}$, with $m_{Z_0}$ the mass of the SM
$Z$ boson before mixing.  The resulting massive eigenstates are
\beq
\left(\begin {array}{c} Z \\ Z_D \end{array}\right) = 
      \left(\begin {array}{cc} \cos\xi &\sin\xi \\ -\sin\xi &\cos\xi \end{array}\right)
     \left(\begin {array}{c} Z_0 \\ Z_{D,0} \end{array}\right),
\eeq
with mixing angle 
\beq\label{eq:mixingangle}
\tan\xi = \frac{1  - \eta^2 \, \sin^2\theta_W  - \delta^2 - \mathrm{Sign}(1-\delta^2)\, \sqrt{4\, \eta^2 \, \sin^2\theta_W + (1  - \eta^2\, \sin^2\theta_W - \delta^2)^2}}{2\, \eta \, \sin\theta_W}\,.
\eeq
The massive eigenvalues are
\beq\label{eq:masses}
m^2_{Z, Z_D} = \frac{m_{Z_0}^2}{2}\left( 1+\delta ^ 2+\eta^2 \, \sin^2\theta_W \pm 
   \mathrm{Sign}(1-\delta^2) \sqrt{(1+\delta^2 +\eta^2 \, \sin^2\theta_W)^2 - 4\, \delta^2} \right)\,.
\eeq
\noindent We can now compute the couplings of the SM fermions and the
DM with the $Z_D$ gauge boson:
\begin{eqnarray}\label{eq:Zcoupl}
g_{Z_D f} & =& \frac{g}{\cos\theta_W}\, \left(- \sin\xi \,( T^3 \,\cos^ 2\theta_W - Y\,\sin^2\theta_W)
     +\eta \, \cos\xi \,\sin\theta_W\, Y \right)\,,  
     \nonumber \\
     g_{Z_D\chi} & = & g_D \cos\xi,
\end{eqnarray}
\noindent where $Y$, $t^3$ are the hypercharge and isospin of the
(Weyl) fermion $f$.  The physical $Z$ boson acquires a (vector-like)
coupling to $\chi$:
\beq\label{eq:ZchichiCoupl}
 g_{Z\chi} = g_D \sin\xi.
\eeq
Note that this coupling is $\epsilon$-suppressed, contrary to the
corresponding coupling of the $Z_D$. In fact, if we expand the
couplings in (\ref{eq:Zcoupl}) and (\ref{eq:ZchichiCoupl}) to leading
order in $\epsilon$ we find
\begin{eqnarray}\nonumber
g_{Z_D f} & \approx& \epsilon\frac{g}{\cos\theta_W}\left(\tan\theta_W\frac{m_Z^2}{m_Z^2-m_{Z_D}^2}(T_3\cos^2\theta_W-Y\sin^2\theta_W)+Y\tan\theta_W\right)\\\nonumber
g_{Z_D\chi}&\approx& g_D,\,\\
\label{eq:zcouplfinal}
g_{Z\chi}&\approx& -\epsilon g_D\tan\theta_W\frac{m_Z^2}{m_Z^2-m_{Z_D}^2}.
\end{eqnarray}
Our simple model can be described by four independent free parameters,
which we take to be $\alpha_D,\epsilon,m_\chi$ and $m_{Z_D}$.

\subsection{Thermal Freezeout and Indirect Detection}
\label{sec:VecID}

   \begin{figure}[!t]
\begin{center}
\includegraphics[scale=0.6]{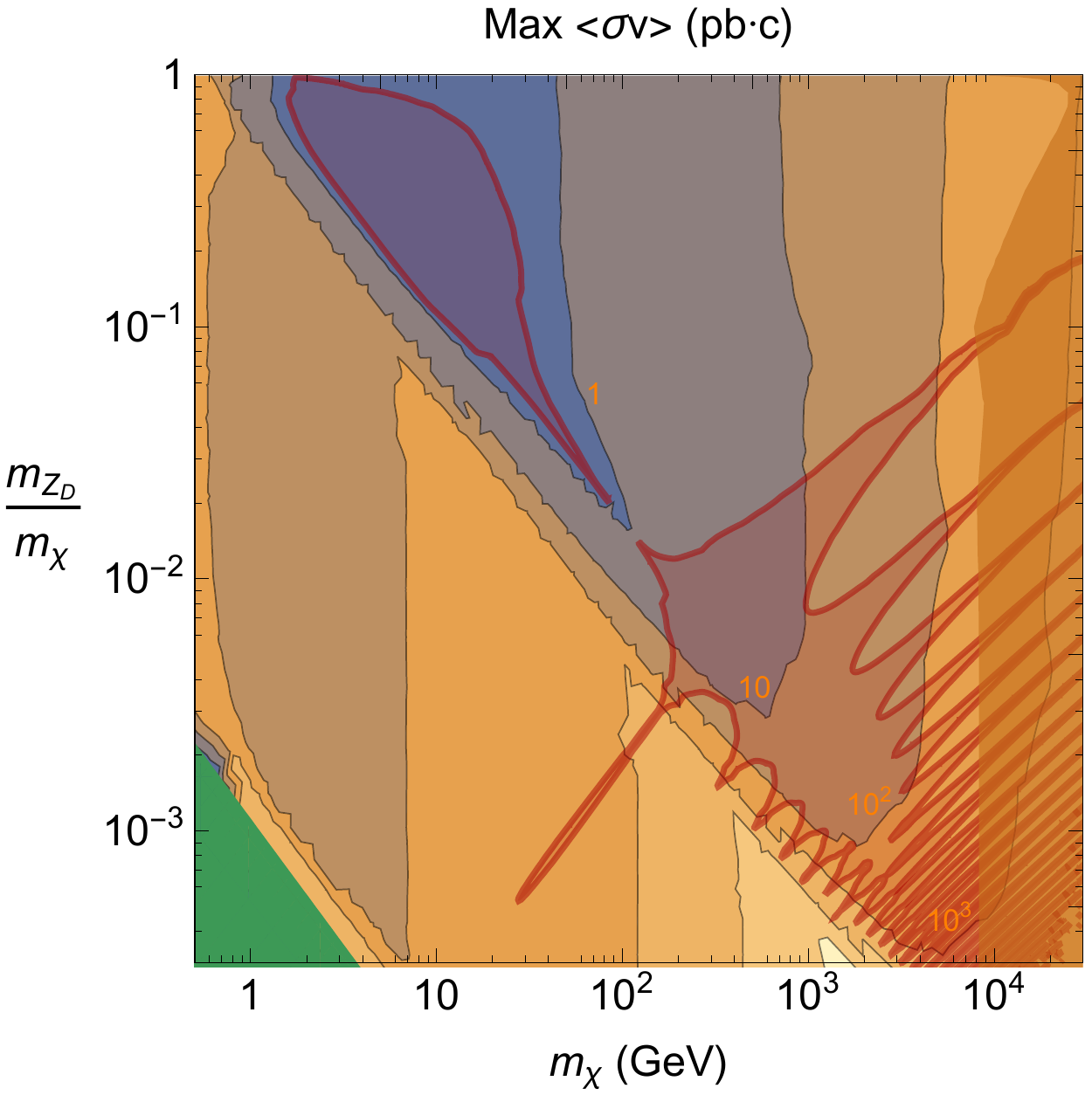}~
\includegraphics[scale=0.6]{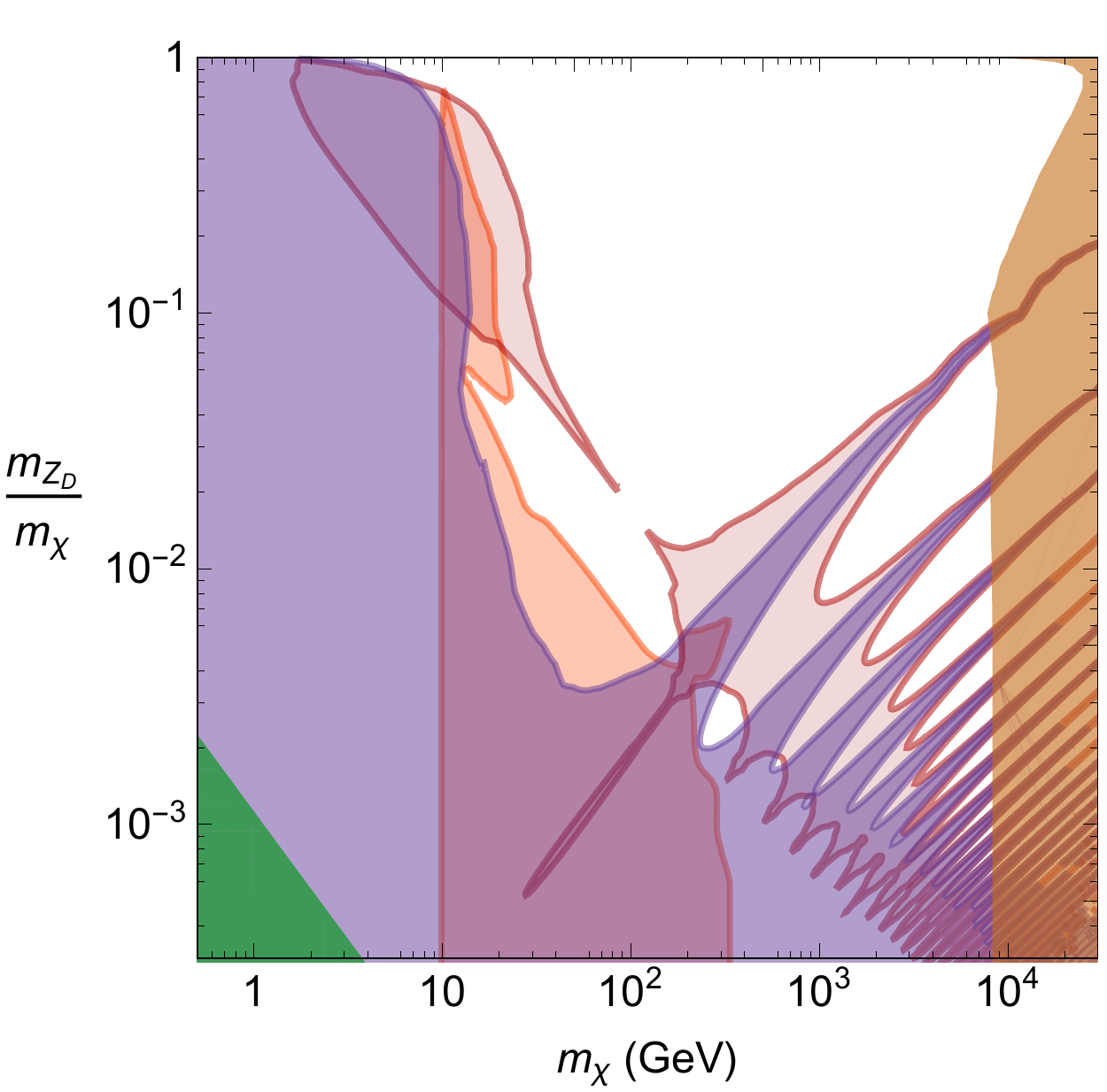}
\end{center}
\caption{{\bf Left:} Maximum annihilation rate (in pb$\cdot$c)
  consistent with the Fermi-LAT $\gamma$-ray measurements of dwarf
  galaxies \cite{Ackermann:2013yva} in the \HSFOV\ model.  The red
  regions indicate where Fermi dwarf measurements exclude the thermal
  relic abundance.  The brown region on the right of the plot
  illustrates where the freezeout coupling as determined with and
  without Sommerfeld enhancement deviates by more than a factor of two
  so that the determined value of $g_D$ becomes inaccurate
  (\ref{eq:SEregion}).  The green region in the lower left is where
  $m_{Z_D}<2m_e$.   {\bf Right:} Indirect detection bounds on the \HSFOV\ parameter space. The
  thermal relic abundance is excluded by Fermi dwarf $\gamma$s (red),
  positrons at AMS-02 (orange) \cite{Elor:2015bho}, and CMB spectral
  distortions from Planck, SPT, ACT, and WMAP (purple).  The
  constraints are sensitive to the precise locations of the Sommerfeld
  resonances in the lower right region, which are here only
  approximately determined. Brown and green regions as in the left
  figure.  }
\label{fig:VectorIndirect}
\end{figure}

When DM is heavier than the dark photon, it can annihilate via
$\bar\chi\chi\to Z_D Z_D$.  This is the only annihilation channel in
the small $\epsilon$ limit, and in this limit the thermally averaged
cross-section for this annihilation,
\beq
\langle{\sigma v}\rangle_0 = 
\frac{g_D^4 (m_\chi^2-m_{Z_D}^2)^{3/2}}{ 4 \pi m_\chi(2 m_\chi^2-m_{Z_D}^2)^2} +\mathcal O({v^2}),
\eeq 
is independent of $\epsilon$.  For sufficiently heavy DM
($m_\chi\gtrsim \mathrm{TeV}$), Sommerfeld enhancement can be
important during freezeout, $\vev{\sigma v} =\vev{\sigma v}_0
\vev{S_0(v)}$, which we implement via a Hulth\'en potential as
described in Appendix~\ref{sec:SE}.  Requiring that this reaction
yields the observed relic abundance as measured by Planck,
$\Omega_{DM} h^2 = 0.1186 \pm 0.0020$ \cite{Ade:2015xua}, fixes
$\alpha_D$ for each choice of $m_\chi, m_{Z_D}$.  The same
annihilation cross-section governs indirect detection signals.

In order to assess the constraints from indirect detection, we utilize
the measurement of dwarf galaxies from the Fermi-LAT and DES
collaborations \cite{Ackermann:2015zua}.  The energy flux of photons
from DM annihilations in an astrophysical source can be expressed as
\beq
\frac{d\Phi^{E}_\gamma}{d E} = \frac{\langle\sigma v\rangle}{16\pi
  m_\chi^2} E \frac{dN}{dE} \, J,
\label{eq:Flux}
\eeq
where we have specialized to Dirac DM.  Here $dN/dE$ is the number
distribution of photons from a single DM annihilation, and $J$ is the
astrophysical $J$-factor, describing the line-of-sight 
density of dark matter in the direction of the source \cite{Geringer-Sameth:2014yza}.

We use the 41 dwarf galaxies within the nominal sample of
\cite{Ackermann:2015zua} to obtain limits on the DM annihilation
cross-section using the procedure outlined in
Appendix~\ref{sec:dwarf}. The corresponding upper bounds on the cross section as a function of the dark matter mass $m_X$ and of the mass ratio $m_{Z_D}/m_X$ are shown in
Fig.~\ref{fig:VectorIndirect}; the red regions show where the thermal
relic abundance is excluded.  These regions appear in two distinct
places: the region in the lower right is where Sommerfeld
enhancements are important, while in the upper left they are not (see
Appendix~\ref{sec:SE} for details).

The flux of positrons observed in the AMS-02 experiment
\cite{Aguilar:2013qda,Elor:2015bho} can constrain photon-poor
annihilation channels.  In order to set constraints, we use the limit
for one-step $e^+e^-$ channels from \cite{Elor:2015bho} and compare
this to $\vev{\sigma v}\times \mathrm{Br}(Z_D\to e^+e^-)$. In
principle, considering all $Z_D$ decay modes would slightly improve
this result,
but achieving this mild improvement is beyond the scope of this work.
The resulting exclusion is shown in orange in the right panel of
Fig.~\ref{fig:VectorIndirect}.
 
Dark matter annihilation during the era of recombination can broaden
the surface of last scattering and distort the CMB anisotropies
through the injection of electrons and photons into the plasma.  For
WIMPs annihilating with a velocity-independent cross-section, the
effect of this energy injection can be accurately encapsulated by a
redshift-independent efficiency parameter $\feff (m_{DM})$, which
depends on the DM mass and the species of particles produced by DM
annihilations \cite{Finkbeiner:2011dx}.  Planck results together with
results from ACT, SPT, and WMAP limit $\feff(m_\chi) \vev{\sigma
  v}/{m_\chi} < 14$ pb c / TeV \cite{Slatyer:2015jla}, allowing for
robust bounds to be placed on dark matter models.  The $\feff$ values
for DM annihilation to pairs of SM particles have been computed in
\cite{Slatyer:2015jla, Madhavacheril:2013cna}.  Due to the rather soft
dependence on $m_\chi$ for all branching ratios except for photons and
leptons, we use the $\feff$ values in \cite{Slatyer:2015jla} evaluated
at $m_\chi/2$ for non-leptonic channels, together with
$\feff^{\pi^\pm}$ from \cite{Madhavacheril:2013cna}.  For leptonic
channels we use $\feff^{VV\to 4\ell} (m_\chi)$ from
\cite{Slatyer:2015jla}.
We derive a net $f^{net}_{\mathrm{eff}}$
\beq
f^{net}_{\mathrm{eff}}(m_\chi,m_{Z_D}) = \sum_{\ell} \mathrm{Br}(Z_D\to\ell\ell) f^{VV\to 4\ell}_{\mathrm{eff}}(m_\chi) + \sum_{X\neq \ell} \mathrm{Br}(Z_D\to XX) f^{XX}_{\mathrm{eff}}\!\!\lp \frac{m_\chi}{2}\rp
\eeq
where $m_{Z_D}$ governs the branching ratios.  The resulting CMB bound 
is shown in purple in the right panel of Fig.~\ref{fig:VectorIndirect}, as a function of the dark matter mass $m_X$ and of the mass ratio $m_{Z_D}/m_X$.

\subsection{Direct Detection}
\label{sec:VecDD}

  \begin{figure}[!t]
\begin{center}
\includegraphics[scale=0.6]{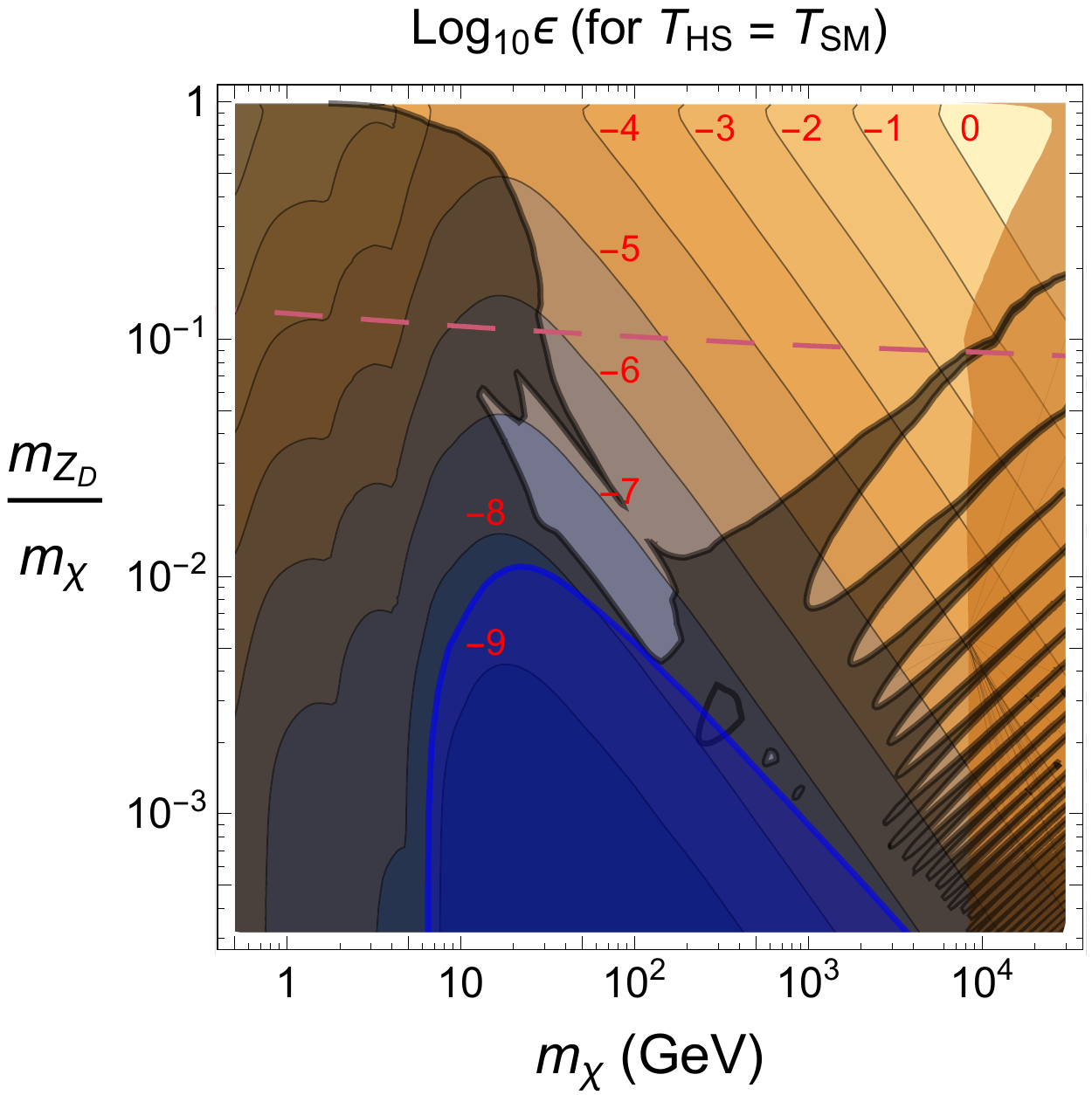}~~~~
\includegraphics[scale=0.6]{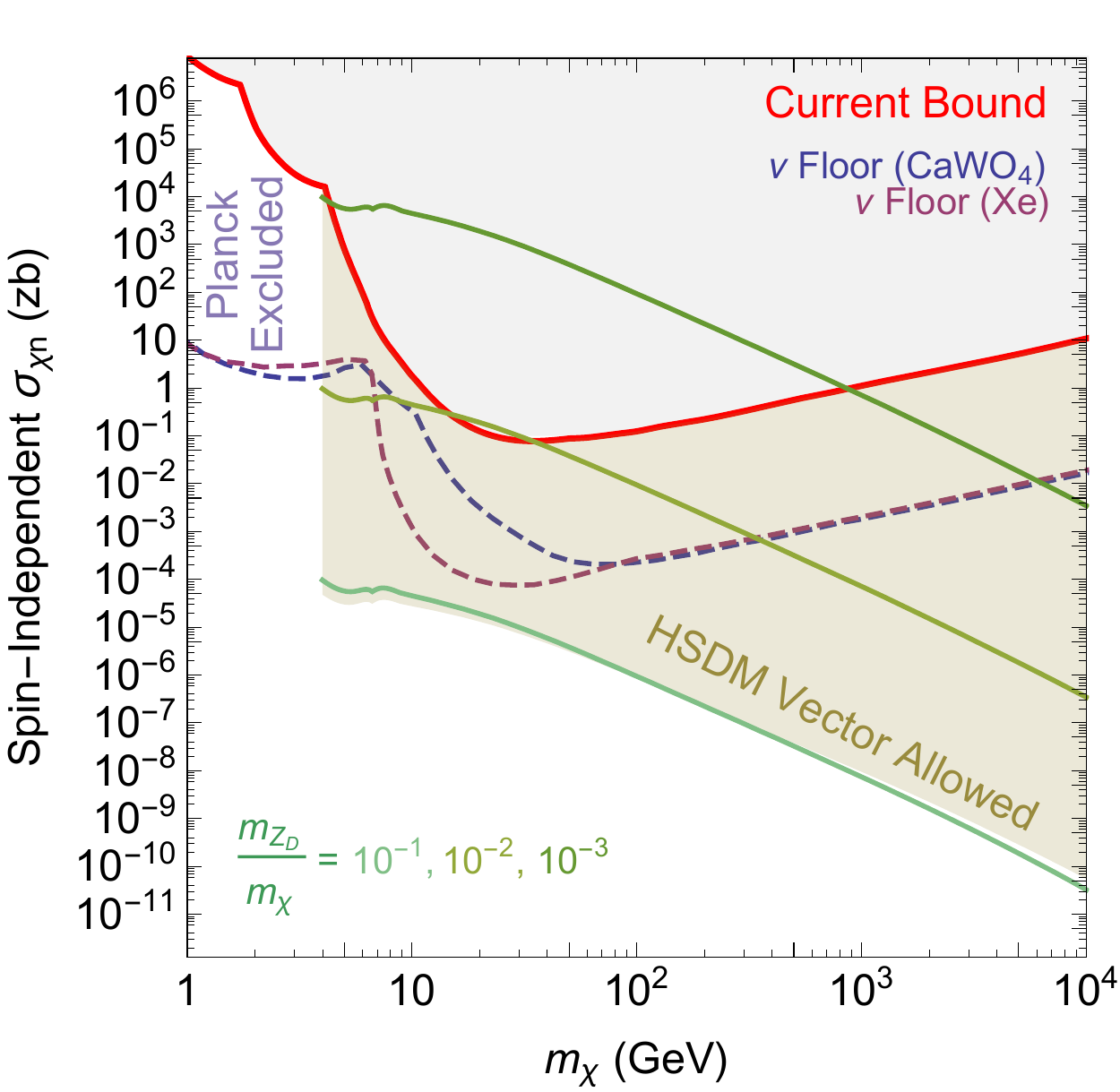}
\end{center}
\caption{{\bf Left:} Contours of the maximum allowed value of $\epsilon$
  consistent with direct detection bounds in the \HSFOV\ model.
  Within the shaded blue region, the two sectors were not in thermal
  equilibrium at freezeout.  The combined region excluded by indirect 
  detection constraints from Fig.~\ref{fig:VectorIndirect} is shown by 
  the black region.  Above the dashed pink line is where $Z_D$ is 
  non-relativistic at the freezeout temperature ($m_{Z_D}/T_f>2.46$).  
  The brown region on the right of the plot illustrates
  where the freezeout coupling as determined with and without
  Sommerfeld enhancement deviates by more than a factor of two 
  (\ref{eq:SEregion}). {\bf Right:} The tan region shows the direct detection
  parameter space for WIMPs next door in the \HSFOV\ model.  At low
  dark matter masses, Planck excludes all dark vector masses and
  couplings.  Shown in green are three specific lower bounds on the
  direct detection cross section for the mass ratios $m_{Z_D}/m_\chi =
  10^{-1}, 10^{-2}$, and $10^{-3}$.  We also show the neutrino floor
  for xenon and CaWO$_4$ (used in CRESST) with dashed purple and blue
  lines, respectively.}
\label{fig:DDVector}
\end{figure}

Direct detection experiments are an excellent test of this minimal
model, and over much of the parameter space place the most stringent
constraints on the portal coupling $\epsilon$.  The spin-averaged,
non-relativistic amplitude-squared for DM to scatter off of a nucleus
is mediated by the exchange of both dark vector and $Z$ bosons, and is
given by
\beq
\abs{\bar{ \mathcal M}^{NR}(E_R)}^2 = \abs{\frac{\mathcal M}{4m_\chi m_N}}^2 = 
 g_D^2 \epsilon^2 A^2 F^2(E_R)    \abs{\frac{f^{({Z_D})}_n}{m_{Z_D}^2+2 m_N E_R}+\frac {f^{(Z)}_n s_W}{m_Z^2-m_{Z_D}^2}}^2,
\eeq
where $A$ is the mass number of the target nucleus, $F^2(E_R)$ is
the Helm nuclear form factor \cite{Helm:1956zz,Lewin:1995rx} as a function of the recoil energy $E_R$, $s_W\equiv\sin\theta$, $m_N$ is the mass of the nucleus, and
$f^{(Z_D)}_n$ and $f^{(Z)}_n$ are given by
\beq
f^{(X)}_n =\frac{1}{A} \big( Z (2 g_{u,X}+g_{d,X}) +(A-Z) (g_{u,X}+2g_{d,X})\big),
\eeq
with $Z$ the atomic number, and $g_{X,u}$ and $g_{X,d}$  the couplings of the boson $X$ with up and
down quarks in (\ref{eq:zcouplfinal}).  Here we have retained
the momentum dependence in the propagator from $Z_D$ exchange, as it
is needed to accurately describe the scattering when 
$m_{Z_D} \lesssim \mu_{\chi N} v_\chi$, where the DM-nucleus reduced mass 
is $\mu_{\chi N} =m_\chi m_N / (m_\chi +m_N)$.

When the scattering amplitude does not depend on the DM velocity, then
the event detection rate per unit detector mass in the experiment can
be expressed as \cite{Fan:2010gt,Freese:2012xd}
\beq
R\lp\bar{\mathcal M}^{NR}(E_R)\rp = \frac{\rho_\chi}{2\pi m_\chi} \int_{0}^\infty dE_R \abs{\bar{\mathcal M}^{NR}(E_R)}^2  \epsilon(E_R) \eta(E_R).
\label{eq:DDrate}
\eeq
where $\rho_\chi$ is the local DM density, $\epsilon(E_R)$ is an experiment-specific selection efficiency, and $\eta(E_R)$ is the mean inverse speed \cite{Freese:2012xd} defined by
\beq
\eta(E_R) = \int_{v>v_{min}(E_R)} \frac{f(v)}{v} d^3v
\eeq
for which, following the experiments, we use the expression in Ref.~\cite{Lewin:1995rx}.\footnote{While this is experimental usage, a more accurate expression for $\eta(E_R)$ can be found in Ref.~\cite{Freese:2012xd}.}  If the amplitude is independent of the recoil energy
to leading order, it is reasonable to approximate $\bar{\mathcal
  M}^{NR}(E_R)\to \bar{\mathcal M}^{NR}(0)$ in (\ref{eq:DDrate}),
allowing for the particle physics contributions to the rate to be
entirely factorized from the experimental and astrophysical
inputs.  Experimental results are typically expressed in terms
of a cross-section that has been factorized in this manner and further
simplified by defining an effective per-nucleon cross-section,
facilitating comparison between different experiments. For the \HSFOV\
model, this DM-nucleon cross-section is
\beq
\sigma_{\chi n}^0 = \frac{ \mu_{\chi n}^2 \abs{\bar{\mathcal M}^{NR}(0)}^2}{\pi A^2} =  \frac 1\pi g_D^2 \epsilon^2 \mu_{\chi n}^2 \abs{\frac{f^{({Z_D})}_n}{m_{Z_D}^2}+\frac {f^{(Z)}_n s_W}{m_Z^2-m_{Z_D}^2}}^2
\label{eq:DDexclusion}
\eeq
where we have defined the nucleon-DM reduced mass $\mu_{\chi n}=
m_\chi m_n/(m_\chi + m_n)$. 

However, for the \HSFOV\ model, $\bar{\mathcal M}^{NR}(E_R)$ \emph{is}
sensitive to the recoil energy once $m_{Z_ D}^2 \lesssim 2 m_N E_R \sim \mu_{\chi N}^2 v_\chi^2$.
In order to correctly account for this important effect, we will
determine the excluded cross-section via
\beq
\sigma_{\chi n} = \sigma_{\chi n}^0 \frac{R\lp\bar{\mathcal M}^{NR}(E_R)\rp}{R\lp\bar{\mathcal M}^{NR}(0)\rp},
\label{eq:DDnet}
\eeq
where the function $R$ is determined separately for each experiment.
Given masses for the DM and dark vector, the relic density constraint
fixes $\alpha_D$. The latest XENON1T \cite{Aprile:2017iyp}, LUX
\cite{Akerib:2016vxi,Akerib:2015rjg}, PandaX-II \cite{Cui:2017nnn},
Super-CDMS \cite{Agnese:2014aze}, CDMSlite \cite{Agnese:2013jaa} and
CRESST-II \cite{Angloher:2015eza} searches then determine the maximum
allowed value of the portal coupling $\epsilon$.  We show these upper
bounds in the left panel of Fig.~\ref{fig:DDVector}.  Sensitivity is greatest at small
values of $m_{Z_D}/m_\chi$, thanks to the $1/m_{Z_D}^4$ behavior of
the nuclear matrix element.  However, the sensitivity saturates when
$m_{Z_D}^2 \lesssim 2 m_N E_{min}$ (the threshold energy of the
experiment), and the propagator in the matrix element is dominated by
the momentum.  Over a sizable region where $m_{Z_D}/m_\chi \lesssim
0.01$ and $m_\chi\sim 10$ GeV, current direct detection limits are
sensitive enough to exclude values of the portal coupling at and below the thermalization floor.
This region
is shown in blue in the left panel of Fig.~\ref{fig:DDVector}; see
Appendix~\ref{sec:KDV} for details of its determination.  Future
direct detection experiments will be able to test this cosmological
origin for DM over a broader range of DM and mediator masses.  In the
right panel of Fig.~\ref{fig:DDVector}, we show the direct detection
parameter space consistent with our \HSFOV\ WIMP next door (in tan).  As Fig.~\ref{fig:DDVector} shows, this cosmological history for DM can predict spin-independent
cross-sections well below the neutrino floor.

\subsection{Accelerator and other mediator constraints}

  \begin{figure}[!th]
\begin{center}

\includegraphics[scale=0.88]{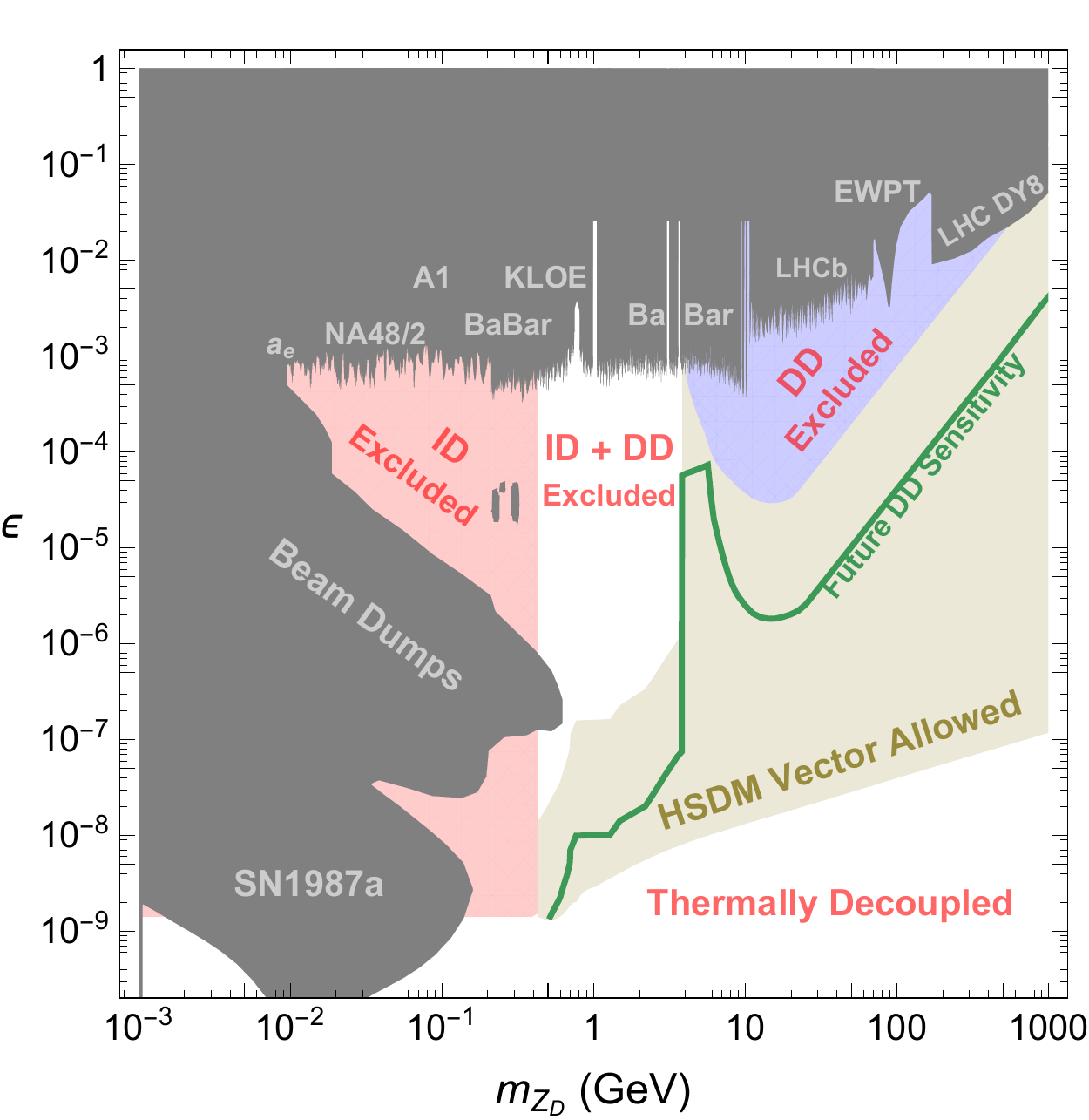}
\end{center}
\caption{Parameter space for WIMPs next
  door in the \HSFOV\ model as a function of dark vector mass and
  portal coupling.  In the red (blue) shaded regions, indirect (direct) 
  detection alone excludes that point in $\{\epsilon,m_{Z_D}\}$ 
  parameter space for all values of $m_\chi > m_{Z_D}$.  In the white
  region, all values of $m_\chi > m_{Z_D}$ are excluded by a combination 
  of direct and indirect detection.  
  The future direct detection sensitivity is determined by assuming
  CRESST-III Phase 2 and DARWIN (Fig.~\ref{fig:DD}) are placing
  their nominal limits.  }
\label{fig:VectorPlots}
\end{figure}

Direct searches for the $Z_D$ are the leading terrestrial signal of
this model.  As summarized in \cite{Hoenig:2014dsa,Curtin:2014cca,
Ilten:2016tkc,Aaij:2017rft} and in
Fig.~\ref{fig:VectorPlots}, there are many constraints on massive
vector bosons kinetically mixed with SM hypercharge.  Most of these results
come from searches for rare meson decays, beam dump
experiments, precision electroweak tests, direct production
at BaBar or the LHC, and Supernova 1987A.  
Fig.~\ref{fig:VectorPlots} shows the parameter space for a vector
portal WIMP next door as a function of $m_{Z_D}$ and $\epsilon$.   
As shown there, Supernova 1987A uniquely probes the thermalization 
floor in a limited range of dark photon masses at around few 
$\times 10^{-2}$ GeV. Furthermore,
especially at low masses, terrestrial searches for dark photons bound $\epsilon$ more tightly than 
the direct detection constraints do alone.

Fig.~\ref{fig:VectorPlots} highlights the unique capability of  
direct detection experiments to probe otherwise challenging regions of 
dark photon parameter space.  Within the blue region, direct detection 
excludes dark photons for any choice of $m_\chi >m_{Z_D}$.   
Indirect detection from CMB experiments cuts off the entire region of 
dark photon parameter space below 400 MeV (red region), while a combination of 
both direct and indirect detect results exclude all values of 
$m_\chi >m_{Z_D}$ in a region up to 4 GeV for 
a range of portal couplings.  Future direct detection experiments, 
DARWIN and CRESST-III Phase 2 \cite{Schumann:2015cpa,Strauss:2016sxp}, 
will greatly cut into this range (green line), even excluding down to the 
thermalization floor near 500 MeV.  

However, while the net impact of these direct mediator searches is generally
subdominant to the combined constraints from direct and indirect
detection for the minimal \HSFOV\ model, it is important to emphasize
that they provide complementary information.  In particular,
Fig.~\ref{fig:VectorPlots} shows that any dark photon discovered in
meson decays or at high-energy colliders is sufficiently strongly
coupled to the SM to populate a dark radiation bath in the early
universe, regardless of the identity of dark matter.  Further, as we
will discuss below, simple extensions to the minimal model can
suppress the direct detection cross-section, thus leaving dark photon
searches as the leading terrestrial test of vector portal WIMPs next
door.

\begin{figure}[!t]
\begin{center}
\includegraphics[scale=1.2]{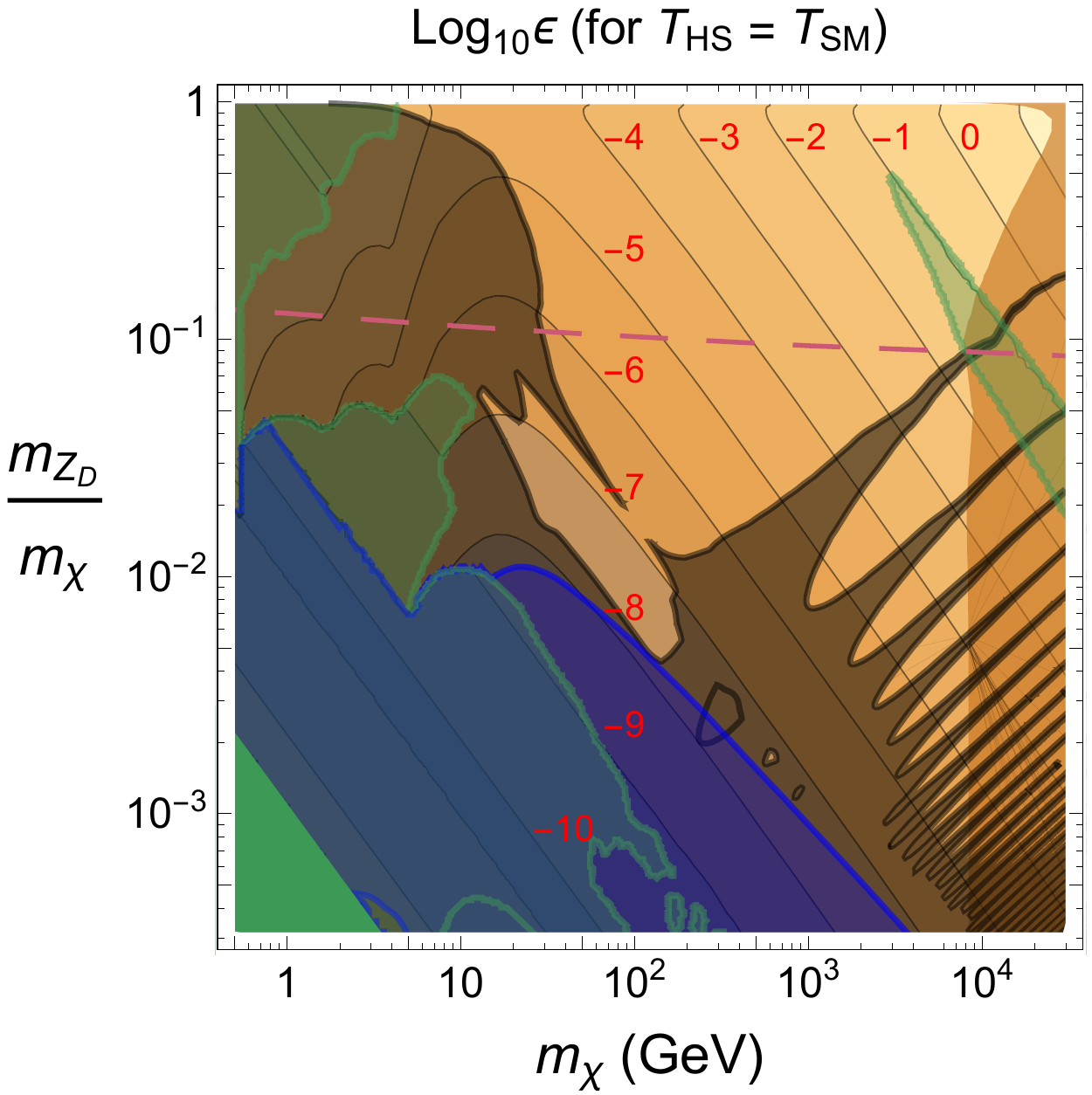}
\end{center}
\caption{Combined constraints on the kinetic mixing coupling
  $\epsilon$ in the vector model.  The black region shows the combined
  indirect detection constraints (see Fig.~\ref{fig:VectorIndirect})
  and excludes all values of $\epsilon$ for the indicated masses given
  $T_{SM}=T_{HS}$ at the time of freezeout. Regions where direct 
  detection constraints are superseded by constraints on the mediator
   (collider, beam dump, supernova, etc.)~are shown in green.  Labelled 
  contours show the maximum value of $\epsilon$ permitted by either
  direct detection or mediator constraints, whichever bound is stronger.   
 The blue region
  indicates where thermalization cannot occur prior to DM freezeout,
   the brown contour illustrates where the freezeout calculation is
  less reliable, and above the pink dashed line the mediator is 
  non-relativistic during freezeout. The solid green region in the lower left is where $m_{Z_D}<2m_e$.}
\label{fig:MasterVector}
\end{figure}

A summary of all constraints is shown in Fig.~\ref{fig:MasterVector}
as a function of $m_\chi$ and $m_{Z_D}/m_\chi$.  As before, we show
the union of $\epsilon$-independent constraints from Fermi dwarfs,
AMS-02 positrons, and the CMB with the black shaded region.  The
shaded green region denotes where the most important bound on
$\epsilon$ comes from collider, beam dump, and supernova searches.
Above the pink dashed line, the mediator was non-relativistic at dark
matter freezeout, and thus the $\epsilon$ floor from thermalization
can be much lower.  As in Fig.~\ref{fig:VectorIndirect}, the solid
green in the lower left corner has $m_{Z_D}<2m_e$, causing issues with
BBN, and the brown region on the right of the plot
illustrates where the freezeout coupling as determined with and
without Sommerfeld enhancement deviate by a factor of 2.

We do not show constraints from mono-$X$ searches at the LHC, since
they are generically weaker than the constraints coming from direct
detection experiments. In particular, the most stringent mono-$X$
constraint arises from the ATLAS and CMS mono-jet searches performed
with Run II data \cite{CMS:2016pod,Aaboud:2016tnv}. Values of
$\epsilon$ as small as $\sim 0.1$ are only probed in a small region of
parameter space at around $m_\chi\sim 100$ GeV and $m_\chi\sim
(0.8-1)m_{Z_D}$.

\subsection{Beyond the minimal model of dark vector interactions}\label{sec:beyondminimalZD}

While we have worked with a minimal two-species model consisting only
of fermionic DM and the vector mediator, the salient features of this
model are representative of the behavior of a broad class of dark
sectors with a vector mediator.  In this section, we briefly discuss
the modifications of the dark sector phenomenology obtained by
introducing new dark degrees of freedom (or altering the assumed
quantum numbers of DM), and argue that our minimal
two-species \HSFOV\ model provides a reasonable general guide to the
characteristic sizes and locations of signals for vector portal WIMPs
next door.

To begin with, any additional relativistic species in the thermal
plasma at freezeout will contribute to the Hubble parameter, thereby
requiring a mild increase of the value of $\alpha_D$ needed to obtain the
thermal relic abundance. The DM relic abundance is proportional to
$\alpha_D^2 \times g_{*S}/\sqrt{g_*}$ at freezeout.  Neglecting the
logarithmic sensitivity of the freezeout temperature to $g_{*S}$, we
can thus simply estimate the effect of adding additional equilibrated
dark species by rescaling $\alpha_D$ to absorb the shift in
$g_{*S}/\sqrt{g_*}$.  This is a minor quantitative effect,
particularly at relatively high DM masses where $g_{*SM}(T_f)\gg 1$.

An excellent motivation for introducing additional dark species is to
provide a dark Higgs mechanism to generate
$m_{Z_D}$ \cite{ArkaniHamed:2008qn, Cline:2014dwa, Bell:2016fqf}.
 Our model assumed a Stueckelberg mechanism
for simplicity. Using a dark Higgs to generate $m_{Z_D}$ is a generic
alternative scenario, but with a dark Higgs comes additional model
dependence, especially through the choice of the DM's U(1)$_D$ quantum
numbers.
When the dark Higgs is light, it can furnish additional annihilation
modes: depending on the spectrum, both the $s$-wave $\chi \bar\chi \to
Z_D h_D$ process  and the $p$-wave $\chi\bar\chi\to
h_D h_D$ process can contribute to freezeout \cite{Bell:2016fqf}. (We continue to assume that the vector portal coupling
dominates the dark sector's interactions with the SM.)  The additional
annihilation modes change the specific value of $\alpha_D$ needed to
obtain the thermal relic abundance, generically by no more than an
$\mathcal{O}(1)$ amount. These additional annihilation modes also
alter the detailed cosmic ray spectrum for indirect detection. When
$p$-wave contributions are important, the expected annihilation cross
section for CMB and galactic signals can be decreased, generically by a factor of no more than $\mathcal{O}(1)$.  Thus introducing these additional annihilation modes
generally changes indirect detection signals quantitatively but not
qualitatively.

On the other hand, a dark Higgs can drastically impact direct detection. A
Stueckelberg mass for the dark photon requires Dirac dark matter, and
thus yields unsuppressed spin-independent direct detection
cross-sections.  However, a dark Higgs mechanism allows the Dirac
spinor to split into two Majorana mass eigenstates, the lighter of
which is dark matter \cite{TuckerSmith:2001hy}.  If the mass splitting is
small so that the Majorana states are nearly degenerate (pseudo-Dirac),
then the leading spin-independent cross-section is now inelastic. Inelastic
scattering is significantly more challenging to observe at direct detection
experiments, but some signals are still possible \cite{Bramante:2016rdh}.   

However, as the mass splitting increases, the dominant direct detection signals come from elastic processes.  These processes can arise at tree level, from the now axial-vector coupling of the DM to $Z_D$. At relatively high dark vector masses, the axial components of the SM--$Z_D$ couplings are sizable, giving rise to spin-dependent cross-sections.  The vector SM--$Z_D$ couplings yield spin-independent cross-sections suppressed by DM velocities or nuclear recoil momenta, giving small but still potentially interesting signal rates \cite{DEramo:2016gos}. Elastic spin-independent cross-sections are also  induced at one loop \cite{Cirelli:2005uq,Essig:2007az,Kopp:2009et,Freytsis:2010ne,Haisch:2013uaa}. The size of this contribution is thus sensitive to the UV field content of the dark sector.  Finally, while the coupling of the dark Higgs to the SM is sub-leading for thermalization, the exchange of the dark and SM Higgses gives a spin-independent cross-section, and, depending on the size of the dark Higgs-SM couplings, could provide the leading direct detection signal; for further discussion of Higgs-portal direct detection, see Sec.~\ref{sec:scalardd} below.   

If there are additional light dark sector species, $\phi$, then $Z_D$ can have
open decay modes within the hidden sector, and when these are active
one generally expects $\mathrm{Br}(Z_D\to \phi\phi)\simeq 1$.  Letting
$Z_D$ decay can eliminate (if $\phi$ is stable) or modify (if $\phi$
is unstable \cite{Elor:2015bho}) cosmic ray and CMB constraints on DM
annihilation, and, often for either case, subject the mediator to the
generally weaker terrestrial searches for $Z_D\to$ invisible
\cite{Lees:2017lec,Banerjee:2017hhz,Aguilar-Arevalo:2017mqx}.  As the 
vector portal coupling is the leading interaction between sectors, 
these new light states will have longer lifetimes than the dark vector, 
which can potentially lead to stringent cosmological constraints.
In particular, as BBN does not allow for an additional radiation 
species with the SM temperature $T_{SM}$, a model that maintains 
equilibrium with the SM through the vector portal has limited prospects 
for including stable dark radiation.  As illustrated by the small green region in the left panel of 
Fig.~\ref{fig:KD}, a vector portal hidden sector equilibrated with the SM cannot
decouple from the SM at sufficiently early times to permit significant
departures of the HS temperature from $T_{SM}$.  Interestingly, a model 
with a portal coupling of $\epsilon\sim2\times10^{-9}$ and DM that 
freezes out after the chiral phase transition may provide a WIMP next 
door that permits additional dark radiation.  As this model could 
eliminate indirect detection signals, it would open up interesting parameter 
space for GeV-scale DM, together with dark radiation signals that could  be 
observable at CMB-S4 \cite{Abazajian:2016yjj}.   Verifying the existence of this dark radiation window, 
via a more detailed calculation of the thermal production rate of dark 
photons from the hadronic plasma near the chiral phase transition, is an 
interesting topic for future studies.   

Our reference model assumes fermionic DM. If DM is instead a
complex scalar, the story is qualitatively unchanged: the leading
annihilation cross-section is $s$-wave, while the leading direct
detection cross-section is spin-independent and unsuppressed.  Once
again, the introduction of a dark Higgs would make only minor changes
to the indirect detection signals, while potentially introducing
sizeable and model-dependent changes to the direct detection signals.

Finally, in our \HSFOV\ model and the variants above, annihilations 
of only one representation of the dark $U(1)$ symmetry are 
important during freezeout.  Introducing more states in different 
representations and allowing coannihilation to be important in 
determining the relic abundance can significantly alter the 
phenomenology and open up different areas of parameter space, 
but represents a much greater departure from the minimal model 
discussed here.

To summarize,  our minimal model provides a good guide to the
essential physics of vector portal WIMPs next door. Many possible 
additions to the dark sector would change signals qualitatively, by 
$\sim \mathcal{O}(1)$ amounts, e.g., through affecting Hubble.
Indirect detection signals are especially robust, as adding additional
annihilation channels, etc., generically changes cosmic ray signals
quantitatively but does not suppress them significantly below 
expectations for an $s$-wave thermal relic. For vector portal models there
is very little scope to eliminate indirect detection signals via $Z_D$
decays to dark radiation.  On the other hand, direct detection signals
are especially sensitive to the origin of dark symmetry breaking.  The
direct detection signals for the minimal model we present are
maximally predictive; in a dark sector with a dark Higgs mechanism,
direct detection signals can be suppressed by model-dependent
amounts.

\section{Higgs portal}
\label{sec:ss}

Here we define a simple reference model for a Higgs portal WIMP next
door, \HSFOH.  We consider a Majorana fermion dark matter, $\chi$, with
a scalar mediator, $S$, that interacts with SM states through a (small)
Higgs portal coupling.  A useful simple model is
\cite{Shelton:2015aqa} (see also \cite{Pospelov:2007mp,Baek:2011aa,Dupuis:2016fda})
\beq\label{eq:SH}
\Lag = \Lag_{kin} -\frac{1}{2} \left(y S\right)\left(\chi \chi +\hc \right) + \frac{\mu_s ^ 2}{2} S^2-\frac {\lambda_s} {4!} S ^ 4-\frac{\epsilon}{2}S ^ 2 |H | ^ 2 - V(|H|),
\eeq
where we use the usual conventions for the Higgs  potential,
\beq\label{eq:potential}
V(H) = -\mu ^ 2 |H |^ 2+\lambda |H | ^ 4.
\eeq
We should also add to this Lagrangian the interaction of the Higgs with 
the quarks, leptons and gauge bosons of the SM. These interactions 
will  be inherited by the dark scalar through its mixing with the SM Higgs. 
 In the Lagrangian in (\ref{eq:SH}--\ref{eq:potential}), we have imposed 
 a discrete symmetry taking $S\to -S$, $\chi\to i\chi$, thus forbidding 
 cubic and linear terms in $S$ as well as a Majorana mass for $\chi$.  
 Imposing this discrete symmetry allows us to expose the essential 
 physics of this theory with the minimum number of parameters.

 In order for the fermions to be massive, both $S$ and $H$ must
 acquire nonzero vacuum expectation values (vevs), $S=v_s +s_0$ and
 $H=\frac1 {\sqrt 2} \lp v_h +h_0 \rp$, where $v_x$ is the vev of $X$
 ($v_h=246$ GeV).  Minimizing the potential gives analytic expressions
 for the vevs,
 \bea
 v_s^2 =&\, \frac{6 \lp2\lambda\mu_s^2 - \epsilon\mu^2 \rp}{2\lambda\lambda_s-3\epsilon^2} ,\\
 v_h^2 =&\, \frac{2 \lambda_s \mu^2 - 6 \epsilon\mu_s^2}{2\lambda\lambda_s-3\epsilon^2}. 
 \eea
The dark matter gets a mass of $m_\chi=y v_s$,  while the scalars have a simple mass matrix,
  \beq
  M = \lp \begin{tabular}{cc}
  $\frac 13 \lambda_s v_s^2$ & $\epsilon v_s v_h$ \\
  $\epsilon v_s v_h$ &   $2 \lambda v_h^2$
  \end{tabular}
  \rp, 
\eeq
yielding the mass eigenvalues
 \beq
 m_{h,s}^2 = \lambda v_h^2 + \frac16 \lambda_s v_s^2 \pm \sqrt{\lp \lambda v_h^2 - \frac16 \lambda_s v_s^2 \rp^2 +  \epsilon^2 v_h^2 v_s^2 },
 \eeq
and a mixing angle defined by
 \beq
 \tan \theta = \frac{ \epsilon v_h v_s}{\lambda v_h^2 - \frac16 \lambda_s v_s^2+\sqrt{\lp \lambda v_h^2 - \frac16 \lambda_s v_s^2 \rp^2 +  \epsilon^2 v_h^2 v_s^2 }}= \frac{ \epsilon v_h v_s}{ m_{h}^2-m_s^2} + \order{\epsilon^3},
 \label{eq:tantheta}
\eeq
where the latter equality holds only for $\frac16 \lambda_s v_s^2\neq
\lambda v_h^2$ and therefore the mass of the scalar $S$ quite
different from 125 GeV. 
In this regime, for small $\epsilon$, $\cos\theta\sim 1$ and
$\tan\theta\sim\sin\theta\propto\epsilon$,  
so either $\sin\theta$ or $\epsilon$ can be viewed as a measure of the
strength of coupling between the SM and dark sectors.
 
We can express the Lagrangian in terms of $v_h=246$ GeV, $m_h=125$ GeV
and the four free parameters, $y$, $m_\chi$, $m_s$, and $\sin\theta$.
In terms of these parameters, the most important couplings for our discussion, to leading order in $\sin\theta$ and for
$\frac16 \lambda_s v_s^2\neq \lambda v_h^2$, are:
 \bea 
 \mathcal{L}_{0}\ni&\, \frac{1}{2} \left(y s\right)\left(\chi \chi +\hc \right) -\frac {3 y^2 m_s^2}{m_\chi^2}\frac{s^4}{4!}  -\frac {3 y m_s^2}{m_\chi}\frac{s^3}{3!}-\frac {3 m_h^2}{v_h^2}\frac{h^4}{4!}  -\frac {3 m_h^2}{v_h} \frac{h^3}{3!} + \frac{m_f}{v_h} h f \bar f  \\
 \mathcal{L}_{\sin\theta}\ni&\, \sin\theta y \frac{m_h^2+2m_s^2}{2 m_\chi} h s^2 - \sin\theta \frac{m_f}{v_h} s f \bar f + \frac 12 \sin\theta y h\lp \chi\chi+\hc\rp .
 \label{eq:scalarInt}
\eea
It will sometimes be useful to refer to a fine-structure-like
constant, $\alpha_Y\equiv\frac{y^2}{4\pi}$.
 
\subsection{Thermal Freezeout and Indirect Detection}

Assuming freezeout of the Majorana dark matter is governed by
interactions entirely within the hidden sector (i.e., effects
proportional to $\epsilon$ can be neglected), three diagrams dominate
the process $\chi\chi\to ss$: $t$- and $u$-channel exchange of $\chi$,
and $s$-channel annihilation through an off-shell $s$.  The
spin-averaged amplitude, integrated over final state phase space, can
be written as
\bea
\frac12 \int d\Omega_{CM} \abs{\bar{\mathcal{M}}}^2  
&= \frac {9 m_s^4 y^4(s-4 m_\chi^2)}{2 m_\chi^2 (s-m_s^2)^2}- \frac{ y^4\lp 2 s m_\chi^2 +16 m_\chi^4 - 16 m_\chi^2 m_s^2 +3 m_s^4\rp}{ s m_\chi^2 -4 m_\chi^2 m_s^2+ m_s^4}  \\
&- y^4\lp s^2\!+\!16 s m_\chi^2\! - \!4 s m_s^2  \!-\!32 m_\chi^4 \!-\! 16 m_\chi^2 m_s^2 \!+\!6 m_s^4\rp \frac{\ln\lp\frac{s-2m_s^2 - s\beta_\chi\beta_s}{s-2m_s^2 + s\beta_\chi\beta_s}\rp}{s \beta_\chi\beta_s  (s-2m_s^2)} \hspace{3mm}   \\
&+ \frac {6 m_s^2 y^4}{ (s-m_s^2)} \left[ \frac{s+2m_s^2
    -8m_\chi^2}{s\beta_\chi\beta_s} \ln\lp\frac{s-2m_s^2 -
    s\beta_\chi\beta_s}{s-2m_s^2 + s\beta_\chi\beta_s}\rp -2 \right] \equiv \Xi
    \label{eq:sFO}
\eea
where $\beta_i=\sqrt{1-4m_i^2/s}$, and the factor of one-half for
identical final state particles is explicit.  This quantity, $\Xi$, is
related to the cross-section by, $\sigma_{cm} = \beta_s \Xi/(16\pi
sv_{rel})$, where the relative velocity is conventionally defined as
$v_{rel}=\abs{\vec v_1 - \vec v_2}$.

Defining $R\equiv \frac{m_s}{m_\chi}$, the thermally averaged
cross-section, to leading order in $v_{rel}^2$, can be written
\cite{Shelton:2015aqa}
\bea
\left<\sigma v_{rel}\right>_1 
&=\! \frac{\!\left< v_{rel}^2\right> y^4 \sqrt{1\!-\!R^2}}{12 \pi m_\chi^2} \!\!\lp\! \frac{72 \!-\! 160 R^2 \!+\! 165 R^4 \!-\! 99 R^6 \!+\! 37 R^8 \!-\! \frac{33 R^{10}}{4} \!+\! \frac{27 R^{12}}{32}}{(2  -R^2)^4(4-R^2)^2} \!\rp \!\!+\! \order{\left< v^4\right>},
\eea
which is $p$-wave suppressed.  This cross-section is then corrected by
the Sommerfeld enhancement (\ref{eq:SEp}) to give $\vev{\sigma v}
=\vev{\sigma v_{rel}}_1 \vev{S_1(v_{rel})}$.

Thanks to the $p$-wave annihilation cross-section, this model does not
yield observable signals from DM annihilations in halos, nor do late
annihilations $\chi\chi\to s s\to {\rm{SM}}$ deposit noticeable amounts of
energy into the CMB. Thus standard indirect detection strategies do
not constrain the minimal \HSFOH\ model.  Dark matter density spikes
surrounding super-massive black holes could potentially boost $p$-wave
DM annihilation to observable rates, yielding a point-like gamma-ray
signal \cite{Shelton:2015aqa}.  Additionally, there are regions of
parameter space at high $m_\chi$ where $m_s \lesssim \alpha_Y^2
m_\chi/4$ and DM annihilations can proceed through bound state
formation and decay.  These processes are $s$-wave, and thus although
they are unimportant during thermal freezeout, they can leave an
imprint on the CMB \cite{An:2016kie}.
Assuming that the velocity of DM at recombination satisfies $v \ll R$
and bound states are accessible, an analytic solution for the bound
state formation rate (via monopole transitions into $S$-wave bound
states) can be written as \cite{An:2016kie}
 \beq
\vev{\sigma v}^M_{s} = \frac{16 \pi \alpha_Y^4}{9 m_\chi^2} \abs{\frac{\Gamma\lp a^+\rp \Gamma\lp a^-\rp}{\Gamma\lp 1+ i \frac{v}{R}\rp}}^2 \sum_{n<\alpha/2\sqrt R }\frac{e^{-4n}}{n^3} \abs{L_{n-1}^1(4n)}^2
\label{eq:BSformation}
\eeq
where $a^\pm = 1 + \frac{i v}{2R}\lp 1\pm\sqrt{1-\frac{4 \alpha
    R}{v^2}}\rp$, and a factor of $1/4$ appears due to only a single
available bound state with spin-0 due to Majorana dark matter
\cite{An:2016kie}.  In practice, the Gamma functions in
(\ref{eq:BSformation}) yield roughly a cosecant function with maxima
at $\alpha_Y/R = m\in \mathbf N$ regularized at the singular points by
a tiny imaginary contribution.  This bound state decay can be used to
bound $p$-wave annihilating dark matter models with $\frac12 \alpha_Y
R^{-\frac12}>1$, using the condition
\beq
 \frac{\feff(m_\chi)
  \vev{\sigma v}^M_s}{m_\chi} < 14 \mbox{ pb c / TeV},
 \label{eq:scalarCMB}
\eeq   
where we use the same treatment as in Sec.~\ref{sec:VecID} to
derive $\feff(m_\chi)$.  The excluded region is plotted in purple in 
Fig.~\ref{fig:DDScalar}.\footnote{For numerical feasibility, we
  plot lines centered on the exclusion with a thickness
  representative of the actual limit.  The difference between this
  plot and a literal plot of the excluded region is nearly
  imperceptible.}

\subsection{Direct Detection}
\label{sec:scalardd}

  \begin{figure}[!t]
\begin{center}
\includegraphics[scale=0.6]{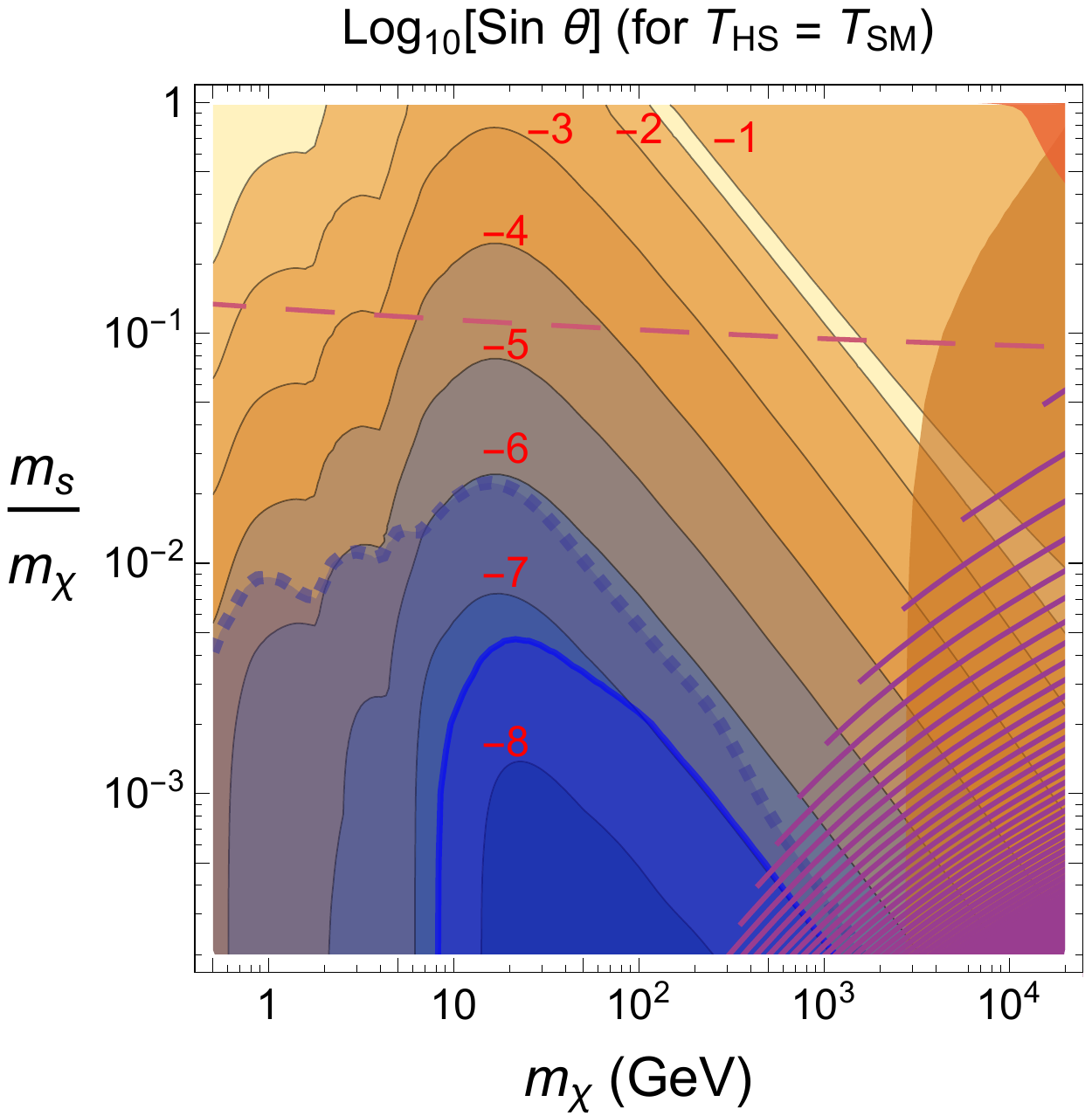}~~~~
\includegraphics[scale=0.61]{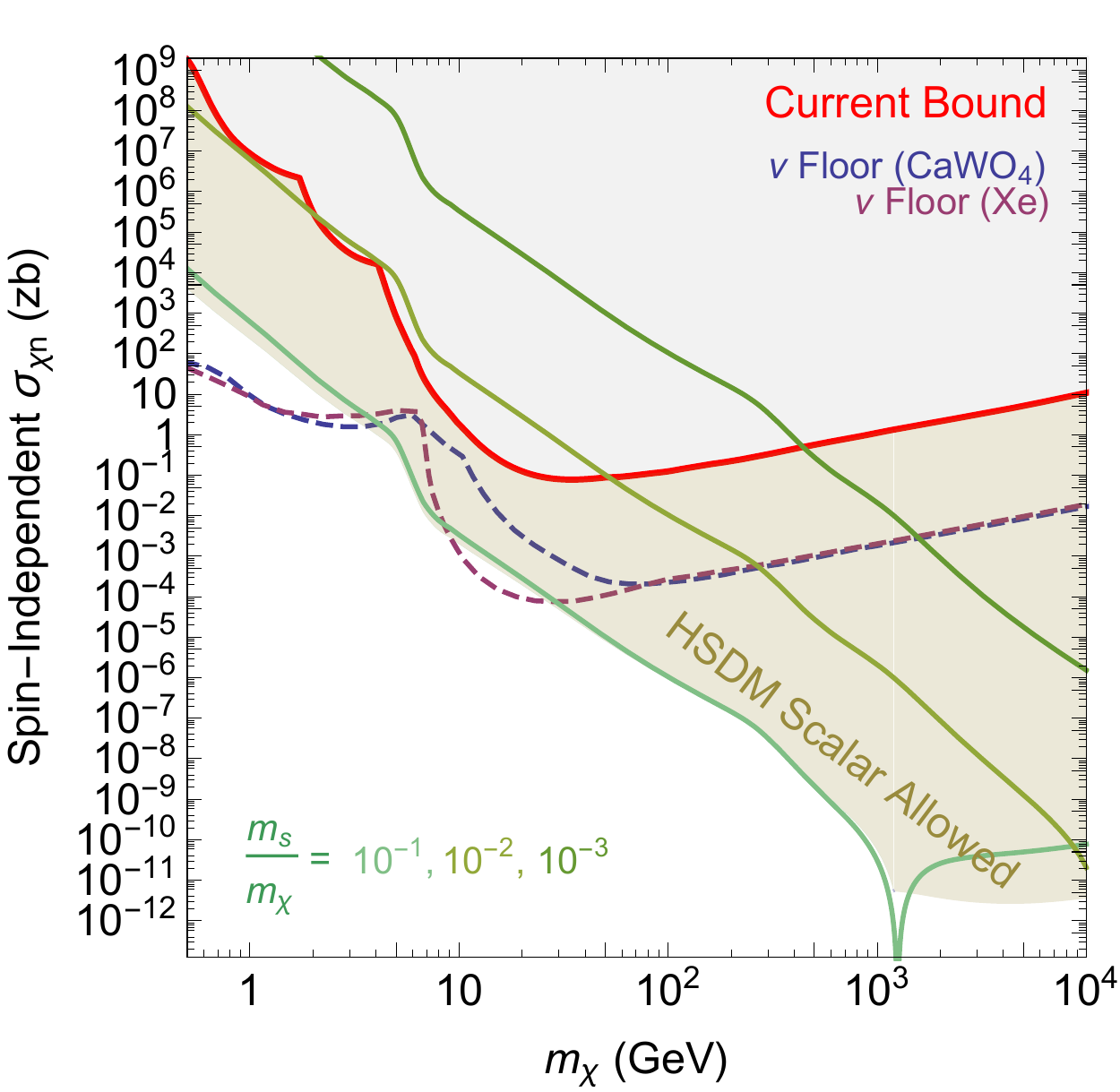}
\end{center}
\caption{{\bf Left:} Contours illustrating the maximum allowed value
  of $\sin\theta$ in the \HSFOH\ model that is consistent with direct
  detection bounds on spin-independent dark matter. 
  The purple lines are constraints from bound state annihilation
  modifying the CMB (\ref{eq:scalarCMB}).  Within the blue region, the
  two sectors were never in thermal equilibrium.  Within the dashed
  blue-gray region, the Higgs portal interaction equilibrated the two
  sectors but then fell out of equilibrium again prior to DM
  freezeout.  Thus in this region the assumption that the temperatures
  of the two sectors are equal at DM freezeout is not consistent. 
   The red region (in the
  upper right corner) illustrates where the model is non-perturbative
  ($\alpha_D>1$), and the brown region on the right of the plot illustrates
  where the freezeout coupling with and without the Sommerfeld
  enhancement deviates by more than a factor of two
  (\ref{eq:SEregion}).  {\bf Right:} The tan region shows the direct detection
  parameter space for WIMPs next door in the \HSFOH\ model.  
    Shown in green are three specific lower bounds on the
  direct detection cross section for the mass ratios $m_{s}/m_\chi =
  10^{-1}, 10^{-2}$, and $10^{-3}$.  We also show the neutrino floor
  for xenon and CaWO$_4$ (used in CRESST) with dashed purple and blue
  lines, respectively.}
\label{fig:DDScalar}
\end{figure}

As in the \HSFOV\ model, the \HSFOH\ model has a leading
spin-independent direct detection cross-section, and direct detection
therefore provides a powerful test of this model.  The direct
detection cross-section is mediated by exchanges of both the dark
scalar and the visible Higgs boson.  In the regime where $m_s \ll
m_\chi$, momentum exchange rather than the dark scalar mass dominates
the dark scalar propagator.  In order to account for this important
recoil energy dependence, we follow the procedure discussed in 
Sec.~\ref{sec:VecDD}.  For the \HSFOH\ model, the spin-averaged,
non-relativistic amplitude-squared for DM-nucleon scattering is
\beq
\abs{\bar{ \mathcal M}^{NR}(E_R)}^2 = \abs{\frac{\mathcal M}{4m_\chi m_N}}^2
\approx  \frac{2 y^2 \sin\theta^2 f^{(s)2}_N A^2 F^2(E_R) m_n^2}{ v_h^2}\abs{\frac{1}{2 m_N E_R +m_s^2}-\frac{1}{m_h^2}}^2,
\label{eq:scalarDDamp}
\eeq
where $m_n$ is the mass of the nucleon  and 
\beq
f^{(s)}_N \equiv  \frac1A\lp Z f^{(s)}_{p} +(A-Z) f^{(s)}_n \rp .
\eeq
Here the nucleon matrix elements are
\bea
f^{(s)}_n =&\, \sum_{q=udscbt} \braket{n}{\frac{m_q}{m_n} \bar q q}{n} =  \sum_{q=udscbt} f^{(s)}_{n,q} \\ =&\, \sum_{q=uds} f^{(s)}_{n,q} + \frac29 \lp1- \sum_{q=uds} f^{(s)}_{n,q}\rp = \frac19 \lp2+7 \sum_{q=uds} f^{(s)}_{n,q} \rp ,
\eea
where heavy quark flavors have been removed in the second line
\cite{Shifman:1978zn}.  In principle $f^{(s)}_n$ and $f^{(s)}_p$ could
differ substantially, but owing to the dominance of heavy-flavor
(i.e., isospin-universal) contributions, they are nearly identical for
Yukawa-coupled scalars. We use here $f^{(s)}_n\sim 0.293$ and $f^{(s)}_p\sim
0.291$ \cite{Hill:2014yxa}. 
If the recoil energy in (\ref{eq:scalarDDamp}) can be ignored, then the DM-nucleon cross-section is 
\beq
\sigma_{\chi n}^0 \approx \frac{2 y^2 \sin^2\theta f^{(s)2}_n m_n^2 \mu_{n\chi}^2}{\pi v_h^2}\abs{\frac1{m_s^2}-\frac{1}{m_h^2}}^2.
\label{eq:DDscalar}
\eeq
However, it is important to retain the recoil energy dependence in the dark boson propagator to accurately describe scattering rates when $m_s\lesssim \mu_{\chi N} v_\chi$; to handle this, we use the procedure discussed previously in Sec.~\ref{sec:VecDD}.
In the left panel of Fig.~\ref{fig:DDScalar}, we show the maximum allowed value of the
mixing angle $\sin\theta$ consistent with the current direct detection
bounds \cite{Aprile:2017iyp,Akerib:2016vxi,Akerib:2015rjg,Cui:2017nnn,Agnese:2015nto,Angloher:2015ewa}.
  The blue region shows where direct detection experiments are probing values of $\sin\theta$ below the thermalization floor.  Additionally, the dashed blue-gray region corresponds to where the direct detection bound
on $\sin\theta$ implies the dark and SM sectors have decoupled and
their formerly equilibrated temperature can drift (\ref{eq:Tdrift}).
Once again, direct detection experiments can probe all the way down to
the cosmological lower bound on the portal coupling in some portions
of parameter space.  The direct detection parameter space consistent
with Higgs portal WIMPs next door is shown in the right panel of
Fig.~\ref{fig:DDScalar}, where the tan region is defined by cutting out
the region where $m_s$ is within 10\% of $m_h$.

\subsection{Accelerator and other mediator constraints}

\begin{figure}[!t]
\begin{center}
\includegraphics[scale=0.8]{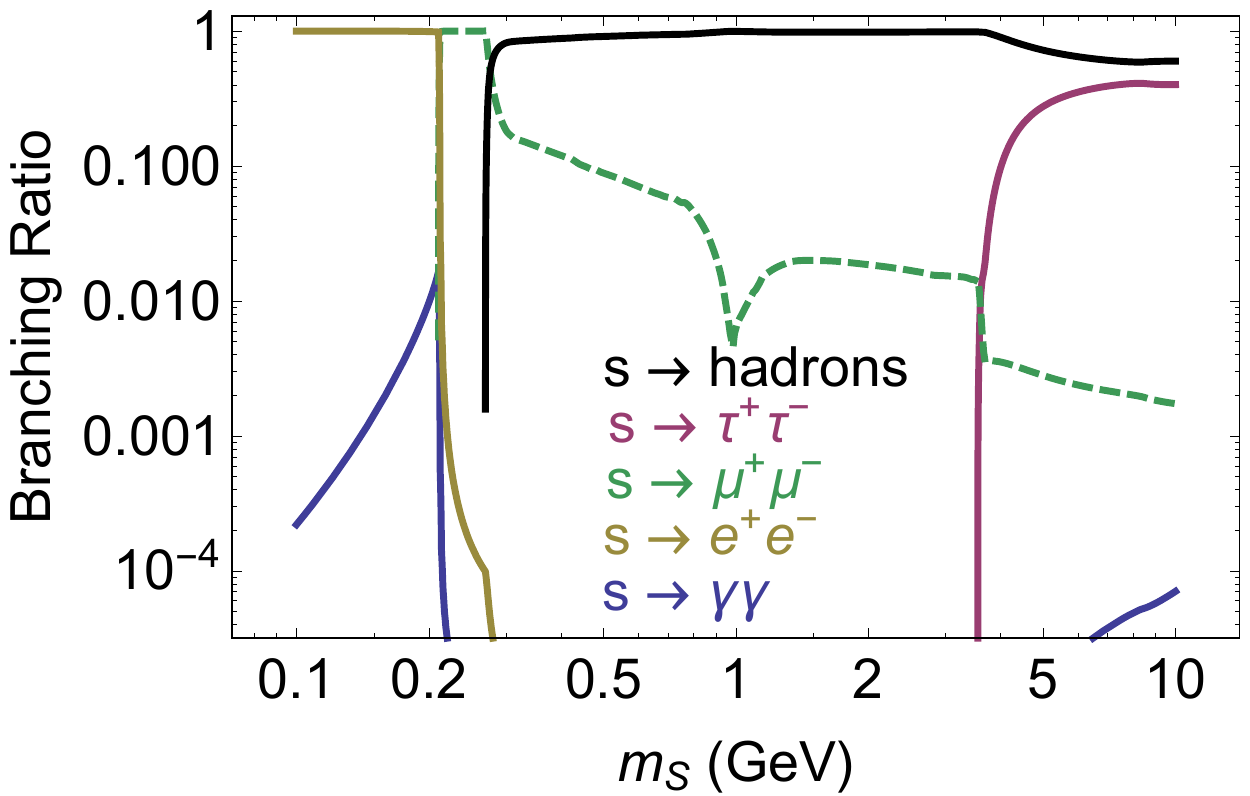}
\end{center}
\caption{Scalar branching ratios in the light hadron region used in
  this work. For masses below $\sim 1.4$ GeV, we use
  Ref. \cite{Donoghue:1990xh}. We use a smooth extrapolation in the region from $1.4\,\mathrm{GeV}$ to $2m_D$.}
\label{fig:ScalarBR}
\end{figure}

\begin{figure}[!t]
\begin{center}
\includegraphics[scale=0.6]{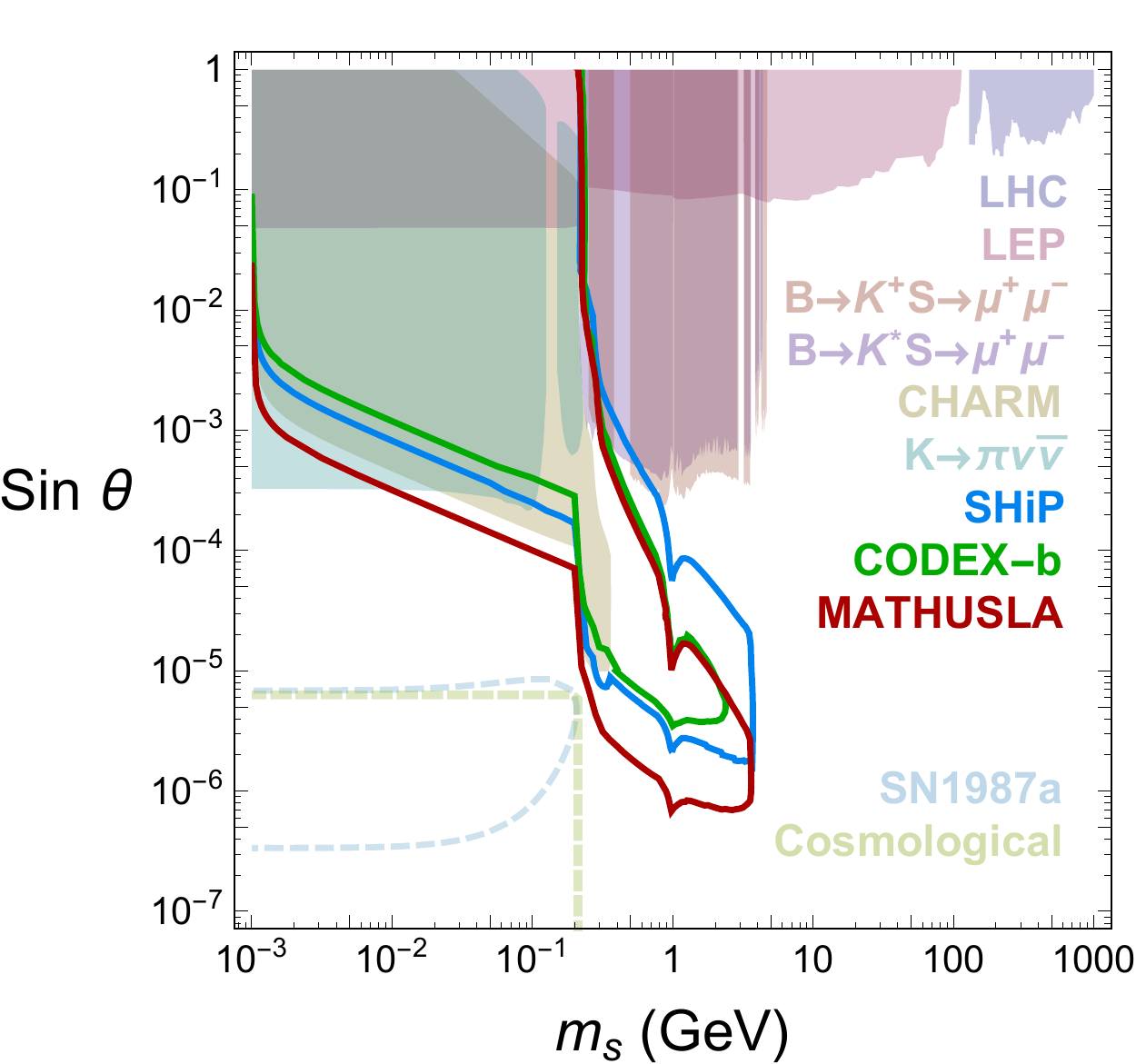}~~~
\includegraphics[scale=0.59]{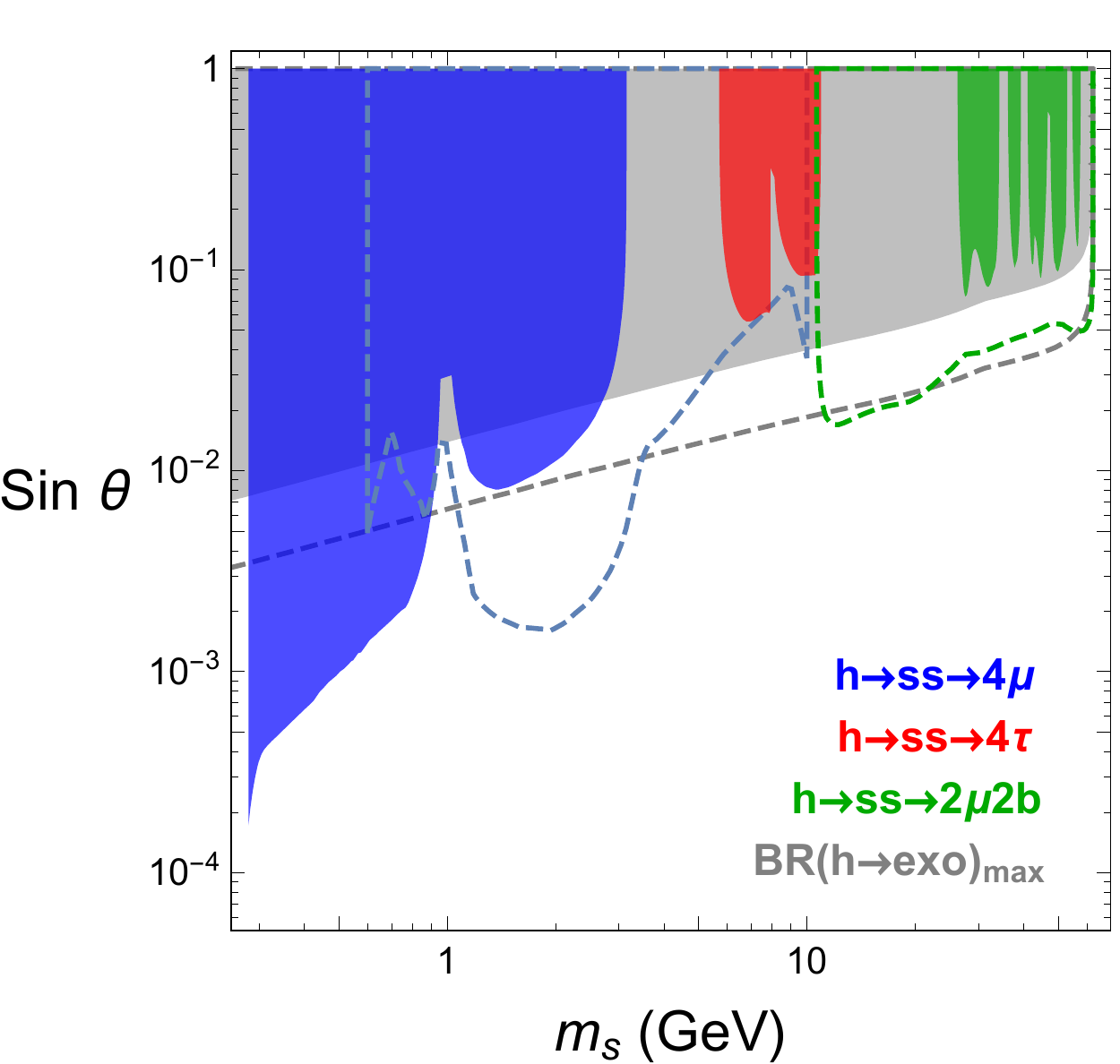}
\end{center}
\caption{ {\bf{Left:}} Summary of the current constraints on a dark Higgs
  portal scalar, $s$, from: direct searches for heavy scalars decaying
  to dibosons at LHC \cite{Khachatryan:2015cwa,CMS:2016ilx,ATLAS:2017spa} (blue-gray); Higgs
  searches at LEP \cite{Buskulic:1993gi,Acciarri:1996um,Barate:2003sz}
  (maroon); $B^+\to K^+(s\to\mu^+\mu^-)$ \cite{Aaij:2016qsm} and
 $B\to K^*(s\to\mu^+\mu^-)$ at LHCb \cite{Aaij:2015tna}
  (brown and purple, respectively); $K\to \pi+ X$, with $X$ invisible at E949
  \cite{Artamonov:2009sz} (teal); and the CHARM experiment at the CNGS
  beam-dump \cite{Bergsma:1985qz} (mustard).  The blue, green, and red
  contours show the projected sensitivity for the SHiP, CODEX-b, and
  MATHUSLA experiments respectively
  \cite{Alekhin:2015byh,Gligorov:2017nwh,Chou:2016lxi,Evans:2017lvd}. In
  the lower left corner of the plot we indicate approximate
  constraints from SN1987A (dashed blue) \cite{Krnjaic:2015mbs} and
  late-time entropy injection (dashed green). {\bf{Right:}}
  Bounds from exotic Higgs decays for fixed $m_\chi=2m_s$. In blue we
  show the region probed by searches for $h\to ss (aa)\to 4\mu$
  \cite{Khachatryan:2015wka}, in red the region probed by searches for
  $h\to ss (aa)\to 4\tau$ \cite{Khachatryan:2015nba} and finally in
  green the region probed by searches for $h\to ss ~(aa)\to 2\mu 2b$
  \cite{CMS:2016cel}. The region shaded in gray produces an exotic
  branching ratio larger than $34\%$. The corresponding dashed lines
  are the possible bounds obtained at the HL-LHC.  While constraints 
  are shown for $m_\chi=2m_s$, all of these bounds on $\sin\theta$ 
  scale roughly proportionally to $\sqrt{m_\chi}$.}
\label{fig:ScalarSummary}
\end{figure}

The leading collider and accelerator searches for the \HSFOH\ model
are again direct searches for the mediator, $s$.  Many different
low-energy and collider observables are sensitive to Higgs-portal
coupled scalars.  Additionally, there are astrophysical and
cosmological constraints on $s$ when it becomes sufficiently light and
long-lived.  Our results for the current experimental reach for a
light mediator, $s$, are presented in Fig.~\ref{fig:ScalarSummary}.  In
order to establish these results we need the SM branching ratios of
the scalar.  These branching ratios are notoriously uncertain for
scalar masses in the range between $2 m_\pi$ and $\sim 4$ GeV (see
\cite{Clarke:2013aya} for more details).  We adopt the hadronic
branching fraction derived in Ref.~\cite{Donoghue:1990xh},
supplemented with a simple extrapolation in the very uncertain 1--4
GeV region, as shown in Fig.~\ref{fig:ScalarBR}.  Features near 1 GeV
in Fig.~\ref{fig:ScalarSummary} are sensitive to this choice, which
is conservative for signals sensitive to muon decays; the overall
lifetime is also important for determining projections for SHiP \cite{Alekhin:2015byh},
MATHUSLA \cite{Chou:2016lxi}, and CODEX-b \cite{Gligorov:2017nwh}.

In the rest of this subsection, we will explain in the detail the
constraints shown in Fig.~\ref{fig:ScalarSummary}. While other
experiments have constrained this parameter space, such as KTeV
\cite{AlaviHarati:2000hs} and NA48/2 \cite{Batley:2011zz}, these
results have been surpassed by the bounds from other experiments and
will not be discussed here.  For high scalar masses 
and large portal couplings, additional constraints from perturbativity and electroweak precision tests can be 
important \cite{Lopez-Val:2014jva,Robens:2016xkb}; however, these 
constraints are model-dependent and not in our main regime of interest, and we do not discuss them further here.

\subsubsubsection{LHC (ATLAS \& CMS): Heavy Higgs Searches} 
There have been many searches looking for Higgs-like bosons at 
the LHC. The strongest limits in the region 
$130\,\mathrm{GeV}<m_s<1000$ GeV come from a search at 
ATLAS and two at CMS in the diboson decay channels 
\cite{Khachatryan:2015cwa,CMS:2016ilx,ATLAS:2017spa}.
      
\subsubsubsection{LEP (OPAL, DELPHI, ALEPH \& L3): Higgs Searches} The
LEP Working Group for Higgs boson searches combined the data from the
four experiments at LEP to place very tight constraints on Higgs
states that are produced in association with a $Z$ from 15-115 GeV
\cite{Barate:2003sz}.  For scalar masses below 15 GeV, the tightest
constraints come from L3 \cite{Acciarri:1996um}, but the ALEPH $h\to
\{\mathrm{invisible}\}$ search \cite{Buskulic:1993gi} sets slightly
stronger limits below the muon threshold where the scalar is detector
stable.
   
\subsubsubsection{LHCb: $B\to K(s\to\mu^+\mu^-)$} LHCb has made many
detailed measurements of the important observables within $B\to
K^{(*)}\mu^+\mu^-$.  Two of these have been specifically interpreted
to constrain light Higgs-mixed scalars, in the $B^0\to K^{*0}(s\to
\mu^+\mu^-)$ \cite{Aaij:2015tna} and the $B^+\to K^+(s\to\mu^+\mu^-)$
channels \cite{Aaij:2016qsm}.  We shift these limits to match our
branching ratios.

\subsubsubsection{E949 \& E787: \boldmath{$K^+ \to \pi^+
    \,+\,$invisible}} The E949 collaboration at BNL has made the most
accurate measurement of $K^+\to \pi^+\nu\bar\nu$
\cite{Artamonov:2009sz} by measuring the decays of stopped $K^\pm$.
In this study, they also reinterpret the results (including the data
from E787) to place 90\% CL upper limits on BR$(K^+\to \pi^+ X, \,
X\to\{$invisible$\})$ for $m_X<125$ MeV and $150$ MeV$<m_X<250$ MeV.
 The outer radius of the barrel veto
is 1.45 m \cite{Adler:2008zza}, and all scalars that decay before this
radius are assumed to be vetoed.  The detector itself has
$2/3\times 4\pi$ solid angle coverage. It is straightforward to
reliably apply these limits on invisible decays to metastable
particles. 
 
The NA62 experiment \cite{Fantechi:2014hqa,NA62TDR}, which will solidly establish BR$(K^+\to \pi^+ \nu\bar\nu)$ at the $\sim 10\%$ level, should be able to significantly
improve this limit (likely by an order of magnitude on
$\sin^2\theta$).  However, the experimental design of NA62 is rather
different from that of E949: notably, E949 measured stopped kaons
while NA62 measures the decays of kaons in flight, so a detailed study
is required in order to reliably assess the reach of this powerful new
experiment \cite{Evans:2017lvd}.

\subsubsubsection{CHARM: proton beam-dump} The CHARM experiment, with
a 35 m decay region located 480 m downstream from the 400 GeV CNGS
proton beam-dump, performed a search for axion-like particles
\cite{Bergsma:1985qz}.  They did not find any.  However, as various
hadrons are amply produced in the $p$-Cu interactions, this null
result can be used to constrain light scalars with long lifetimes
\cite{Bezrukov:2009yw}.  The approximate number of scalars that would
be expected to decay within the detector can be estimated as
 \beq
 N_{ev} \approx N_{s,prod} \, \mbox{BR}(s\to \mbox{\{obs\}})  \, e^{-\frac{\ell_{dec}}{\gamma c\tau_s}} \lp1- e^{-\frac{\ell_{det}}{\gamma c\tau_s}} \rp,
 \eeq
 where $\ell_{dec} =480$ m is the distance from the fixed target to the decay
 region of the experiment, $\ell_{det}=35$ m is the length of the decay
 region of the detector, $\tau_s$ is the lifetime of $s$,
 $\mbox{BR}(s\to \mbox{\{obs\}})$ is the total branching ratio of $s$
 into final states that would be detected at the experiment (in the
 case of CHARM, this is $\gamma\gamma$, $e^+e^-$, and $\mu^+\mu^-$),
 $N_{s,prod}= \sum_M N_M \mbox{BR}(M\to s+X)$ is the number of scalars
 produced in rare meson (e.g., $K$, $D$, $B$) decays, and $N_M$ is the
 number of mesons of species $M$ produced.  For the CNGS beam,
 $N_{K^\pm}=2 N_{K_L}=6.2\times10^{16}$ and $N_B=2.6\times10^{10}$.

 The proposed SHiP experiment could also achieve excellent sensitivity
 to light scalars.  At SHiP, unlike at CHARM, many kaons will be
 stopped. However, the overall luminosity is expected to be much
 higher, resulting in $N_B=3.2 \times10^{13}$,
 $N_{K^\pm}=2.9\times10^{16}$, and $N_{K_L}=1.4\times10^{15}$
 \cite{Anelli:2015pba}. SHiP should be sensitive to any of the scalar
 final states, which greatly increases the total branching ratio into
 visible states above the pion threshold.  Moreover, the SHiP
 experiment would have $\ell_{dec}=64$ m and $\ell_{det}=50$ m
 \cite{Anelli:2015pba} resulting in a much improved sensitivity.  We
 follow Ref.~\cite{Lanfranchi:2243034} in producing our estimated
 sensitivity from SHiP.
 
 \subsubsubsection{MATHUSLA and CODEX-b} The proposed MATHUSLA
 \cite{Chou:2016lxi} and CODEX-b \cite{Gligorov:2017nwh} experiments
 can be sensitive to long-lived scalars emitted in rare meson decays
 produced at the LHC.  To estimate the reach of both experiments, we
 use the distributions of $B$-mesons produced in Pythia 8.223 \cite{Sjostrand:2007gs},
 decaying to scalars following Ref.~\cite{Evans:2017lvd}.  The event rate
 is normalized to $\sigma_{b\bar b} =0.5$ mb.  For MATHUSLA, the
 detector is treated as a $200\times200\times20$ m box located 100 m
 above the interaction point and 100 m in the beam direction
 \cite{Chou:2016lxi}.  A 95\% exclusion contour is shown assuming a
 flat 75\% detection efficiency with no appreciable background for 3
 ab$^{-1}$ of data.  For CODEX-b, we consider a 10 m cubic detector
 starting from 5 m in the beam direction ($z$) and offset by 15 m in
 $x$, and centered 2 m below the interaction point in $y$.

 A third recently proposed experiment, FASER \cite{Feng:2017uoz} could 
 have sensitivity to extremely forward, boosted scalars produced in $B$ 
 decays \cite{Feng:2017vli}.  While we do not reproduce the sensitivity here, 
 the experiment would be expected to exclude additional territory below the 
 $\tau^+\tau^-$ threshold in between the sensitivity of SHiP and LHCb 
  \cite{Feng:2017vli}.
   
 \subsubsubsection{Supernova 1987A}  The observed
 duration of the neutrino pulse from supernova 1987A places a
 restriction on how much energy it could have radiated into light
 scalars that escape the core of the supernova
 \cite{Krnjaic:2015mbs,Ishizuka:1989ts}.  We follow the treatment in
 \cite{Krnjaic:2015mbs} to estimate this constraint.  The total power
 radiated into scalars per unit volume is
 \beq
 P_s \approx \frac{22}{\lp15\pi\rp^3} \lp \frac{\sin\theta f_n^{(s)} m_n f_{\pi nn}^2 T_{SN}^2}{v_h m_\pi^2}\rp^2 G\lp\frac{m_\pi}{p_F}\rp p_F^5  \, \xi\lp\frac{m_s}{T_{SN}}\rp,
 \label{eq:SNpower}
\eeq
where the pion-nucleon coupling from low-energy scattering data is
$f_{\pi nn} \approx 1.0$ \cite{Perez:2016aol}, $T_{SN}$ is the core temperature, the nucleon Fermi
momentum is given by $p_F = \lp\frac{3\pi^2 \rho_{SN}}{2 m_n}\rp^{\frac 13}$ in terms of the core density $\rho_{SN}$,
$G(x)$ is a function that contributes an $\order{1}$ factor (see
\cite{Ishizuka:1989ts} for the full expression), and $\xi(u) =
\frac{1}{2\zeta_3}\int_u^\infty dx\, x \sqrt{x^2-u^2} \lp e^x
-1\rp^{-1}$ was introduced in \cite{Krnjaic:2015mbs} to account for
finite mass effects.  A scalar only contributes to the energy loss if
it escapes the core.  The probability for a produced scalar to escape
is
 \beq
 \epsilon_{esc} \approx e^{-\frac{R_{SN}}{\gamma c\tau_s}} e^{-\frac{R_{SN}}{\lambda_s}},
 \label{eq:SNesc}
\eeq
where the mean free path of a scalar, $\lambda_s$, can be approximated
using the principle of detailed balance in terms of the equilibrium
abundance of $s$ at the core temperature, $T_{SN}$, giving $\lambda_s \approx
\rho_{s,eq}(m_s,T_{SN})/P_s$.
 
The total power radiated from the supernova in light scalars can be
written as
 \beq
 P_{rad} = P_s V_{SN} \epsilon_{esc}  < P_{max},
 \label{eq:SNradtot}
\eeq
where $V_{SN}$ is the volume of the supernova core.  The maximum
allowed power radiated is $P_{max} \approx 3\times 10^{52}$ erg/s $=
1.2 \times 10^{31}$ GeV$^2$ \cite{Raffelt:1996wa}.  To produce the
disfavored region shown in Fig.~\ref{fig:ScalarSummary}, we use the
parameters of the fiducial model in \cite{Chang:2016ntp}: $V_{SN} =
4/3 \,\pi R_{SN}^3$ with $R_{SN}=10$ km, $T_{SN}=30$ MeV, and
$\rho_{SN}=3.0 \times10^{14}$ g/cm$^3$.

However, several simplifying assumptions have been made here, 
such as the treatment of the stellar nuclei as degenerate, and a 
less approximate treatment (such as that done for dark photons in 
\cite{Chang:2016ntp}) would produce refined results.  As 
emphasized in \cite{Chang:2016ntp}, the robustly excluded region 
is smaller than the region excluded by the fiducial model, thanks 
to the uncertainty on the properties of the progenitor star, i.e., 
$T_{SN}$, $R_{SN}$, and $\rho_{SN}$.  Unlike in the case of the 
vector model, a na\"ive application of 
(\ref{eq:SNpower}--\ref{eq:SNradtot}) for the different progenitor 
star models in \cite{Chang:2016ntp} leave no regions that are 
robustly excluded. For these reasons, we present the fiducial 
SN1987A bound with a dashed line in  Fig.~\ref{fig:ScalarSummary}.   

\subsubsubsection{Cosmological constraints} Finally, for the lowest
values of $m_s$ in the WIMP next door parameter space, the scalar can
become cosmologically long-lived as all of its accessible decay modes
are suppressed by tiny Yukawa couplings.  Here, the deposition of
macroscopic amounts of entropy into the SM during and after BBN can
lead to unacceptably large decreases of the neutrino temperature
relative to the photon temperature (as measured by
$N_{\mathrm{eff}}$ \cite{Flacke:2016szy,Poulin:2015opa}), or unacceptably large disagreements between the
BBN and CMB determinations of the baryon-to-photon ratio $\eta_B$.  A
precise determination of these constraints requires a careful
treatment of the temperature evolution of the scalars, which in this
regime have thermally decoupled from the SM prior to their decay and
thus undergo cannibal behaviour \cite{Pappadopulo:2016pkp}, and is
beyond the scope of this paper.  We have indicated the region where we
estimate these cosmological constraints to be important with a dashed
boundary in Fig.~\ref{fig:ScalarSummary}.
 
\subsubsubsection{LHC searches for exotic Higgs decays} ATLAS and CMS
have developed a program of searches for Higgs decays to two singlet
scalars, $s$, or pseudoscalars, $a$. These decays give rise to the
signatures $h\to ss ~(aa)\to 4\tau$ \cite{Khachatryan:2015nba}, $h\to
ss ~(aa)\to 2\mu2\tau$ \cite{CMS:2016cqw,Aad:2015oqa}, $h\to ss
(aa)\to 4\mu$ \cite{Khachatryan:2015wka}, $h\to ss ~(aa)\to 2\mu 2b$
\cite{CMS:2016cel}, and $h\to ss ~(aa)\to 4b$
\cite{Aaboud:2016oyb}. Furthermore, searches for invisible Higgs
decays constrain the invisible width of the Higgs boson to be smaller
than $24\%$ \cite{Khachatryan:2016whc}. Finally, the ATLAS and CMS Run
I combination of Higgs coupling measurements constrain the total
exotic width of the Higgs boson to be smaller than $34\%$
\cite{ATLASCMSCombination}. We show these constraints on the right
panel of Fig.~\ref{fig:ScalarSummary}, having fixed $m_\chi=2m_s$. Presently, the  constraints from
global fits of Higgs properties (gray region) are generically stronger than the
constraints from direct searches for exotic Higgs decays, with the
exception of searches for $h\to ss (aa)\to 4\mu$ for masses below
$\sim 5$ GeV (see the blue region in Fig.~\ref{fig:ScalarSummary}
right). In the figure, we also show the possible prospects for probing
the decays $h\to ss(aa)\to2\mu2b$ (dashed green line), and $h\to
ss(aa)\to4\mu$ (dashed blue line) with 3000 fb$^{-1}$ LHC data as
studied in \cite{Curtin:2014pda} and \cite{Curtin:2014cca},
respectively.\footnote{Note that Ref. \cite{Curtin:2014cca} considers
  only the prospects for searches using four isolated muons, while the
  experimental search \cite{Khachatryan:2015wka} also captures final
  states with collimated muons.}  Also shown is the expected bound on
the exotic Higgs width from Higgs fits (dashed gray line). At the
HL-LHC, the $2\mu2b$ signature is also expected to set more stringent
bounds on light (pseudo-)scalars than the indirect bound on the exotic
Higgs width.  As the Higgs branching ratio into scalars is proportional to 
$ \alpha_Y\sin^2\theta/m_\chi^2$ (\ref{eq:scalarInt}), and we also have 
that the correct relic abundance gives roughly $\alpha_Y\propto m_\chi$ 
increasing the dark matter mass globally weakens the $\sin\theta$ 
bounds in Fig.~\ref{fig:ScalarSummary} right by approximately $\sqrt{m_\chi /2 m_{s}}$.\\

\begin{figure}[!t]
\begin{center}
\includegraphics[scale=1.2]{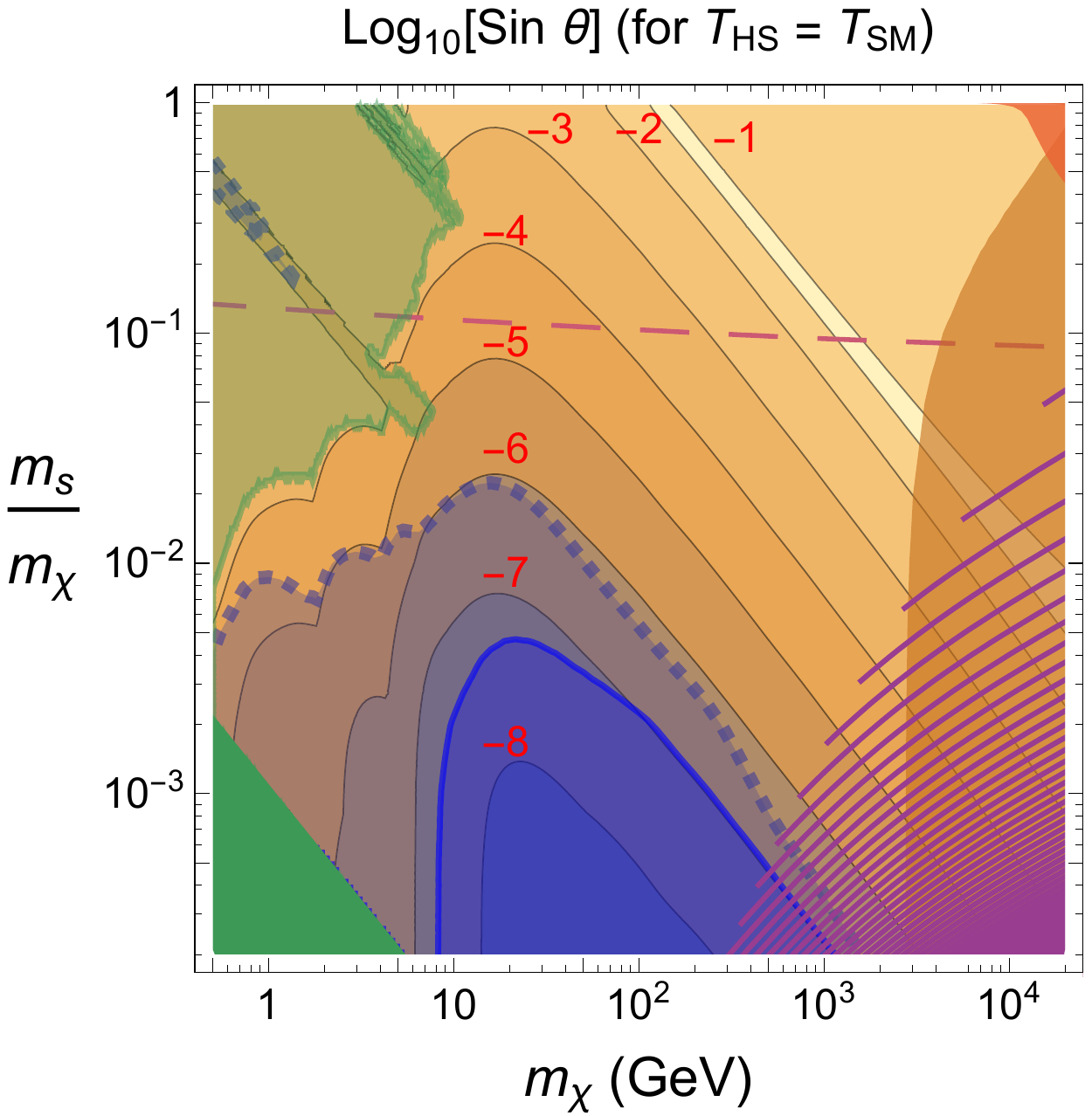}
\end{center}
\caption{Combined constraints on $\sin\theta$ in the \HSFOH\ model.
  The CMB (purple) constraints from bound state production exclude all
  values of $\sin\theta$ under the assumption that $T_{SM}=T_{HS}$ at
  the time of freezeout.  Regions where other constraints on the
  mediator $s$ (collider, beam dump, cosmology, etc.) supersede the
  direct detection constraints shown in Fig.~\ref{fig:DDScalar} are
  shown in green.  Other colors are as in Fig.~\ref{fig:DDScalar}.  }
\label{fig:MasterScalar}
\end{figure}

A summary of all constraints is shown in Fig.~\ref{fig:MasterScalar} as a function of $m_\chi$ and $m_s/m_\chi$, where we present the contours for $\sin\theta$ allowed by direct detection experiments. Also shown in the plot are the CMB constraints from bound state production (purple lines) and the regions where other constraints on the mediator, $s$, (collider, beam dump, cosmology, etc.) supersede the direct detection constraints (shaded green region). In blue, we indicate the region where the HS and the SM sectors were never in thermal equilibrium, and within the dashed blue-gray region the Higgs portal interaction equilibrated the two sectors but then fell out of equilibrium again prior to DM freezeout. As evident from the figure, over a sizable region of parameter space, current direct detection limits are sensitive enough to exclude values of the scalar portal coupling close to the thermalization floor.

\subsection{Beyond the minimal model of dark scalar interactions}

As with the vector portal model, the phenomenological features of our
\HSFOH\ model are representative of the behavior of more complicated
Higgs portal WIMPs next door (see Sec.~\ref{sec:beyondminimalZD}),
under the assumption that dark matter is a fermionic state (and that
$CP$ is conserved).

One consequence of
our decision to minimize the number of free parameters is that the
mass of the dark matter is directly tied to the VEV of the scalar and
therefore to the dark-visible Higgs mixing angle $\sin\theta$.  This
choice specifies a relationship between the portal coupling,
$\epsilon$, and the scalar-Higgs mixing angle, $\sin\theta$, for any
values of of $m_\chi$, $m_s$ once the Yukawa coupling $y = m_\chi/v_s$ is determined through the relic abundance.  Relaxing this assumption, e.g., by
considering a model with a less stringent symmetry structure, allows
the DM to have a bare mass $m_0$ independent of the scalar VEV.
 Unless there
is some theoretical reason to expect $m_0$, to be at the
same scale as $y v_s$, one would generally expect either one or the
other to dominate.  The case $m_\chi\sim yv_s\gg m_0$ is well-described by
the results shown above.  However even when $m_\chi\sim m_0 \gg y v_s$, the value
of the Yukawa coupling $y$ determined in our freezeout calculation will not change much: freezeout is
dominated by the $t$- and $u$-channel processes in (\ref{eq:sFO}), which are independent of $v_s$.
Thus the thermal relic result for $y$ is largely determined by
$m_\chi$, regardless of the origin of this mass. 
Therefore, at a fixed $m_\chi$, $m_s$, and $\sin\theta$, introducing an 
independent bare DM mass results in a smaller $v_s$, and thus a 
\emph{larger} value for $\epsilon$ (\ref{eq:tantheta}).    As discussed in
Appendix~\ref{sec:KD},  most of the scattering rates important
for thermalization are dependent on $\sin\theta$,  but the process $ss\to
f\bar f$ depends instead directly on $\epsilon$.  Decoupling the DM
mass from $v_s$ thus makes the $ss\to f\bar f$ process more important 
relative to the other processes.   As can be seen from 
Fig.~\ref{fig:rateexample} right, the $ss\to f\bar f$ processes peak near 
30 GeV and die off sharply afterwards, so that adding a bare DM mass means that for $T \sim 30$ GeV, thermalization 
becomes more efficient for the same value of $\sin\theta$.

Viewed in terms of
$\epsilon$, the case treated in detail here where $m_\chi = y v_s$ 
results in the largest parameter space above the thermalization floor, i.e., it allows
thermalization for smaller $\epsilon$.  The case where $m_\chi\gg y v_s$
has smaller $\sin\theta$ for a fixed $\epsilon$ and thus typically
requires larger $\epsilon$ to thermalize.  Insofar as direct detection
cross-sections depend on $\sin\theta$, and for most temperatures
thermalization is controlled by $\sin\theta$, the ability of direct detection
experiments to probe the thermalization floor is largely unaffected by
the introduction of a bare Majorana DM mass. The exception is in
regions $T\sim 30$ GeV where thermalization is dominated by the $ss\to f\bar f$
process so that the thermalization floor is located at smaller values of
$\sin\theta$. 
 
It is worth bearing in mind that our reference model, strictly
speaking, predicts the scale of its own symmetry breaking phase
transition.  As discussed in Appendix~\ref{sec:KD}, this phase
transition happens long before DM freezeout in our reference model;
however, this is an important caveat to keep in mind when considering
dark Higgs sectors with a different symmetry structure.

Unlike in the vector case, making the dark matter Dirac instead of
Majorana only provides quantitative shifts and no qualitative changes,
as this change does not alter the leading spin-independent matrix
element for direct detection.  A much more substantial change arises
when one considers scalar dark matter: in this case the leading
annihilation channel $\chi\chi\to s s$ is now $s$-wave, and indirect
detection signals become important and constraining \cite{Abdullah:2014lla,Martin:2014sxa}. 
(An unsuppressed $s$-wave annihilation cross-section could also arise
for fermionic DM in the presence of CP-violation \cite{Kahlhoefer:2017umn}, or if the dark
sector contains a light pseudo-scalar $a$ in addition to the light
scalar $s$, such that $\chi\chi\to a s$ can be an important
annihilation channel \cite{Nomura:2008ru}.)  Broadly
speaking, the indirect detection signals and constraints in the
presence of an $s$-wave annihilation cross-section are generally
similar to the results found for the \HSFOV\ model.  In particular,
constraints from the CMB are nearly identical, while limits from
cosmic ray searches are qualitatively similar, achieving sensitivity
to annihilation cross sections at the same order of magnitude.

Again, adding additional light states can allow for invisible mediator
decays (and, in the presence of a leading $s$-wave annihilation cross
section, mediator decays into stable dark states would allow indirect
detection signals to be re-suppressed).  Also as before, these
additional dark sector states can face strong cosmological
constraints; these can become especially acute for low-mass unstable
dark sector states, which can become very long-lived when the Higgs
portal is the leading interaction between sectors.  Importantly, Higgs
portal WIMPs next door are amenable to larger temperature drifts
between the SM and hidden sector temperatures (see right panel of
Fig.~\ref{fig:KD}), which allows more scope for relativistic species
at BBN.  Stable dark radiation species would give visible
signals in CMB-S4.

To summarize, one of the major differences between vector portal and 
Higgs portal WIMPs next door is that the Higgs portal models offer more 
opportunities to include dark radiation.  There is
slightly more model-dependence in the detailed location of the
thermalization floor, as different choices in constructing the dark
Higgs sector can alter the relationship between $\sin\theta$ and
$\epsilon$.   Straightforward extensions and variations of our minimal \HSFOH\ model
allow for the introduction of an $s$-wave annihilation cross-section
and thus reintroduce indirect detection signals and constraints, while
leaving direct detection signals qualitatively undisturbed.

\section{Summary and conclusions}
\label{sec:conclusions}

In this paper, we have comprehensively assessed the current constraints
on and discovery prospects for a  class of minimal hidden sector freezeout models,
where the DM relic abundance is set by the freezeout of a single
DM species, $\chi$, into a dark mediator, $\phi$.  We consider the
well-motivated scenario where the leading interaction between the dark
sector and the SM is renormalizable, for two simple reference
models, \HSFOV, where the mediator is a dark photon kinetically mixed
with SM hypercharge, and \HSFOH, where the mediator is a dark scalar
that mixes with the SM Higgs boson. In 
both cases, the interaction with the SM renders the mediator 
cosmologically unstable.  Experiments across the cosmic, intensity, and 
energy frontiers provide complementary information about the nature 
and cosmic history of hidden sector DM in these reference models, and leading signals can
can differ substantially from models of more traditional WIMP-like dark matter. 

We have carefully considered the cosmology of these HSFO models. 
When the interaction between the SM and dark sectors is sufficiently 
strong to ensure that the two sectors achieve thermal equilibrium prior  
to dark matter freezeout, the cosmology is highly predictive. 
When the leading interaction between the dark sector and the SM is 
renormalizable, this minimal cosmology is additionally UV-insensitive: 
the scattering that works to equilibrate the two sectors becomes 
increasingly important, relative to the Hubble rate, as the universe 
expands.  Thus, as we emphasize, our simple models of HSFO define 
a minimal and robust dark cosmology. 
We define the {\em WIMP next door} as dark matter that freezes out 
from a dark radiation bath in thermal equilibrium with the SM.  One 
major consequence for WIMPs next door is the existence
of a new cosmological lower bound on the portal coupling, the thermalization floor 
$\epsilon_{min}(T_{f})$: the minimum value of $\epsilon$ that allows the
dark sector to reach thermal equilibrium with the SM before DM freezes out at the temperature
$T_{f}$.  
We provide an initial computation of this thermalization floor for both Higgs and vector portal couplings.  
This bound is significantly more stringent than the bound from BBN 
over the vast majority of parameter space, and can be terrestrially 
interesting. While obtained in the context of our minimal models, 
these results should generally serve as a good guide for thermally 
populated DM in more general `next door' hidden sectors.

WIMPs next door provide a sharply predictive scenario for hidden
sector DM.  Requiring that the dark radiation bath attains thermal 
equilibrium with the SM prior to DM freezeout enforces a relationship 
between the temperatures of the two sectors, so that the parameter
space is bounded and clearly defined.  Both the DM and mediator
masses are bounded from below by BBN and from above by
perturbativity, while the coupling of the mediator to the SM is
bounded from below by $\epsilon_{min}(T_{f})$. 

Dark matter direct detection experiments can access a significant
portion of the parameter space for WIMPs next door, and in some
regions can probe all the way down to the thermalization floor
$\epsilon_{min}(T_{f})$.  Low-mass ($m\lesssim 10$ GeV) WIMPs next
door with unsuppressed $s$-wave annihilation cross-sections (like in the 
\HSFOV\ model) can be robustly excluded from their impact on the CMB.  
On the other hand, WIMPs next door with a velocity-suppressed $p$-wave 
annihilation cross-section (like in the \HSFOH\ model) predict interesting 
direct detection signals for DM candidates in the low-mass range that offer 
an attractive target for low-threshold direct detection experiments.  
Additionally, unlike in the case of standard WIMPs, a notable fraction of the viable 
WIMP next door parameter space dwells underneath the coherent 
neutrino scattering floor, providing a target for future direct detection 
experiments that would need some ability to distinguish these signals 
from the neutrino background.  

The leading accelerator signature of WIMPs next door is the production 
of dark {\em mediators}.  A variety of experiments currently constrain both 
kinetically-mixed vectors and Higgs-mixed scalars.  
Both existing experiments, such as LHCb and NA62, and proposed 
experiments, like SHiP, MATHUSLA, CODEX-b, and FASER, project sensitivity to 
significant regions of unexplored territory,  and will either lead to a revolutionary discovery or greatly improve the  
constraints on this parameter space.  At higher masses, the multi-purpose LHC 
experiments have the best opportunities to discover the mediators in exotic 
Higgs decays (Higgs portal) or through direct production (vector portal).  By contrast the traditional LHC mono-$X$ searches give little hope of finding DM.

One avenue for future work is improving on our estimates 
of the thermalization floor, most critically through the SM's two phase transitions.  In particular, a
 careful treatment of dark photon production through the chiral phase transition could be important for understanding windows for dark radiation and thus low mass ($m\lesssim 10$ GeV) vector portal WIMPs next door.

\section*{Acknowledgments}
\noindent
We are grateful to H.~An, G.~Kribs, A.~Long, S.~McDermott, B.~di
Micco, T.~Robens, V.~Martinez Outschoorn, M.~Williams, Y.~Zhang, and J.~Zupan for useful
discussions.  The work of JS and JAE is supported in part by DOE grant
DE-SC0015655. JS acknowledges additional support from DOE grant
DE-SC0017840.  SG acknowledges support from the University of
Cincinnati. SG is supported by a National Science Foundation CAREER Grant No. PHY-1654502. JAE and SG thank the Aspen Center for Physics, which is
supported by National Science Foundation grant PHY-1066293, where part
of this work was performed. SG is grateful to the hospitality of the
Kavli Institute for Theoretical Physics in Santa Barbara, CA,
supported in part by the National Science Foundation under Grant
No. NSF PHY11-25915, where some of the research reported in this work
was carried out.

\appendix

\section{Thermal (De)coupling}
\label{sec:KD}

In this appendix, we describe in detail our estimate of the
minimum portal coupling necessary to thermalize the hidden sector with
the Standard Model in the early universe, and point out some ways to
improve on our treatment.  We always assume that the particles of the
hidden sector (i.e., the DM, $\chi$, and the dark mediator, $\phi$)
rapidly thermalize among themselves and can be characterized by a
single temperature.  As we focus on the regime where the mediator
forms a radiation bath at the time of equilibration, this assumption
is well justified.

We begin with some general comments.  First, we are mainly interested
in the thermal interaction rates between particles at temperatures
$T\gtrsim m$, where classical statistics do not apply.  Once final
state blocking and/or enhancement factors can no longer be neglected,
the evaluation of collision terms becomes significantly more
technically involved.  Fortunately, the SM thermal bath is dominated
by fermions: empirically, classical statistical
(``Maxwell-Boltzmann'') treatments of relativistic scattering processes
involving fermions provide a reasonable approximation to
the full quantum statistical expressions, agreeing within a factor of
$\lesssim 2$ (see, e.g.,~\cite{Adshead:2016xxj}).  Thus we employ
classical statistics to evaluate the rates for $2\too 2$ scattering
processes like $\phi f \to (g/\gamma) f$, $\phi f\to h f$, and the
crossed processes $\phi (g/\gamma)\to f\bar f$, $\phi h\to f\bar f$,
etc.

The other major simplifying approximation we make is to neglect $2\too
2$ scatterings with EW gauge bosons.  This is a good approximation
thanks in large part to the sheer numerical dominance of quarks in the
SM plasma, combined with $\alpha_s > \alpha_2$ and the larger color
factors present in QCD scattering amplitudes.  The processes $\phi f
\to (W,Z) f'$ and their crosses are thus numerically unimportant
compared to $\phi f \to g f$ at high temperatures at our level of
precision.  At temperatures $T\ll m_W$, the $W$, $Z$ masses render
these scatterings irrelevant.  Meanwhile all-bose processes such as
$\phi V \to V V$ are only important for a small range of temperatures
at and below the electroweak crossover $T_c\approx 160$ GeV
\cite{DOnofrio:2014rug} before Boltzmann suppression kicks in. A study
of dark mediator production from electroweak boson scattering in this regime is interesting, but
involves a careful treatment of (evolving, nonperturbative) thermal
masses, and is beyond the scope of this paper.  Above the electroweak
crossover, the leading scattering processes that mediate
thermalization have a different structure, as discussed further for
each model below.

We incorporate three-loop running of $\alpha_s$ above the chiral phase
transition (everywhere in this work, we use $\Lambda_{QCD}\equiv300$ 
MeV).  However, below the QCD phase transition, $2\too3$ pion processes, 
 e.g., $\pi^+\pi^0 \to \pi^+\pi^0 Z_D$, dominate.  Due to the qualitative 
 similarities, these are lumped into our $2\too2$ processes in the 
 discussion below. In order to estimate these processes, 
we expand the chiral Lagrangian to leading order in $\{p,m_\pi\}/4\pi f_\pi$ to 
compute the relevant cross-sections. As the thermally averaged cross-section receives important contributions from values of $s$ 
where this expansion is no longer reliable, we introduce a simple regulator that ensures  $\sigma(s)$ has 
physical high $s$ behavior ($\sigma(s\gg m_\pi) \propto s^{-1}$).  As the 
extremely broad QCD $\sigma$ $(f_0)$ resonance would be expected to 
perform the bulk of the unitarization of pion scattering, we define our 
multiplicative regulator to be
\beq
\mbox{Reg}(s) = \left\{
\begin{tabular}{ll}
1 & $s \leq m_\sigma$ \\
$ \left[ \frac{m_\sigma^2 \Gamma_\sigma^2}{(s-m\sigma^2)^2+m_\sigma^2 \Gamma_\sigma^2}\right]^\xi$ & $s > m_\sigma$
\end{tabular}\right.,
\eeq
where the exponent $\xi$ is chosen to enforce the desired UV behavior and 
$m_\sigma$ ($\Gamma_\sigma$) are set to 500 (600) MeV.  While this 
assumption is grounded in physical expectations, it is a somewhat arbitrary 
choice and a different regulator could modify the results significantly.  Further, 
while pion processes are included, kaons and other light QCD resonances, 
have been neglected.  These states have a higher Boltzmann suppression, 
which should generally make their contribution subleading, but their contribution 
may still be considerable for scalar production thanks to the larger strange 
Yukawa coupling.  Lastly, near the phase transition the strong self-interactions of the hadronic plasma likely make significant thermal corrections to the mediator production rates.  For these reasons, our $\epsilon_{min}(T_{eq})$ 
results in this region are much more uncertain than the roughly factor of two 
precision that we have elsewhere.  In Figs.~\ref{fig:rateexample} 
and \ref{fig:KD}, we shade  temperatures near both the electroweak ($T\sim 160$ GeV) and QCD ($T\sim 300$ MeV) phase transitions to  highlight the large uncertainties 
in these areas.  

At the level of precision we are using, we can check whether a process
is in equilibrium simply by comparing it to the Hubble rate, requiring
\beq
 \Gamma_{int}(T) > H(T) = \sqrt{\frac{4\pi^3 g_*(T)}{45}} \frac{T^2}{M_{pl}}.
 \label{eq:Kdcond}
\eeq
Here the effective number of relativistic degrees of freedom $g_*(T)$ 
includes degrees of freedom from the dark mediator as well as those 
of the SM.  We have checked that this simple equilibration
criterion reproduces the results of a full numerical treatment of
the energy transfer rate between sectors to within an
$\mathcal{O}(1)$ factor; see also \cite{Kuflik:2017iqs,
   evansshelton}.

    \begin{figure}[t]
\begin{center}
\raisebox{0.1cm}{\includegraphics[scale=0.6]{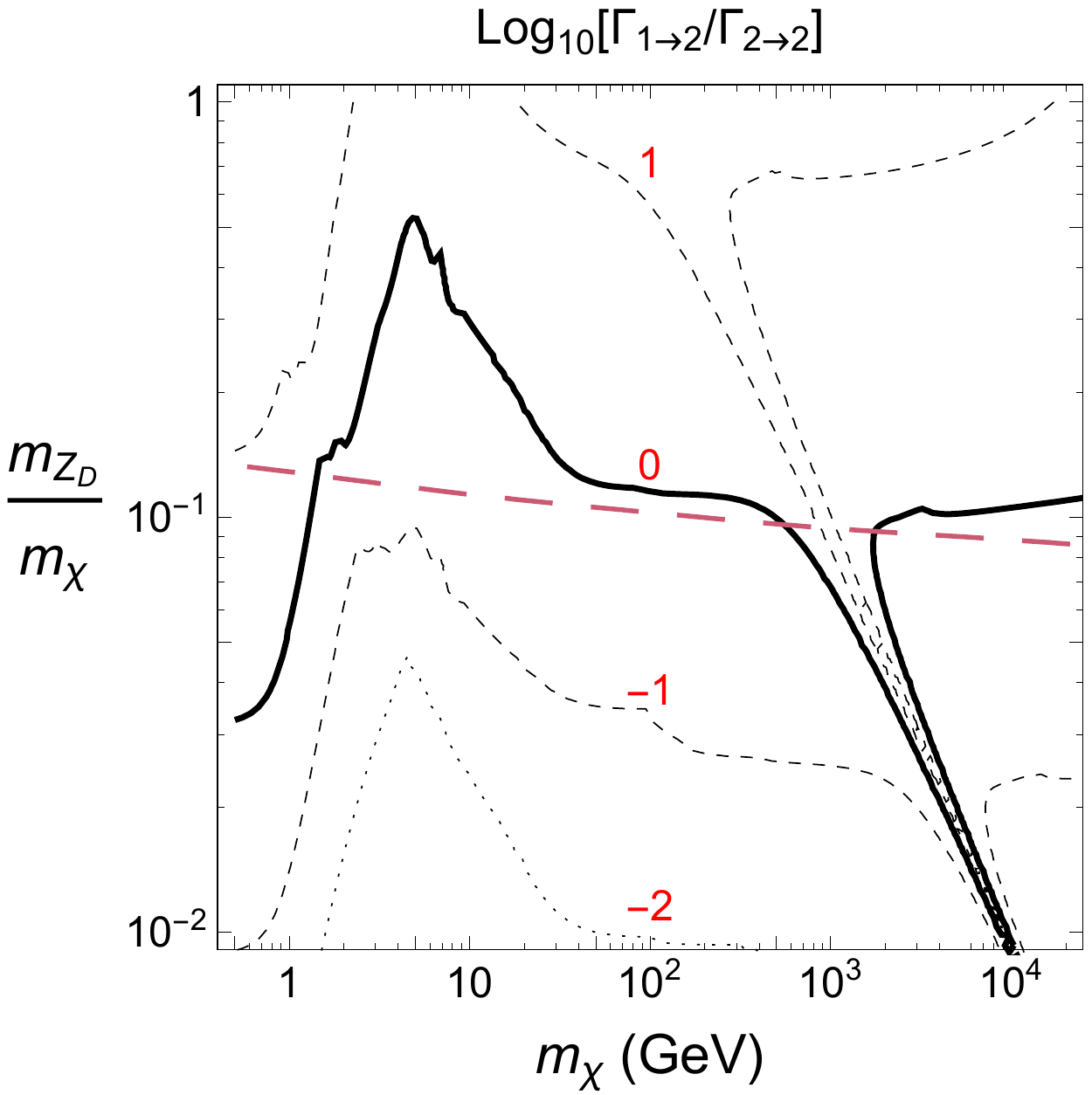}}~~~~
\includegraphics[scale=0.60]{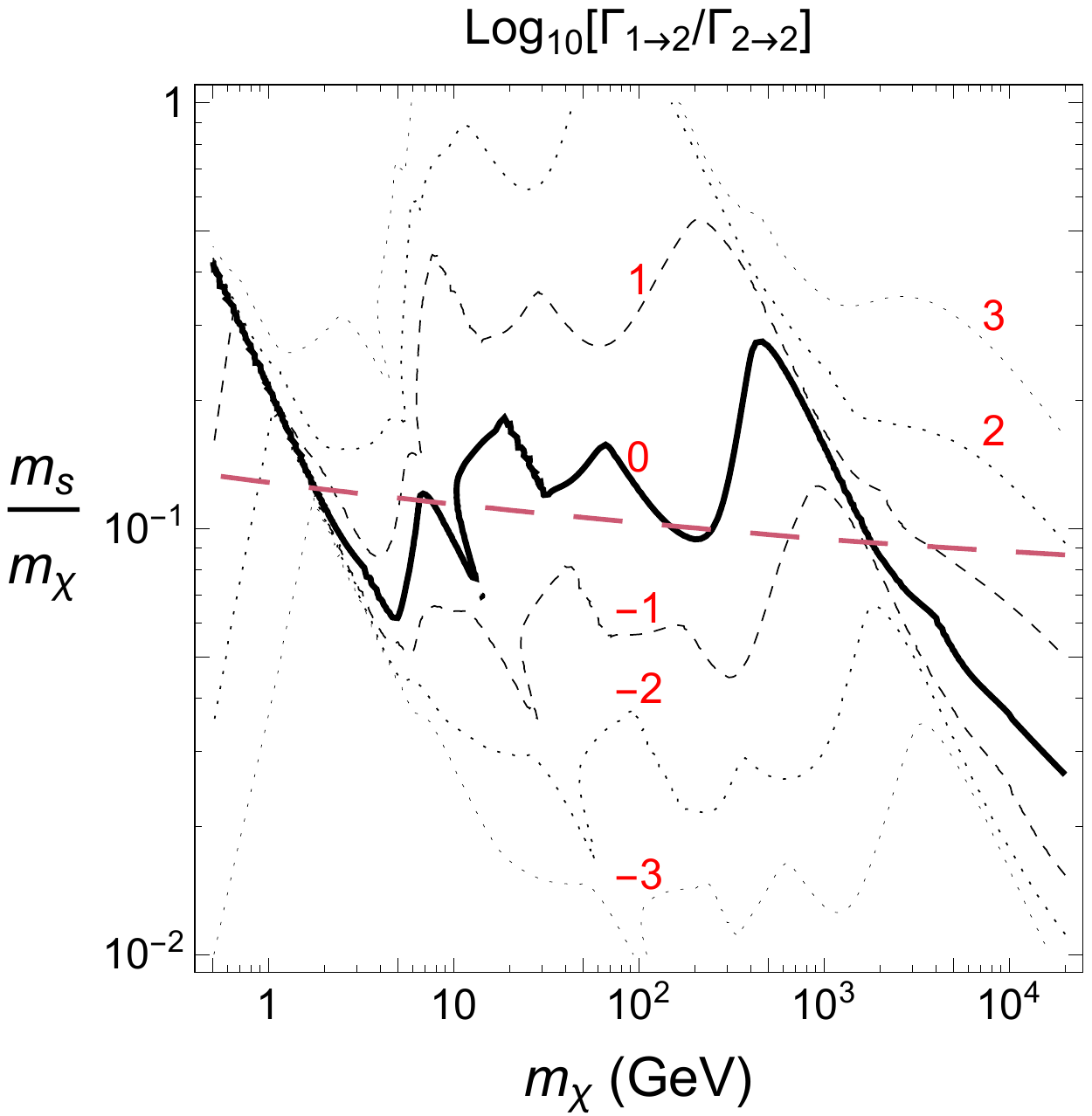}
\end{center}
\caption{ Contours showing the ratio of $1\too 2$ to $2\too 2$
  scattering rates in our minimal models.  In regions below the pink
  dashed line, the dark mediator furnishes a radiation bath at the
  time of DM freezeout.  In the region where $\Gamma_{1\too 2}\gtrsim\Gamma_{2\too 2}$,
 the true value of $\epsilon_{min}(T_{eq})$ is lower than the value we
 quote from $2\too2$ processes alone.
   {\bf Left:} \HSFOV.  For dark vector masses near the $Z$-pole, 
   the $1\too2$ grows very large. 
   {\bf Right:} \HSFOH.  }
\label{fig:rateratio}
\end{figure}

The ratio of scattering rates $\Gamma_{1\too 2} /\Gamma_{2\too 2}$ in
both \HSFOV\ and \HSFOH\ models is shown in Fig.~\ref{fig:rateratio}.
Regions below the dashed pink line in Fig.~\ref{fig:rateratio} have a
radiation bath at freezeout. The figure demonstrates that $2\too
2$ scattering processes dominate thermalization in almost all of this
region.

\subsection{Vector Decoupling}
\label{sec:KDV}
 
\begin{figure}[!t]
\begin{center}
\includegraphics[scale=0.6]{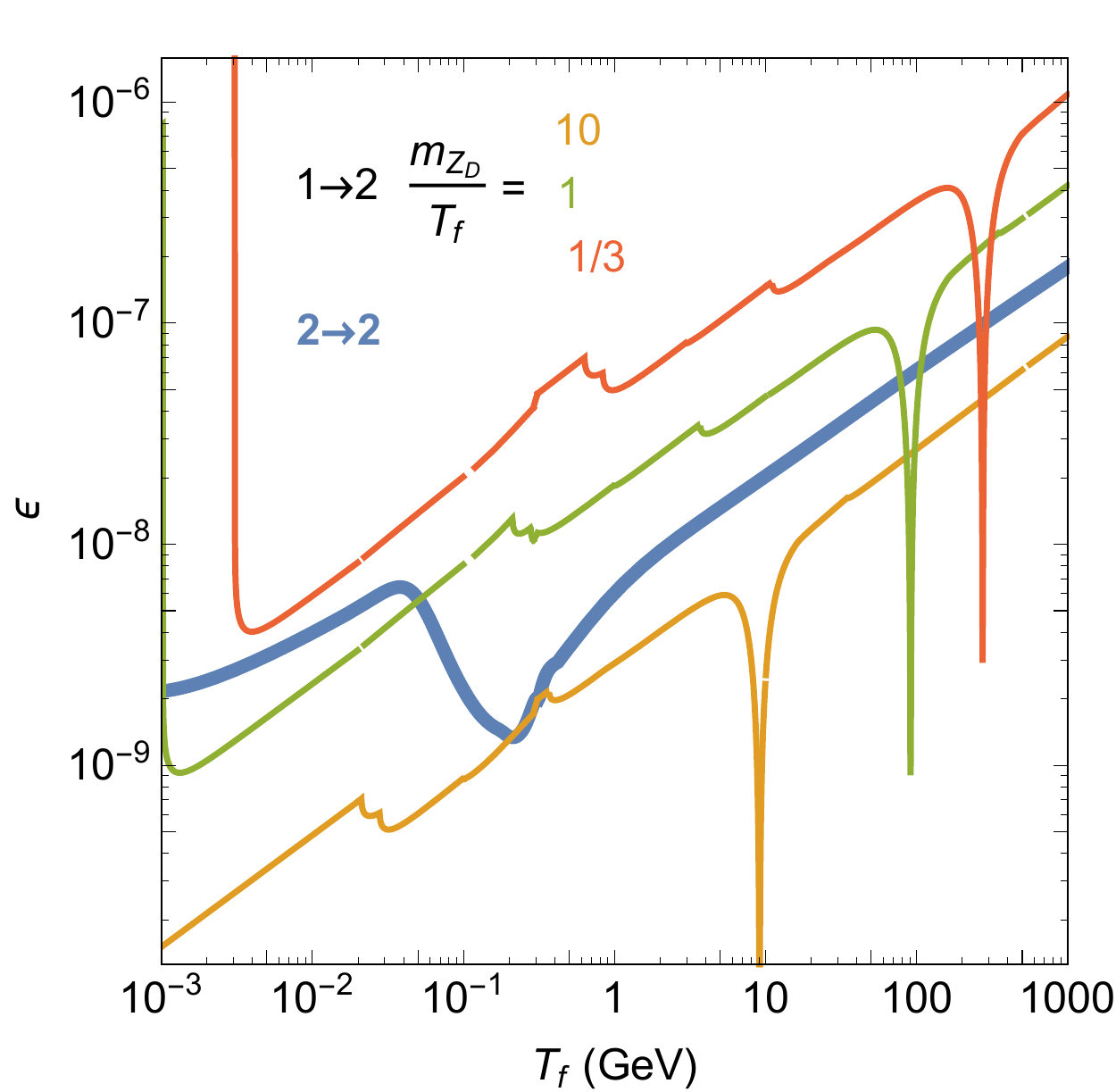}~~~
\includegraphics[scale=0.6]{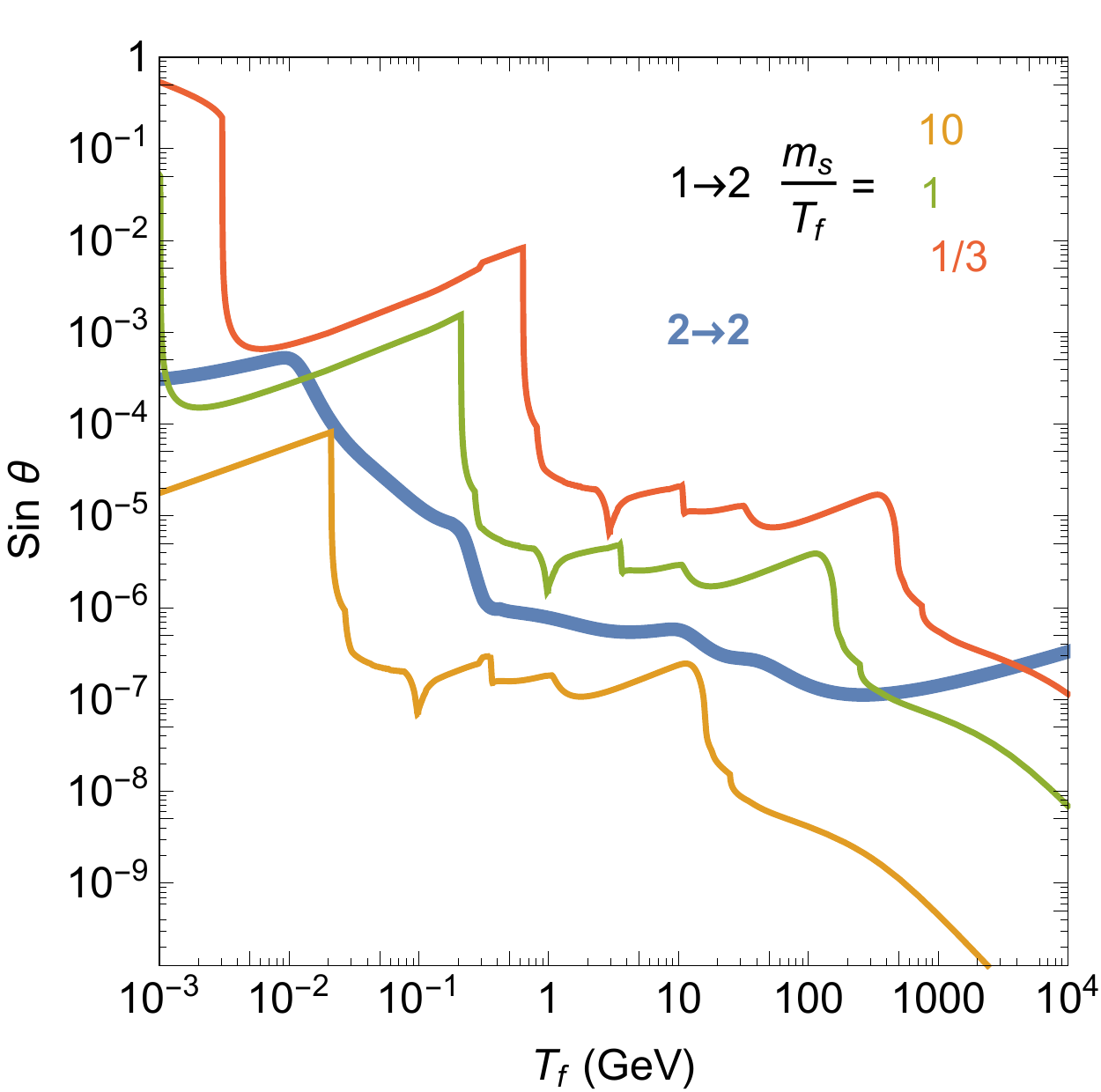}
\end{center}
\caption{Comparison of the
minimum allowed value for the portal coupling for only the $2\too 2$ processes
and only the $1\too 2$ process for fixed values of $m_{\phi}/T_{f}$. {\bf{Left:}} The
  \HSFOV\ model for three choices of $m_{Z_D}/T_{f}$ where
  $m_\chi/T_f \sim 20$.  For $m_{Z_D}\lesssim T_f$, $1\too 2$
  processes are unimportant except at very small masses and for
  $m_{Z_D}\sim m_Z$.  For $m_{Z_D}\gg T_f$, $1\too 2$ processes can
  dominate everywhere.  {\bf{Right:}} As left figure, but for the
  \HSFOH\ model.  Here when $m_s\lesssim T_f$, $1\too 2$ processes
  are unimportant except at very small masses and for very large
  masses, where the very large decay rate into $W$ and $Z$ bosons
  takes over.  Again, for $m_s \gg T_f$, $1\too 2$ processes are
  dominant.  }
\label{fig:2to2v1to2}
\end{figure}

In our vector model, the $2\too 2$ rate is dominated by the scattering
processes $Z_D f \to g/\gamma f$, $Z_D g/\gamma \to f \bar f$, $Z_D f
\to h f$, and $Z_D h \to f \bar f$.  For each of these cross-sections,
$\sigma_{12\to34}$, the thermal average given by Maxwell-Boltzmann
statistics is \cite{Edsjo:1997bg}
  \beq
\vev{\sigma_{12\to34} v}_T n^{eq}_2 =  \int_{\sqrt{s_{\mathrm{min}}}}^\infty \frac{g_2 \lp s^2 -2s(m_1^2+m_2^2)-(m_1^2-m_2^2)^2\rp K_1\lp\frac{\sqrt s}{T}\rp}{8\pi^2 m_1^2 K_2\lp\frac{m_1}{T}\rp}  \sigma_{12\to34}(s)  d\sqrt{s},
\eeq
where $\sqrt{ s_{\mathrm{min}}}=\mbox{Max}[m_1+m_2,m_3+m_4]$ and $K_n$ is the modified Bessel function of the second kind. This expression for the cross section is related
to the total reaction rate by
\beq
\Gamma_{int,2\to 2}^{Z_D}(T) = \sum_{X,Y,Z}  \vev{\sigma_{Z_DX\to YZ} v}_T n^{eq}_X.
\eeq
If $\Gamma_{int,Z_D}(T) < H(T)$ for all $T>T_f$, then the two
sectors are not in thermal equilibrium at the time of freezeout and
may have completely different temperatures.  The resulting value of
$\epsilon_{min}(T_f)$ required for the sum of these $2\too 2$
scattering rates to equilibrate the dark sector with the SM is
indicated by the blue curve in the left panel of Fig.~\ref{fig:2to2v1to2}.
Individual contributions to the scattering rate are shown in
Fig.~\ref{fig:rateexample}.  In this model, $T_f \sim m_\chi /
(30-20)$, with the larger splitting for $m_\chi\sim 10$ TeV and the
smaller splitting for $m_\chi\sim 1$ GeV.\footnote{We use
  $2Y_{eq}(T_f)\equiv Y_0(T_f)$ as the definition of our
  freezeout temperature, as determined numerically in our freezeout
  calculation.}

The rate of inverse decay processes can potentially be larger than the
$2\too 2$ rate.  We continue to neglect final state blocking and
stimulated emission factors in evaluating this rate.  Here this
approximation is reasonable as the $1\too 2$ rate becomes most
important in comparison with the $2\too 2$ rate as $T\sim m_{Z_D}$.
We include all SM species in evaluating this rate, including EW gauge
bosons. We compute the $1\too 2$ rate as
\beq
\Gamma_{int, 1\to 2}^{Z_D}= \frac{1}{n_{Z_D,eq}}\int \frac{d^3p}{(2\pi)^3} f_{Z_D,eq}(E) \frac{m_{Z_D}}{E}\Gamma_{Z_D},
\eeq
using the Bose-Einstein distribution $f_{eq}(E)$ for the $Z_D$.  This yields 
\beq
\Gamma_{int,1\to 2}^{Z_D}(T) =\Gamma_{Z_D} \times  \left\{ 
\begin{tabular}{lcr}
 $\displaystyle{\frac{\pi^2}{12 \zeta_3}\frac{m_{Z_D}}T}$ && $ T \gg m_{Z_D}$ \\
& \hspace{5mm} \phantom{.} &\\
 $\displaystyle{\frac{K_1\lp\frac{m_{Z_D}}T\rp}{K_2\lp\frac{m_{Z_D}}T\rp}}$ && $T \lesssim m_{Z_D} $ 
 \end{tabular}
 \right.
\eeq
where $ \Gamma_{Z_D}\propto \epsilon^2$ is the zero-temperature width
of the dark vector. Here in the top line we have used the
Bose-Einstein result for the relativistic $n_{Z_D, eq}$, while the
bottom line gives the Maxwell-Boltzmann result.  At low temperatures,
the Bessel function ratio asymptotes to unity.  As the interaction
rate for the $2\too 2$ processes scales as $\Gamma_{int, 2 \too 2}^{Z_D}
\propto T$ in the UV and $\Gamma_{Z_D}\propto m_{Z_D}$, this results
in a rough parametric scaling of
\beq
\frac{\Gamma_{int,2 \to2}^{Z_D}}{\Gamma_{int,1\to 2}^{Z_D}}\approx \{\mbox{few}\}\times \{\alpha_s\mbox{ or } \alpha_{EM}\} \frac{T^2}{m_{Z_D}^2},
\label{eq:parscal}
\eeq
so decays and inverse decays are typically unimportant for 
thermalization unless $T_f\lesssim m_{Z_D}$.  A comparison of the
minimum allowed $\epsilon$ value for only the $2\too 2$ processes
(which do not depend on the vector mass) and only the $1\too 2$
process for fixed values of $m_{Z_D}/T_{f}$ is shown in the left panel of
Fig.~\ref{fig:2to2v1to2}.  As the width of the dark photon
rapidly increases for fixed $\epsilon$ at $m_{Z_D}\sim m_Z$, much
smaller values of $\epsilon$ can thermalize the two sectors in this
region.  Below the QCD confinement scale, $2\too 3$ pion processes 
briefly dominate, but after they become subdominant near temperature 
of $\sim50$ MeV, $2\too 2$ process becomes proportional to 
$\alpha_{EM}$, and decays and inverse decays become relatively 
more important for thermalization.  For $m_{Z_D}\sim m_\chi \gg 
T_f$, the large width of $Z_D$ allows for this to dominate the 
thermalization.

The narrow green region in Fig.~\ref{fig:KD} corresponds to a 
scenario where at higher temperatures near 200 MeV, the hidden sector 
and standard model were in thermal equilibrium, but have since 
decoupled.  The temperatures of the two sectors are then allowed to 
drift apart.  For the vector model, this is a very narrow region of 
parameter space, so we defer our detailed discussion of this 
interesting region to the scalar decoupling section near (\ref{eq:Tdrift}).

Finally, depending on the origin of the dark vector mass, the dark
vector model may implicitly contain a symmetry-breaking phase
transition where the mass of the dark vector is generated.  In this
case, when both SM and dark sectors are in the unbroken phase, the
leading scattering process responsible for bringing the two sectors
into thermal equilibrium is $f\bar f \to \chi\chi$ \cite{Chu:2011be}.
As this process depends on the dark Yukawa coupling, and $\alpha_D\ll
\alpha_S$, for the purposes of determining the minimal portal coupling
that can yield thermalization, the unbroken UV interaction rate is
unimportant compared to the interaction rate after both electroweak symmetry breaking
 and dark symmetry breaking.

\subsection{Scalar Decoupling}
\label{sec:KDS}

The dark scalar model predicts the critical
temperature, $T_c$, of its phase transition from the symmetric vacuum
($\langle S\rangle = 0$) to the broken vacuum where $S$ develops a
VEV.  This phase transition occurs comfortably prior to DM freezeout,
$T_c \gg T_f$, as we now demonstrate.  This model exhibits a
second-order phase transition, so $T_c$ occurs when the second
derivative of the thermal potential at the origin changes sign.  To
estimate the critical temperature and understand its relation to other
mass scales in the dark sector, it suffices to consider the one-loop
approximation to the thermal effective potential for $S$, yielding
\bea
V''(0)&=&\!\!\!\!\!\!
\frac{1}{2}\left( \frac{3m_s^4y^2}{16\pi^2 m_\chi^2}
\left[1+\gamma_E+\ln\left(\frac{m_s}{ 4\pi T}\right) \right]  -\frac{y^2 m_\chi^2}{4\pi^2}+y^2T^2\left(\frac{1}{6}+\frac{m_s^2}{4 m_\chi^2}\right)-m_s^2\right. \;\;\\
&& \left. -\,T \frac{3 y^2 m_s^2}{4\pi m_\chi^2} \sqrt{ \abs{T^2\left(\frac{y^2}{12}+\frac{y^2m_s^2}{8 m_\chi^2}\right)-\frac{m_s^2}{2}}}  \mbox{ Sign}\!\left[ y^2 T^2\left(\frac{1}{6}+\frac{m_s^2}{4 m_\chi^2}\right)-m_s^2\right]\right).
\label{eq:phasetransition}
\eea
Setting (\ref{eq:phasetransition}) to zero, we can estimate $T_c$.  If
$T_c\gg T_f$, then DM freezeout occurs during the broken phase.  At
small scalar masses, the largest contributions to
(\ref{eq:phasetransition}) are $V''(0) \sim \frac1{12}
y^2T^2-\frac{1}{8\pi^2}y^2 m_\chi^2$, so that $T_c \sim \sqrt{\frac32}
\frac{m_\chi}{\pi}\sim 0.4 m_\chi \gg T_f$.  For larger scalar
masses, $V''(0) \sim \left(\frac{1}{12}+\frac{m_s^2}{8
    m_\chi^2}\right) y^2T^2 -\frac 12m_s^2$, and the critical
temperature can be much higher than $m_\chi$.

Until the electroweak phase transition, the scalar's only tree-level
interactions with the SM are with the Higgs multiplet.  In the
unbroken phase, above both the electroweak and dark sector phase
transitions, there is thus a single process controlling the
equilibration of the two sectors, $ss\to hh$\footnote{The crossed
  process $sh\to sh$ also assists thermalization by contributing to
  the energy transfer rate between sectors.  While this process does
  not change the number of dark particles, there exist rapid
  number-changing $s$ self-interactions that serve this function.  The
  $sh\to sh$ scattering rate is $\sim 10\%-20\%$ larger than the
  $ss\to hh$ scattering rate, but provides a highly subleading
  contribution to the energy transfer rate once the temperatures of
  the two sectors are similar.  Thus it is sufficient to estimate
  thermalization based on the scattering $ss\to hh$.}.  As this
process involves only bosons with masses $m_i\ll T$, to accurately
determine the interaction rate it is necessary to use Bose-Einstein
statistics.

Using the techniques of \cite{Birrell:2014uka}, the thermally averaged
scattering rate can be expressed as an integral over the total CM
energy-squared, $s$, and $p$, the magnitude of the three-momentum of
the CM frame in the rest frame of the plasma.  Defining, as usual, the
scattering rate $\Gamma_{UV}$ as the collision term divided by the
(equilibrium) number density of one of the initial state particles, we
have
\barray
\Gamma_{UV} &=& \frac{\epsilon^2 }{256\, \zeta(3)T } \frac {1}{(2\pi)^3}  \int_{s_0}^{\infinity} ds\int_0^{\infinity}
\frac{dp}{p^0}  \,\frac{1}{ \sinh^2 (\beta p^0/2)} 
 \left[\ln\left(\frac{\cosh (\beta (p^0+\beta_1 |\vec p|)/2)-1}{\cosh (\beta
       (p^0-\beta_1 |\vec p|)/2)-1}\right) \right] \nonumber \\
&&\times \left[\ln\left(\frac{\cosh (\beta (p^0+\beta_2 |\vec p|)/2)-1}{\cosh (\beta
       (p^0-\beta_2 |\vec p|)/2)-1}\right) \right].
\label {eq:rate}
\earray
Here $p^0 \equiv \sqrt{s + p^2}$, $\beta = 1/T$, and $\beta_i \equiv
\sqrt{1-4 m_i^2/s}$. We evaluate this integral numerically.  It
diverges as the lower limit on $s$ is taken to
zero, $s_0\to 0$, reflecting the divergence in the Bose-Einstein
distribution $f(E)$ as $E\to 0$.  This divergence is regulated by the
thermal masses of the scattering particles, $s_0 = 4 \max (m_H(T)^2,
m_s(T)^2)$.  With $m_i(T)\propto T$, we find $\Gamma_{UV}\propto T$,
as we must.  The size of the UV scattering rate thus depends
indirectly on couplings internal to the two sectors through their role
in determining the thermal masses of $S$ and $H$.  Smaller thermal
masses cut off the divergence at a lower value of $s_0$, and hence
increase the rate.  The relatively large couplings of the SM Higgs to
the top quark, electroweak gauge bosons, and (to a lesser extent)
itself ensure that $m_H(T)$ determines $s_0$, making $\Gamma_{UV}$
relatively insensitive to the detailed couplings of $s$ within the
hidden sector.  We find that, for a fixed value of $\epsilon$, the two
sectors will thermalize in the unbroken phase only if they could also
thermalize within the broken phase as well.  In other words, the lower
bound on $\epsilon$ that we find from requiring $\Gamma_{UV}(T) =
H(T)$ at high temperatures is subdominant to the lower bound found by
requiring $\Gamma_{IR}(T) = H(T)$ at temperatures below the phase
transitions.  Thus to understand the process of thermalization through
the $s$-$h$ interaction, it suffices to study the scattering rates in
the broken phase.
  
At temperatures below the electroweak phase transition, but before the
dark phase transition, the dominant processes are $ss\to f \bar f$
processes via a SM-like Higgs mediator.  By far the most important
contribution here is $ss\to b\bar b$, which for temperatures near
$T\sim 30$ GeV benefit from the $s$-channel enhancement for transit
through a nearly on-shell Higgs (regulated with a width set to the SM
value $\Gamma_h=4.15$ MeV, with thermal effects
neglected).\footnote{The treatment of $s s\to f\bar f$ thus also
  includes the important process where $s s\to h$; as this process is
  not considered separately, there is no problem with double
  counting.}  As this process depends on $\epsilon$ rather than
$\sin\theta$,
\beq
\epsilon = \sin \theta  \frac{ym_h^2}{m_\chi v_h} +\order{\sin \theta \frac{m_s^2}{m_h^2},\sin^3 \theta},
\eeq
decreasing the dark matter mass enhances this rate, and for
temperatures $T = \mathcal{O}(10\,\mathrm{GeV})$ it becomes the
dominant factor in determining whether the two sectors have ever 
been in equilibrium; see Fig.~\ref{fig:rateexample}.
  
After both the SM Higgs and the hidden sector scalar have VEVs, many
processes can contribute to thermalization.  Here the dominant ones
are $sh\to f\bar f$, $sf\to h f$, $sg/\gamma\to f\bar f$, $sf\to
g/\gamma f$, as well as the $ss\to f \bar f$ processes discussed
above, which are unaffected at $\order{\sin\theta}$ by the dark phase
transition and thus remain important in the broken phase. For temperatures $T\ll m_t$, mediators may also be produced through $sg\too gg$.  This rate is logarithmically divergent, and we estimate it by cutting off the log divergence with a finite thermal mass for the gluon.  Our estimate indicates $sg\too gg$ is a subdominant contribution.  Below the QCD phase transition, $\pi\pi\too\pi\pi s$  
  processes dominate briefly.  Our treatment of these thermal scattering rates
follows that outlined in the case of the vector model above.

Once again, decays and inverse decays become important when
$m_s\gtrsim T_f$, as can be seen in the right panel of Fig.~\ref{fig:2to2v1to2}.
Our treatment of the $1\too 2$ scattering rates in the Higgs portal
model follows the treatment described for the vector portal
above. Again, we include all SM contributions to the scalar width.

As the hidden sector scalar couples to SM fermions with strength
proportional to the fermion masses, the $2\too 2$ interaction rate
$\Gamma$ drops rapidly after crossing a fermion mass threshold as
these massive particles drop out the thermal bath.  Thus, it is
possible that for some other $T>T_f$, (\ref{eq:Kdcond}) is
satisfied, but not at $T_f$,  so that the SM and the hidden sector
were at one point in thermal equilibrium, but have since decoupled.
When this happens, their temperatures drift apart as
\beq
T^{HS}=\lp \frac{g^{SM}_{*S}(T^{SM}) g^{HS}_{*S}(T_D)}{g^{SM}_{*S}(T_D) g^{HS}_{*S}(T^{HS})} \rp^{1/3} T^{SM}.
\label{eq:Tdrift}
\eeq  
This region is shown in green in the right panel of Fig.~\ref{fig:KD}.  Tracking
the detailed temperature evolution of the hidden sector in this
region, which can involve cannibal behavior when the scalar is
sufficiently massive and long-lived \cite{Farina:2016llk}, is
interesting, but beyond the scope of this paper.

\section{Sommerfeld Enhancement}
\label{sec:SE}

When DM can interact via the long-range exchange of light mediators,
the annihilation rate can exhibit a large enhancement over the
tree-level rate, especially at low DM velocities
\cite{Sommerfeld:1931,Hisano:2002fk,Hisano:2003ec,Hisano:2004ds}.
This Sommerfeld enhancement is most pronounced when three basic
scenarios are satisfied: the dark fine-structure constant,
  $\alpha_D$, is large; the DM velocity, $v$, is small; and
the mediator is much lighter than the dark matter, $R =
\frac{m_{\phi}}{m_{DM}}\ll 1$, all of which can be realized by heavy,
thermal dark matter with a light mediator. It is common to define the
Sommerfeld enhancement through the factorized formula
\cite{Cirelli:2007xd,ArkaniHamed:2008qn,Tulin:2013teo},
 \beq
 \sigma v = S(v) \lp\sigma v\rp^{tree},
\eeq  
where $\sigma v$ is the full cross-section, $\lp \sigma v \rp^{tree}$ is the
tree-level cross-section, and $S(v)$ is the velocity-dependent
Sommerfeld enhancement. 
    
To evaluate the Sommerfeld enhancement to DM annihilations, we make
use of the analytic approximation obtained by replacing
the Yukawa potential with the Hulth\'en potential
\cite{Cassel:2009wt},
\beq
 V_{\mbox{\footnotesize Yukawa}} = - \alpha_D \frac{e^{-m_\phi r}}{r} \approx
 V_{\mbox{\footnotesize Hulth\'en}} = \alpha_D \delta \frac{e^{-\delta r}}{1-
   e^{-\delta r}} ,~~~~~\mbox{ where } \delta = \frac{\pi^2 m_\phi}{6} .
\eeq
For $s$-wave DM annihilation, the Sommerfeld enhancement can then be
written as \cite{Tulin:2013teo}
\beq
S_0(\alpha_D, R,v) = \frac{2\pi \alpha_D}{v} \frac{\sinh\left[\frac{6 v}{\pi R}\right]}{\cosh\left[\frac{6 v}{\pi R}\right] - \cosh\left[\sqrt{\frac{36 v^2}{\pi^2 R^2}-\frac{24 \alpha_D}{R}}\right]}.
\label{eq:SE}
\eeq
For all choices of $\alpha_D$ and $R$, $S_0(\alpha_D, R,v)$ increases
monotonically with decreasing $v$.  For $p$-wave processes, the
Sommerfeld enhancement is \cite{Cassel:2009wt, Tulin:2013teo}
  \beq
S_1(\alpha_D, R,v) = \frac{36 v^2+\lp\pi^2 R- 6\alpha_D \rp^2}{36 v^2+(\pi^2 R)^2} S_0(\alpha_D, R,v).
\label{eq:SEp}
\eeq
These analytic 
results from the Hulth\'en potential provide a good
approximation to scattering from the true Yukawa potential
\cite{Cassel:2009wt} except in the resonant regime where
disagreements can become numerically larger.  

We incorporate the Sommerfeld effect in two different ways.  First,
the Sommerfeld enhancement can become important during freezeout,
especially at large DM masses. In this case, the increased annihilation 
from the Sommerfeld effect will reduce the size of the coupling constant 
necessary to achieve the correct relic abundance.  We include 
Sommerfeld enhancement during freezeout by numerically solving the 
equation
 \beq
\label{eq:sefo}
 \frac{dY}{dx} =-\frac 1{x^2}\frac{s(m_\chi)}{H(m_\chi)} \vev{\sigma v} \lp Y^2 - Y^2_{eq}(x) \rp,
 \eeq
using an
adaptive fifth-order Cash-Karp Runge-Kutta technique.  Here 
$ \vev{\sigma v}_{s} = \sigma_0 \vev{S_0(v)}$ and 
$ \vev{\sigma v}_{p} = \sigma_1v_c(x)^2 \vev{S_1(v)}$, 
where $Y\equiv n_\chi/s$, $x\equiv m_\chi/T$, and $v_c(x) =
\sqrt{\frac{6}{x}}$.
We further use the approximations
 \beq
 \vev{S_0(v)}\approx  S_0(v_c(x) ) \; \mbox{ and }  \vev{S_1(v)}  \approx S_1(v_c(x) ).  
 \label{eq:SEapprox}
\eeq
In Figs.~\ref{fig:VectorIndirect}, \ref{fig:DDVector},
\ref{fig:MasterVector}, \ref{fig:DDScalar}, and \ref{fig:MasterScalar}
we will denote (in brown) the region where thermal freezeout with and
without the inclusion of the Sommerfeld enhancement gives results for
the dark fine structure constant that disagree by more than a factor
of 2,
  \beq
\alpha_D |_{\mbox{\scriptsize no }SE} > 2 \alpha_D |_{SE}.
  \label{eq:SEregion}
\eeq
Outside of this region, the approximation used in (\ref{eq:SEapprox})
proves very accurate.  Deep within this region, the true coupling is
typically smaller than that predicted by the freezeout calculation of
(\ref{eq:sefo}), and in our approximation Sommerfeld resonances will
be improperly positioned, by an even larger amount than from the 
use of the Hulth\'en potential.  We further note that for very
large couplings and/or very near resonances, the Sommerfeld
enhancement as estimated in (\ref{eq:SE}) can violate partial wave
unitarity \cite{Blum:2016nrz}, and, again, the condition
(\ref{eq:SEregion}) reliably insulates us from this region.
 
The Sommerfeld enhancement also may greatly affect the indirect
detection of dark matter.  Both today and at the era of recombination, 
dark matter moves very slowly and the Sommerfeld enhancement 
can substantially increase the annihilation rate.  As the 
Sommerfeld enhancement decreases monotonically with increasing 
velocities, we will conservatively err on the side of assuming larger 
velocities.  Dark matter in the Milky Way have relative velocities 
on the order of $10^{-3}$, and we use the conservative value of 
\beq
v_{GC} = 1.7 \times 10^{-3},
\eeq 
which corresponds to relative velocities of 500 km/s for determining 
the Sommerfeld enhancement for AMS-02.  Dark matter in 
the smaller dwarf galaxies have characteristic velocities on the order 
of $10^{-5}$--$10^{-4}$ \cite{Bringmann:2016din}.  In our treatment of 
dwarf galaxies, we will conservatively use a uniform relative velocity 
across dwarfs of
\beq
v_{dwarf} = 10^{-4}, 
\eeq 
in computing the Sommerfeld enhancement.  Especially in the case 
of faint dwarf spheroidal galaxies with a small half-light radius, such 
as Draco II and Segue I recently discovered by the Dark Energy 
Survey \cite{Abbott:2005bi,Bechtol:2015cbp}, this choice of 
characteristic velocity could significantly underestimate the 
Sommerfeld enhancement.  Lastly, at the time of recombination, the 
characteristic velocity of dark matter was still dictated by its 
red-shifted temperature rather than by virialization within a structure.  In 
particular, after the dark matter decouples from the thermal bath
in the early universe at $T_{KD}$, its velocity can be expressed as
\beq
v_{CMB} \approx \sqrt{\frac{6 T_{KD}}{m_\chi}} \frac{T_{CMB}}{T_{KD}} \lesssim 2 \times 10^{-7} \lp \frac{100 \mbox{ GeV}}{m_\chi}\rp^{1/2}.
\eeq 
where $T_{CMB}\sim 0.27$ eV, and we have imposed the bound 
from Lyman-alpha forest data requiring $T_{KD}\gtrsim 100$ eV 
\cite{Vogelsberger:2015gpr}; the parameter space of interest in this 
work yields values for $T_{KD}$ well above this lower 
limit.\footnote{See Ref.~\cite{Bringmann:2016ilk} for discussion of 
related models in the low $T_{KD}$ regime.}  For simplicity, in this 
work we fix
\beq
v_{CMB}=10^{-7},
\eeq
which for mass ratios of interest falls well into the
regime where the Sommerfeld enhancement does not grow any 
further with decreasing velocity.

\section{Bounds from dwarf galaxies}
\label{sec:dwarf}

 \begin{figure}[!t]
\begin{center}
\includegraphics[scale=0.6]{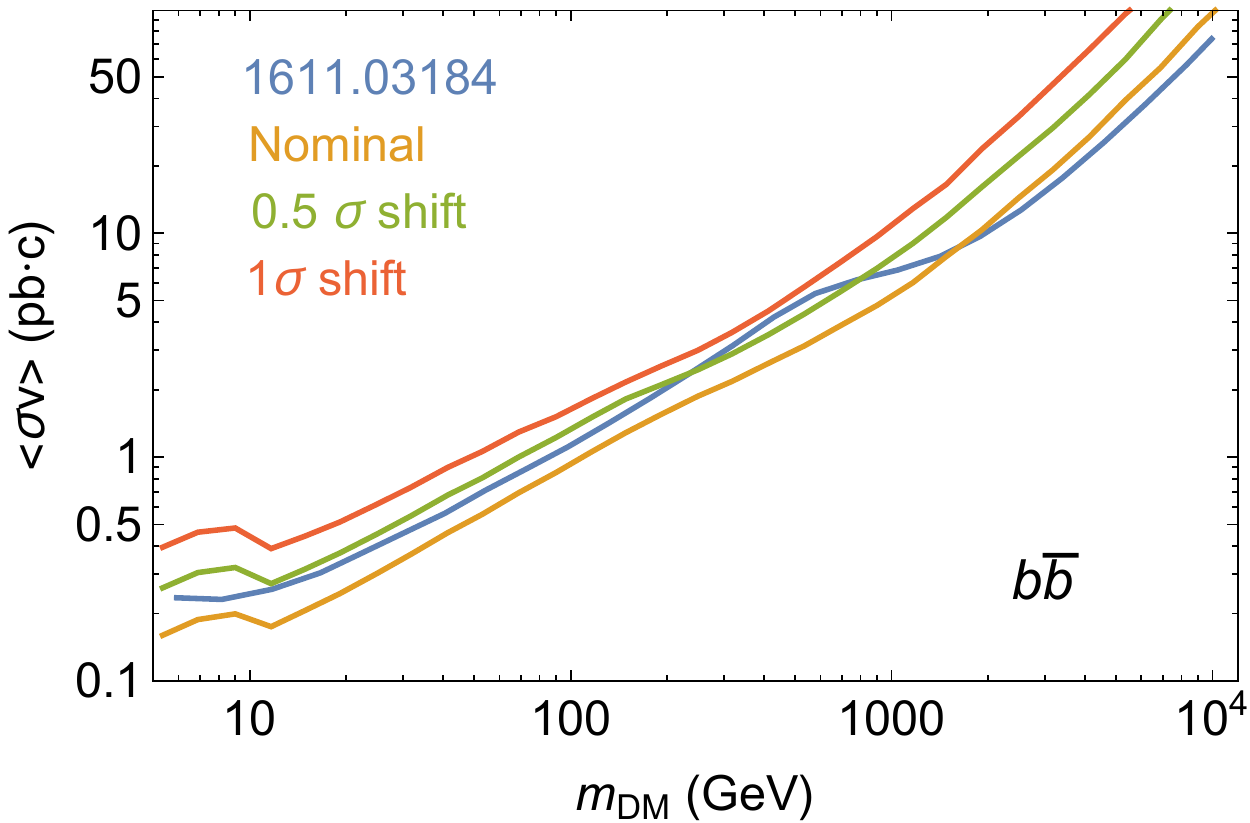}~~~
\includegraphics[scale=0.6]{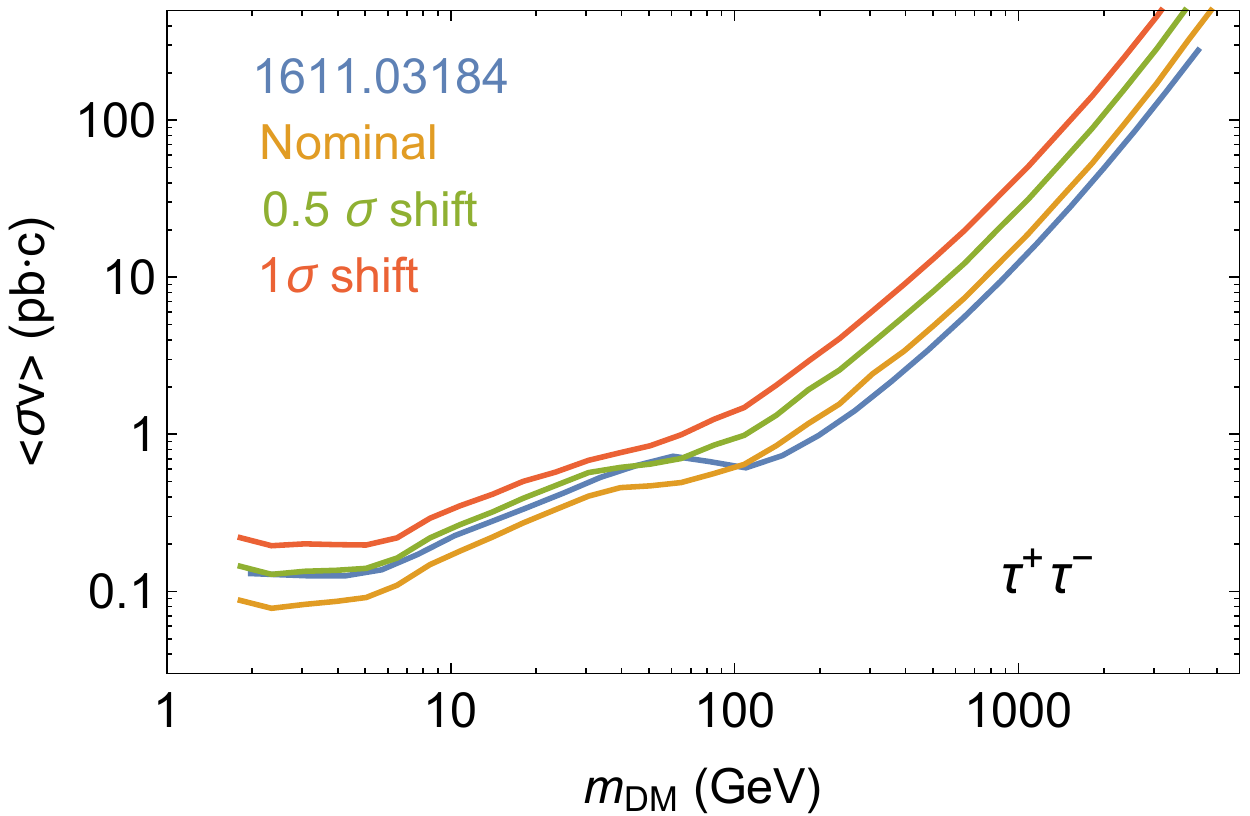}
\end{center}
\caption{ The bound on dark matter annihilation cross-section using different fraction systematic uncertainty shifts in central values compared to those from Fermi-LAT (in blue).  Given the value accurate reproduction for 0.5$\sigma$ shift, we use this to approximate a proper treatment of the correlations in systematic uncertainties.   {\bf{Left:}} $b\bar b$.  {\bf{Right:}}  $\tau^+\tau^-$.
}  
\label{fig:DwarfVal}
\end{figure}

In this Appendix we discuss the procedure we use to set limits on our
DM models from Fermi's search for DM annihilations in dwarf galaxies.
We consider the 41 dwarf galaxies within the nominal sample of
\cite{Ackermann:2015zua}.  The Fermi collaboration provides a
log-likeihood ratio (LLR) for a signal + background assumption to
background only as a function of the injected signal for each of the
41 dwarfs in each of the 24 common energy bins.  To use these LLRs in
combination to constrain a different signal model, it is necessary to
account for the correlation of the systematic uncertainty on the
$J$-factors between dwarfs.  As this information is not provided, we
model it by considering a $0.5\sigma$ downward shift from the na\"ive
central value (using $\sigma_{unmeasured}=0.6$), as this was
determined to replicate constraints on the $b\bar b$ and
$\tau^+\tau^-$ annihilation models fairly reliably at lower masses,
while slightly underestimating constraints at higher masses (see
Fig.~\ref{fig:DwarfVal}).  In particular, the region of \HSFOV\
parameter space currently excluded by Fermi dwarfs is reliably
determined by this choice.

After this shift, for each energy bin the 41 dwarf measurements are
combined to form a net LLR as a function of the signal injected into
that bin.  For each point in the $m_\chi$ vs $m_{Z_D}$ parameter
space, a dark matter annihilation process, $\chi\bar\chi \to Z_D Z_D
\to \{\mbox{all}\}$, is generated within Pythia 8 (v8223)
\cite{Sjostrand:2007gs}\footnote{For $2m_\pi< m_{mediator}< 1.6$ GeV,
  we follow \cite{Liu:2014cma} to assign the meson decay paths for
  light vector mediators.  Above 1.6 GeV, we switch to a parton level
  decay.  We further include a 3-body decay of the vector into $Z_D\to
  \ell^+\ell^-\gamma$ down from the $Z_D\to \ell^+\ell^-$ rate by a
  factor of $\alpha_{em}/4\pi$.}
and the resulting gamma ray spectrum is tabulated into the 24 energy
bins.  The population of these energy bins scales with $\vev{\sigma
  v}$.  Combining these bins into a $\chi^2$ with one d.o.f.~(the
overall DM annihilation rate) places limits on $\vev{\sigma v}$ which
are shown in Fig.~\ref{fig:VectorIndirect}.


\bibliography{DarkSectors}
\bibliographystyle{JHEP}


\end{document}